\documentclass[english]{svmult}
\usepackage[T1]{fontenc}
\usepackage[latin9]{inputenc}
\usepackage{color}
\usepackage{babel}
\usepackage{varioref}
\usepackage{prettyref}
\usepackage{float}
\usepackage{url}
\usepackage{amsmath}
\usepackage{amssymb}
\usepackage{graphicx}
\usepackage[unicode=true,
 bookmarks=true,bookmarksnumbered=true,bookmarksopen=true,bookmarksopenlevel=2,
 breaklinks=false,pdfborder={0 0 1},backref=page,colorlinks=true]
 {hyperref}

\makeatletter

\floatstyle{ruled}
\newfloat{algorithm}{tbp}{loa}[chapter]
\providecommand{\algorithmname}{Algorithm}
\floatname{algorithm}{\protect\algorithmname}

  \newenvironment{svmultproof2}{\begin{proof}}{\smartqed\qed\end{proof}}

\usepackage{mathptmx}       
\usepackage{helvet}         
\usepackage{courier}        
\usepackage{type1cm}        

\usepackage{etex}
\usepackage{bm}
\usepackage{amsfonts}
\renewcommand{\vec}[1]{\boldsymbol{#1}}
\newcommand{\mat}[1]{\vec{#1}}

\newcommand{\RR}{\mathbb{R}}

\newcommand{\NN}{\mathbb{N}}

\DeclareMathOperator{\rank}{rank}

\DeclareMathOperator{\spn}{span}
\DeclareMathOperator{\supp}{supp}
\DeclareMathOperator{\diag}{diag}

\newcommand{\ran}[1]{\mathcal{R}\left({#1}\right)}

\newcommand{\nullsp}[1]{\mathcal{N}\left({#1}\right)}

\usepackage{color} 
\usepackage{amssymb} 
\usepackage{amsmath} 
\usepackage{pstricks}

\newrefformat{fact}{Fact \ref{#1}}
\newrefformat{prop}{Proposition \ref{#1}}
\newrefformat{def}{Definition \ref{#1}}
\newrefformat{cor}{Corollary \ref{#1}}
\newrefformat{app}{Appendix \ref{#1}}
\newrefformat{sub}{Subsection \ref{#1}}
\newrefformat{alg}{Algorithm \ref{#1}}
\newrefformat{rem}{Remark \ref{#1}}

\usepackage{pgf,tikz}
\usepackage{mathrsfs}
\usetikzlibrary{arrows}

\@ifundefined{showcaptionsetup}{}{%
 \PassOptionsToPackage{caption=false}{subfig}}
\usepackage{subfig}
\makeatother

\begin{document}

\title*{On the global-local dichotomy in sparsity modeling}

\author{Dmitry Batenkov, Yaniv Romano and Michael Elad}

\institute{Dmitry Batenkov \at Department of Mathematics, Massachusetts Institute
of Technology, Cambridge, MA 02139, USA. \email{batenkov@mit.edu}
\and Yaniv Romano \at Department of Electrical Engineering, Technion
- Israel Institute of Technology, 32000 Haifa, Israel. \email{yromano@tx.technion.ac.il}
\and Michael Elad \at Department of Computer Science, Technion -
Israel Institute of Technology, 32000 Haifa, Israel. \email{elad@cs.technion.ac.il}}
\maketitle

\abstract{The traditional sparse modeling approach, when applied to inverse
problems with large data such as images, essentially assumes a sparse
model for small overlapping data patches. While producing state-of-the-art
results, this methodology is suboptimal, as it does not attempt to
model the entire global signal in any meaningful way \textendash{}
a nontrivial task by itself. In this paper we propose a way to bridge
this theoretical gap by constructing a global model from the bottom
up. Given local sparsity assumptions in a dictionary, we show that
the global signal representation must satisfy a constrained underdetermined
system of linear equations, which can be solved efficiently by modern
optimization methods such as Alternating Direction Method of Multipliers
(ADMM). We investigate conditions for unique and stable recovery,
and provide numerical evidence corroborating the theory.}

\begin{minipage}[t]{1\columnwidth}%
\end{minipage}

\abstract*{The traditional sparse modeling approach, when applied to inverse
problems with large data such as images, essentially assumes a sparse
model for small overlapping data patches. While producing state-of-the-art
results, this methodology is suboptimal, as it does not attempt to
model the entire global signal in any meaningful way \textendash{}
a nontrivial task by itself. In this paper we propose a way to bridge
this theoretical gap by constructing a global model from the bottom
up. Given local sparsity assumptions in a dictionary, we show that
the global signal representation must satisfy a constrained underdetermined
system of linear equations, which can be solved efficiently by modern
optimization methods such as Alternating Direction Method of Multipliers
(ADMM). We investigate conditions for unique and stable recovery,
and provide numerical evidence corroborating the theory.}


\global\long\def\G{\Gamma}

\global\long\def\O{\Omega}

\section{Introduction}

\subsection{The need for a new local-global sparsity theory}

The sparse representation model \cite{elad2010sparseand} provides
a powerful approach to various inverse problems in image and signal
processing such as denoising \cite{elad2006image,mairal2009non},
deblurring \cite{yu_solving_2012,dong2013nonlocally} and super-resolution
\cite{yang2010image,romano2014single}, to name a few \cite{mairal2014sparse}.
This model assumes that a signal can be represented as a sparse linear
combination of a few columns (called atoms) taken from a matrix termed
dictionary. Consecutively, given a signal, the sparse recovery of
its representation over a dictionary is called sparse-coding or pursuit.
Due to computational and theoretical aspects, when treating high dimensional
data most of the existing sparsity-inspired methods utilize local-patched-based
representations rather than the global ones, i.e. they divide a signal
into small overlapping blocks (patches), reconstruct these patches
using standard sparse recovery techniques, and subsequently average
the overlapping regions \cite{bruckstein_sparse_2009,elad2010sparseand}.
While this approach leads to highly efficient algorithms producing
state-of-the-art results, it is fundamentally limited because the
basic sparse model applies to patches only, and does not take into
account the dependencies between them.

As an attempt to tackle this flaw, methods based on the notion of
\emph{structured sparsity} \cite{eldar_block_2009,huang2010benefit,huang2011learning,kyrillidis2015structured,tropp2006algorithms}
started to appear; for example, in \cite{mairal2009non,dong2013nonlocally,romano2014single}
the observation that a patch may have similar neighbors in its surroundings
(often termed the self-similarity property) is injected to the pursuit,
leading to improved local estimations. Another possibility to consider
the dependencies between patches is to exploit the multi-scale nature
of the signals \cite{mairal2008learning,sulam2014image,papyan2016multiscale}.
A different direction is suggested by the EPLL \cite{zoran_learning_2011,sulam2015expected,papyan2016multiscale},
which encourages the patches of the final estimate (i.e., after the
application of the averaging step) to comply with the local prior.
Also, a related work \cite{romano2015patch,romano2015boosting} suggests
promoting the local estimations to agree on their shared content (the
overlap) as a way to achieve a coherent reconstruction of the signal.

Recently, an alternative to the traditional patch-based prior was
suggested in the form of the convolutional, or shift-invariant, sparse
coding (CSC) model \cite{grosse2012shift,bristow2013fast,heide2015fast,gu2015convolutional,thiagarajan2008shiftinvariant,rusu2014explicit}.
Rather than dividing the image into local patches and process each
of these independently, this approach imposes a specific structure
on the global dictionary \textendash{} a concatenation of banded circulant
matrices \textendash{} and applies a global pursuit. A thorough theoretical
analysis of this model was proposed very recently in \cite{papyan2016working_1,papyan2016working_2,vardan2016convolutional},
providing a clear understanding of its success.

The empirical success of the above algorithms indicates the great
potential of reducing the inherent gap that exists between the independent
local processing of patches and the global nature of the signal at
hand. However, a key and highly desirable part is still missing \textendash{}
a theory which would suggest how to modify the basic sparse model
to take into account mutual dependence between the patches, what approximation
methods to use, and how to efficiently design and learn the corresponding
structured dictionary.

\subsection{Content and organization of the paper}

In this paper we propose a systematic investigation of the signals
which are implicitly defined by local sparsity assumptions. A major
theme in what follows is that the presence of patch overlaps reduces
the number of degrees of freedom, which, in turn, has theoretical
and practical implications. In particular, this allows more accurate
estimates for uniqueness and stability of local sparse representations,
as well as better bounds on performance of existing sparse approximation
algorithms. Moreover, the global point of view allows for development
of new pursuit algorithms, which consist of local operation on one
hand, while also taking into account the patch overlaps on the other
hand. Some aspects of the offered theory are still incomplete, and
several exciting research direction emerge as well.

The paper is organized as follows. In \prettyref{sec:Local-global-sparsity}
we develop the basic framework for signals which are patch-sparse,
building the global model from the ``bottom up'', and discuss some
theoretical properties of the resulting model. In \prettyref{sec:Pursuit-algorithms}
we consider the questions of reconstructing the representation vector,
and of denoising a signal in this new framework. We describe ``globalized''
greedy pursuit algorithms \cite{pati1993orthogonal} for these tasks,
where the patch disagreements play a major role. We show that the
frequently used Local Patch Averaging (LPA) approach is in fact suboptimal
in this case. In \prettyref{sec:Examples} we describe several instances/classes
of the local-global model in some detail, exemplifying the preceding
definitions and results. The examples include piecewise-constant signals,
signature-type (periodic) signals, and more general bottom-up models.
In \prettyref{sec:Numerical-experiments} we present results of extensive
numerical experiments, where in particular we show that one of the
new globalized pursuits, based on the ADMM algorithm, turns out to
have superior performance in all the cases considered. We conclude
the paper in \prettyref{sec:Discussion} by discussing possible research
directions.

\section{\label{sec:Local-global-sparsity}Local-global sparsity}

We start with the local sparsity assumptions for every patch, and
subsequently provide two complimentary characterizations of the resulting
global signal space. On one hand, we show that the signals of interest
admit a global ``sparse-like'' representation with a dictionary
of convolutional type, and with additional linear constraints on the
representation vector. On the other hand, the signal space is in fact
a union of linear subspaces, where each subspace is a kernel of a
certain linear map. Finally we connect the two points of view by showing
that the original local dictionary must carry a combinatorial structure.
Concluding this section, we provide some theoretical analysis of the
properties of the resulting model, in particular uniqueness and stability
of representation. For this task, we define certain measures of the
dictionary, similar to the classical spark, coherence function, and
the Restricted Isometry Property, which take the additional dictionary
structure into account. 

\subsection{Preliminaries}
\begin{definition}[Spark of a matrix]
Given a dictionary $D\in\RR^{n\times m}$, the \emph{spark} of $D$
is defined as the minimal number of columns which are linearly dependent:
\begin{equation}
\sigma\left(D\right):=\min\left\{ j:\;\exists s\subset\left[1,\dots,m\right],\;\left|s\right|=j,\;\rank D_{s}<j\right\} .\label{eq:spark-def}
\end{equation}
\end{definition}
Clearly $\sigma\left(D\right)\leqslant n+1$.

\begin{minipage}[t]{1\columnwidth}%
\end{minipage}
\begin{definition}
Given a vector $\alpha\in\RR^{m}$, the $\ell_{0}$ pseudo-norm is
the number of nonzero elements in $\alpha$:
\[
\|\alpha\|_{0}:=\#\left\{ j:\;\alpha_{j}\neq0\right\} .
\]
\end{definition}
\begin{minipage}[t]{1\columnwidth}%
\end{minipage}
\begin{definition}
\label{def:tropp-mu1}Let $D\in\RR^{n\times m}$ be a dictionary with
normalized atoms. The $\mu_{1}$ coherence function (Tropp's Babel's
function) is defined as
\[
\mu_{1}\left(s\right):=\max_{i\in\left[1,\dots,m\right]}\max_{S\subset\left[1,\dots m\right]\setminus\left\{ i\right\} ,\;\left|S\right|=s}\sum_{j\in S}\left|\langle d_{i},d_{j}\rangle\right|.
\]
\end{definition}
\begin{minipage}[t]{1\columnwidth}%
\end{minipage}
\begin{definition}
\label{def:rip-classical}Given a dictionary $D$ as above, the Restricted
Isometry constant of order $k$ is the smallest number $\delta_{k}$
such that
\[
\left(1-\delta_{k}\right)\|\alpha\|_{2}^{2}\leqslant\|D\alpha\|_{2}^{2}\leqslant\left(1+\delta_{k}\right)\|\alpha\|_{2}^{2}
\]
for every $\alpha\in\RR^{m}$ with $\|\alpha\|_{0}\leqslant k$.
\end{definition}
\begin{minipage}[t]{1\columnwidth}%
\end{minipage}

For any matrix $M$, we denote by $\ran{M}$ the column space (range)
of $M$.

\subsection{Globalized local model}

In what follows we treat one-dimensional signals $x\in\RR^{N}$ of
length $N$, divided into $P=N$ overlapping patches of equal size
$n$ (so that the original signal is thought to be periodically extended).
The other natural choice is $P=N-n+1$, but for simplicity of derivations
we consider only the periodic case. 

So we define for each $i=1,\dots P$
\begin{equation}
R_{i}:=\begin{bmatrix}\vec{0} & \dots & \vec{0} & Id_{n\times n} & \vec{0} & \dots & \vec{0}\end{bmatrix}\in\RR^{n\times N},\label{eq:ri-def}
\end{equation}
the operator extracting $i$-th patch from the signal. 
\begin{definition}
\label{def:model-original-def}Given local dictionary $D\in\RR^{n\times m}$,
sparsity level $s<n$, signal length $N$, and the number of overlapping
patches $P$, the \emph{globalized local-sparse }model is the set
\begin{equation}
{\cal M}={\cal M}\left(D,s,P,N\right):=\left\{ x\in\RR^{N},\;R_{i}x=D\alpha_{i},\;\|\alpha_{i}\|_{0}\leqslant s\;\forall i=1,\dots,P\right\} .\label{eq:patch-sparsity-constraint}
\end{equation}
\end{definition}
This model suggests that each patch, $R_{i}x$ is assumed to have
an $s$-sparse representation $\alpha_{i}$, and this way we have
charactatized the global $x$ by describing the local nature of its
patches.

Next we derive a ``global'' characterization of ${\cal M}$. Starting
with the equations
\[
R_{i}x=D\alpha_{i},\qquad i=1,\dots,P,
\]
and using the equality $Id=\frac{1}{n}\sum_{i=1}^{P}R_{i}^{T}R_{i}$,
we have a representation
\begin{eqnarray*}
x & = & \frac{1}{n}\sum_{i=1}^{P}R_{i}^{T}R_{i}x=\sum_{i=1}^{P}\left(\frac{1}{n}R_{i}^{T}D\right)\alpha_{i}.
\end{eqnarray*}
Let the global ``convolutional'' dictionary $D_{G}$ be defined
as the horizontal concatenation of the (vertically) shifted versions
of $\frac{1}{n}D$, i.e.
\begin{equation}
D_{G}:=\left[\left(\frac{1}{n}R_{i}^{T}D\right)\right]_{i=1,\dots P}\in\RR^{N\times mP}.\label{eq:DG-def}
\end{equation}
Let $\G\in\RR^{mP}$ denote the concatenation of the local sparse
codes, i.e.
\[
\G:=\begin{bmatrix}\alpha_{1}\\
\alpha_{2}\\
\vdots\\
\alpha_{P}
\end{bmatrix}.
\]
Given a vector $\G$ as above, we will denote by $\tilde{R}_{i}$
the operator of extracting its $i$-th portion\footnote{Notice that while $R_{i}$ extracts the $i$-th patch from the signal
$x$, the operator $\tilde{R_{i}}$ extracts the representation $\alpha_{i}$
of $R_{i}x$ from $\G$. }, i.e. $\tilde{R_{i}}\G\equiv\alpha_{i}$.

Summarizing the above developments, we have the global convolutional
representation for our signal as follows:
\begin{equation}
x=D_{G}\G.\label{eq:global-representation-conv}
\end{equation}
Next, applying $R_{i}$ to both sides of \eqref{eq:global-representation-conv}
and using \eqref{eq:patch-sparsity-constraint}, we obtain
\begin{equation}
D\alpha_{i}=R_{i}x=R_{i}D_{G}\G.\label{eq:stripe-equality}
\end{equation}
Let $\O_{i}:=R_{i}D_{G}$ denote the $i$-th stripe from the global
convolutional dictionary $D_{G}$. Thus \eqref{eq:stripe-equality}
can be rewritten as
\begin{equation}
\underbrace{\begin{bmatrix}\vec{0} & \dots & \vec{0} & D & \vec{0} & \dots & \vec{0}\end{bmatrix}}_{:=Q_{i}}\G=\O_{i}\G,\label{eq:q-omega}
\end{equation}
or $\left(Q_{i}-\O_{i}\right)\G=0$. Since this is true for all $i=1,\dots,P$,
we have shown that the vector $\G$ satisfies
\[
\underbrace{\begin{bmatrix}Q_{1}-\O_{1}\\
\vdots\\
Q_{P}-\O_{P}
\end{bmatrix}}_{:=M\in\RR^{nP\times mP}}\G=0.
\]

Thus, the condition that the patches $R_{i}x$ agree on the overlaps
is equivalent to the global representation vector $\G$ residing in
the null-space of the matrix $M$.

An easy computation provides the dimension of this null-space (see
proof in \nameref{app:proof-dimkerM}), or in other words the overall
number of degrees of freedom of admissible $\G$.
\begin{lemma}
\label{lem:dim-ker-M}For any frame $D\in\RR^{n\times m}$ (i.e. a
full rank dictionary), we have
\[
\dim\ker M=N\left(m-n+1\right).
\]
\end{lemma}
Note that in particular for $m=n$ we have $\dim\ker M=N$, i.e. every
signal admits a unique representation $x=D_{G}\G$ where $\G=\left(D^{-1}R_{1}x,\dots,D^{-1}R_{P}x\right)^{T}$.
\begin{definition}
Given $\G=\left[\alpha_{1},\dots,\alpha_{P}\right]^{T}\in\RR^{mP}$,
the $\|\cdot\|_{0,\infty}$ pseudo-norm is defined by
\[
\|\G\|_{0,\infty}:=\max_{i=1,\dots,P}\|\alpha_{i}\|_{0}.
\]
\end{definition}
Thus, every signal complying with the patch-sparse model, with sparsity
$s$ for each patch, admits the following representation.
\begin{theorem}
\label{thm:globalized-characterization}Given $D,s,P$, and $N$,
the globalized local-sparse model \eqref{eq:patch-sparsity-constraint}
is equivalent to
\begin{equation}
{\cal M}=\left\{ x\in\RR^{N}:\;x=D_{G}\G,\;M\G=0,\;\|\G\|_{0,\infty}\leqslant s\right\} .\label{eq:new-characterization}
\end{equation}
\end{theorem}
\begin{svmultproof2}
If $x\in{\cal M}$ (according to \eqref{eq:patch-sparsity-constraint}),
then by the above construction $x$ belongs to the set defined by
the RHS of \eqref{eq:new-characterization} (let's call it ${\cal M}^{*}$
for the purposes of this proof only). In the other direction, assume
that $x\in{\cal M}^{*}$. Now $R_{i}x=R_{i}D_{G}\G=\Omega_{i}\G$,
and since $M\G=0$, we have $R_{i}x=Q_{i}\G=D\tilde{R_{i}}\G$. Denote
$\alpha_{i}:=\tilde{R_{i}}\G$, and so we have that $R_{i}x=D\alpha_{i}$
with $\|\alpha_{i}\|_{0}\leqslant s$, i.e. $x\in{\cal M}$ by definition. 
\end{svmultproof2}

What are the values of $s$ we are interested in? In addition to the
natural requirement that $s<n$ (as in \prettyref{def:model-original-def}),
we would like to have uniqueness of sparse representations. We say
that $\alpha_{i}$ is a \emph{minimal} representation of $x_{i}$
if $x_{i}=D\alpha_{i}$ such that the matrix $D_{\supp\alpha_{i}}$
has full rank \textendash{} and therefore the atoms participating
in the representation are linearly independent. While we treat uniqueness
in more detail in \prettyref{sub:uniqueness-stability}, at this point
we would like to restrict the discussion to minimal patch representations.
Notice that $\alpha_{i}$ might be a minimal representation but not
a unique one with minimal sparsity.
\begin{definition}
Given a signal $x\in{\cal M}$, let us denote by $\rho\left(x\right)$
the set of all locally sparse \emph{and minimal }representations of
$x$:
\[
\rho\left(x\right):=\left\{ \G\in\RR^{mP}:\;\|\G\|_{0,\infty}\leqslant s,\;x=D_{G}\G,\;M\G=0,\;D_{\supp\tilde{R_{i}}\G}\text{ is full rank}.\right\} 
\]
\end{definition}
Let us now go back to the definition \eqref{eq:patch-sparsity-constraint}.
Consider a signal $x\in{\cal M}$, and let $\G\in\rho\left(x\right)$.
Denote $s_{i}:=\supp\tilde{R_{i}}\G$. Then we have $R_{i}x\in\ran{D_{s_{i}}}$
and therefore we can write $R_{i}x=P_{s_{i}}R_{i}x$, where $P_{s_{i}}$
is the orthogonal projection operator onto $\ran{D_{s_{i}}}$. In
fact, since $D_{s_{i}}$ is full rank, we have $P_{s_{i}}=D_{s_{i}}D_{s_{i}}^{\dagger}$
where $D_{s_{i}}^{\dagger}=\left(D_{s_{i}}^{T}D_{s_{i}}\right)^{-1}D_{s_{i}}^{T}$
is the Moore-Penrose pseudoinverse of $D_{s_{i}}$. 
\begin{definition}
Given a support sequence ${\cal S}=\left(s_{1},\dots,s_{P}\right)$,
define the matrix $A_{{\cal S}}$ as follows:
\[
A_{{\cal S}}:=\begin{bmatrix}\left(I_{n}-P_{s_{1}}\right)R_{1}\\
\left(I_{n}-P_{s_{2}}\right)R_{2}\\
\vdots\\
\left(I_{n}-P_{s_{P}}\right)R_{P}
\end{bmatrix}\in\RR^{nP\times N}.
\]
\end{definition}
\begin{minipage}[t]{1\columnwidth}%
\end{minipage}

The map $A_{{\cal S}}$ measures the local patch discrepancies, i.e.
how ``far'' is each local patch from the range of a particular subset
of the columns of $D$.
\begin{definition}
Given a model ${\cal M},$ denote by $\Sigma_{{\cal M}}$ the set
of all valid supports, i.e.
\[
\Sigma_{{\cal M}}:=\left\{ \left(s_{1},\dots,s_{P}\right):\;\exists x\in{\cal M},\;\Gamma\in\rho\left(x\right)\;s.t.\;\forall i=1,\dots,P:\;s_{i}=\supp\tilde{R_{i}}\G\right\} .
\]
\end{definition}
With this notation in place, it is immediate to see that the global
signal model is a union of subspaces.
\begin{theorem}
\label{thm:uos-equiv}The global model is equivalent to the union
of subspaces
\[
{\cal M}=\bigcup_{{\cal S}\in\Sigma_{{\cal M}}}\ker A_{{\cal S}}.
\]
\end{theorem}
\begin{remark}
Contrary to the well-known Union of Subspaces model \cite{blumensath_sampling_2009,lu_theory_2007},
the subspaces $\left\{ \ker A_{{\cal S}}\right\} $ do not have in
general a sparse joint basis, and therefore our model is distinctly
different from the well-known block-sparsity model \cite{eldar_block_2009,eldar_robust_2009}.

An important question of interest is to estimate $\dim\ker A_{{\cal S}}$
for a given ${\cal S}\in\Sigma_{{\cal M}}$. One possible solution
is to investigate the ``global'' structure of the corresponding
signals (as is done in \prettyref{sub:Piecewise-constant-signals}
and \prettyref{sub:Signature-Dictionary}), while another option is
to utilize information about ``local connections'' (\prettyref{sub:local-deps}
and \prettyref{sub:arbitrary-graphs}). 
\end{remark}

\subsection{\label{sub:local-deps}Local support dependencies }

In this section we highlight the importance of the local connections
(briefly mentioned above) between the neighboring patches of the signal,
and therefore between the corresponding subspaces containing those
patches. This in turn allows to characterize $\Sigma_{{\cal M}}$
as the set of all ``realizable'' paths in a certain dependency graph
derived from the dictionary $D$. This point of view allows to describe
the model ${\cal M}$ using only the intrinsic properties of the dictionary,
in contrast to \prettyref{thm:uos-equiv}.

First we show the equivalence of the condition $M\G=0$ to equality
of pairwise overlaps.
\begin{definition}
\label{def:stsb-def}Define the ``extract from top/bottom'' operators
$S_{T}\in\RR^{\left(n-1\right)\times n}$ and $S_{B}\in\RR^{\left(n-1\right)\times n}$:
\begin{align*}
S_{T(op)} & =\begin{bmatrix}I_{n-1} & \vec{0}\end{bmatrix},\quad S_{B(ottom)}=\begin{bmatrix}\vec{0} & I_{n-1}\end{bmatrix}.
\end{align*}
\end{definition}
The following result is proved in \nameref{app:proof1}.
\begin{lemma}
\label{lem:equiv-constraint}Let $\G=\left[\alpha_{1},\dots,\alpha_{P}\right]^{T}$.
Under the above definitions, the following are equivalent:

\begin{enumerate}
\item $M\G=0;$
\item For each $i=1,\dots,P,$ we have $S_{B}D\alpha_{i}=S_{T}D\alpha_{i+1}$.
\end{enumerate}
\end{lemma}
\begin{definition}
Let the matrix $M_{*}\in\RR^{\left(n-1\right)P\times mP}$ be defined
as
\[
M_{*}:=\begin{bmatrix}S_{B}D & -S_{T}D\\
 & S_{B}D & -S_{T}D\\
 &  & \ddots & \ddots\\
\\
\end{bmatrix}.
\]
\end{definition}
\begin{corollary}
The global model is equivalent to
\[
{\cal M}=\left\{ x\in\RR^{N}:\;x=D_{G}\G,\;M_{*}\G=0,\;\|\G\|_{0,\infty}\leqslant s\right\} .
\]
\end{corollary}
\begin{proposition}
\label{prop:rank-condition}Let $0\neq x\in{\cal M}$ and $\Gamma\in\rho\left(x\right)$
with $\supp\G=\left(s_{1},\dots,s_{P}\right)$. Then for $i=1,\dots,P$
\begin{equation}
\rank\left[S_{B}D_{s_{i}}\;-S_{T}D_{s_{i+1}}\right]<\left|s_{i}\right|+\left|s_{i+1}\right|\leqslant2s,\label{eq:basic-rank-inequality}
\end{equation}
where by definition $\rank\emptyset=-\infty$.
\end{proposition}
\begin{svmultproof2}
$x\in{\cal M}$ implies by \prettyref{lem:equiv-constraint} that
for every $i=1,\dots P$ 
\[
\left[S_{B}D\quad-S_{T}D\right]\begin{bmatrix}\alpha_{i}\\
\alpha_{i+1}
\end{bmatrix}=0.
\]
But
\[
\left[S_{B}D\quad-S_{T}D\right]\begin{bmatrix}\alpha_{i}\\
\alpha_{i+1}
\end{bmatrix}=\left[S_{B}D_{s_{i}}\quad-S_{T}D_{s_{i+1}}\right]\begin{bmatrix}\alpha_{i}|s_{i}\\
\alpha_{i+1}|s_{i+1}
\end{bmatrix}=0,
\]
and therefore the matrix $\left[S_{B}D_{s_{i}}\quad-S_{T}D_{s_{i+1}}\right]$
must be rank-deficient. Note in particular that the conclusion still
holds if one (or both) of the $\left\{ s_{i},s_{i+1}\right\} $ is
empty.
\end{svmultproof2}

The preceding result suggests a way to describe all the supports in
$\Sigma_{{\cal M}}$ .
\begin{definition}
\label{def:thegraph}Given a dictionary $D$, we define an abstract
directed graph ${\cal G}_{D,s}=\left(V,E\right)$, with the vertex
set
\[
V=\left\{ \left(i_{1},\dots,i_{k}\right)\subset\left[1,\dots,m\right]:\quad\rank D_{i_{1},\dots,i_{k}}=k<n\right\} ,
\]
and the edge set
\[
E=\biggl\{\left(s_{1},s_{2}\right)\in V\times V:\quad\rank\left[S_{B}D_{s_{1}}\quad-S_{T}D_{s_{2}}\right]<\min\left\{ n-1,\left|s_{1}\right|+\left|s_{2}\right|\right\} \biggr\}.
\]
In particular, $\emptyset\in V$ and $\left(\emptyset,\emptyset\right)\in E$
with $\rank\left[\emptyset\right]:=-\infty$.
\end{definition}
\begin{remark}
It might be impossible to compute ${\cal G}_{D,s}$ in practice. However
we set this issue aside for now, and only explore the theoretical
ramifications of its properties.
\end{remark}
\begin{definition}
The set of all directed paths of length $P$ in ${\cal G}_{D,s}$,
not including the self-loop $\underbrace{\left(\emptyset,\emptyset,\dots\emptyset\right)}_{\times P}$,
is denoted by ${\cal C_{G}}\left(P\right)$.
\end{definition}
\begin{minipage}[t]{1\columnwidth}%
\end{minipage}
\begin{definition}
A path ${\cal S}\in{\cal C_{G}}\left(P\right)$ is called \emph{realizable}
if $\dim\ker A_{{\cal S}}>0$. The set of all realizable paths in
${\cal C_{G}}\left(P\right)$ is denoted by ${\cal R_{G}}\left(P\right)$.
\end{definition}
Thus we have the following result.
\begin{theorem}
\label{thm:graph-necessary-cond}Suppose $0\neq x\in{\cal M}$. Then
\begin{enumerate}
\item Every representation $\G=\left(\alpha_{i}\right)_{i=1}^{P}\in\rho\left(x\right)$
satisfies $\supp\G\in{\cal C_{G}}\left(P\right)$, and therefore 
\begin{equation}
\Sigma_{{\cal M}}\subseteq{\cal R_{G}}\left(P\right).\label{eq:containment-sigmaM}
\end{equation}
\item The model ${\cal M}$ can be characterized ``intrinsically'' by
the dictionary as follows:
\begin{equation}
{\cal M}=\bigcup_{{\cal S}\in{\cal R_{G}}\left(P\right)}\ker A_{{\cal S}}.\label{eq:charact-paths}
\end{equation}
\end{enumerate}
\end{theorem}
\begin{svmultproof2}
Let $\supp\G=\left(s_{1},\dots,s_{P}\right)$ with $s_{i}=\supp\alpha_{i}$
if $\alpha_{i}\neq\vec{0}$, and $s_{i}=\emptyset$ if $\alpha_{i}=\vec{0}$.
Then by \prettyref{prop:rank-condition} we must have that 
\[
\rank\left[S_{B}D_{s_{i}}\;-S_{T}D_{s_{i+1}}\right]<\left|s_{i}\right|+\left|s_{i+1}\right|\leqslant2s.
\]
Furthermore, since $\G\in\rho\left(x\right)$ we must have that $D_{s_{i}}$
is full rank for each $i=1,\dots,P$. Thus $\left(s_{i},s_{i+1}\right)\in{\cal G}_{D,s}$,
and so $\supp\G\in{\cal R_{G}}\left(P\right)$. Since by assumption
$\supp\G\in\Sigma_{{\cal M}}$, this proves \eqref{eq:containment-sigmaM}. 

To show \eqref{eq:charact-paths}, notice that if $\supp\Gamma\in{\cal R_{G}}\left(P\right)$,
then for every $x\in\ker A_{\supp\G}$ we have $R_{i}x=P_{s_{i}}R_{i}x$,
i.e. $R_{i}x=D\alpha_{i}$ for some $\alpha_{i}$ with $\supp\alpha_{i}\subseteq s_{i}$.
Clearly in this case $\left|\supp\alpha_{i}\right|\leqslant s$ and
therefore $x\in{\cal M}$. The other direction of \eqref{eq:charact-paths}
follows immediately from the definitions.
\end{svmultproof2}

\begin{definition}
The dictionary $D$ is called ``$\left(s,P\right)$-good'' if 
\[
\left|{\cal R_{G}}\left(P\right)\right|>0.
\]
\end{definition}
\begin{theorem}
\label{thm:measure-zero-for-dicts}The set of ``$\left(s,P\right)$-good''
dictionaries has measure zero in the space of all $n\times m$ matrices.
\end{theorem}
\begin{svmultproof2}
Every low rank condition defines a finite number of algebraic equations
on the entries of $D$ (given by the vanishing of all the $2s\times2s$
minors of $\begin{bmatrix}S_{B}D_{s_{i}}, & S_{T}D_{s_{j}}\end{bmatrix}$
). Since the number of possible graphs is finite (given fixed $n,m$
and $s$), the resulting solution set is a finite union of semi-algebraic
sets of low dimension, and hence has measure zero.
\end{svmultproof2}

The above considerations suggest that the good dictionaries are hard
to come by; we provide explicit constructions in \prettyref{sec:Examples}.

Now suppose the graph ${\cal G}$ is known (or can be easily constructed).
Then this gives a simple procedure to generate signals from ${\cal M}$,
presented in \prettyref{alg:gen-sig-model}.

\begin{algorithm}
\begin{enumerate}
\item Construct a path ${\cal S}\in{\cal C_{G}}\left(P\right)$.
\item Construct the matrix $A_{{\cal S}}$.
\item Find a nonzero vector in $\ker A_{{\cal S}}$.
\end{enumerate}
\caption{Constructing a signal from ${\cal M}$ via ${\cal G}$}

\label{alg:gen-sig-model}
\end{algorithm}

An interesting question arises: given ${\cal S}\in{\cal C_{G}}\left(P\right)$,
can we say something about $\dim\ker A_{{\cal S}}$? In particular,
when is it strictly positive (i.e. when ${\cal S}\in{\cal R_{G}}\left(P\right)$?)
While in general the question seems to be difficult, in some special
cases this number can be estimated using only the properties of the
local connections $\left(s_{i},s_{i+1}\right)$, by essentially counting
the additional ``degrees of freedom'' when moving from patch $i$
to patch $i+1$. We discuss this in more details in \prettyref{sub:arbitrary-graphs}
(in particular see \prettyref{prop:dimkerAs-bound-arbitrary-graph}),
while here we show the following easy result.
\begin{proposition}
\label{prop:dimkerA}For every ${\cal S}\in{\cal R_{G}}\left(P\right)$
we have
\[
\dim\ker A_{{\cal S}}=\dim\ker M_{*}^{\left({\cal S}\right)}.
\]
\end{proposition}
\begin{svmultproof2}
Notice that
\[
\ker A_{{\cal S}}=\left\{ D_{G}^{\left({\cal S}\right)}\G_{{\cal S}},\;M_{*}^{\left({\cal S}\right)}\G_{{\cal S}}=0\right\} =im\left(D_{G}^{\left({\cal S}\right)}\bigl|_{\ker M_{*}^{\left({\cal S}\right)}}\right),
\]
and therefore $\dim\ker A_{{\cal S}}\leqslant\dim\ker M_{*}^{\left({\cal S}\right)}$.
Furthermore, the map $D_{G}^{\left({\cal S}\right)}\bigl|_{\ker M_{*}^{\left({\cal S}\right)}}$
is injective, because if $D_{G}^{\left({\cal S}\right)}\G_{{\cal S}}=0$
and $M_{*}^{\left({\cal S}\right)}\G_{{\cal S}}=0$, we must have
that $D_{s_{i}}\alpha_{i}|_{s_{i}}=0$ and, since $D_{s_{i}}$ has
full rank, also $\alpha_{i}=0$. The conclusion follows.
\end{svmultproof2}

\subsection{\label{sub:uniqueness-stability}Uniqueness and stability}

Given a signal $x\in{\cal M}$, it has a globalized representation
$\G\in\rho\left(x\right)$ according to \prettyref{thm:globalized-characterization}.
When is such a representation unique, and under what conditions can
it be recovered when the signal is corrupted with noise?

In other words, we study the problem
\[
\min\|\G\|_{0,\infty}\qquad s.t.\;D_{G}\G=D_{G}\G_{0},\;M\G=0\qquad\left(P_{0,\infty}\right)
\]
and its noisy version
\[
\min\|\G\|_{0,\infty}\qquad s.t.\;\|D_{G}\G-D_{G}\G_{0}\|\leqslant\varepsilon,\;M\G=0\qquad\left(P_{0,\infty}^{\varepsilon}\right).
\]

For this task, we define certain measures of the dictionary, similar
to the classical spark, coherence function, and the Restricted Isometry
Property, which take the additional dictionary structure into account.
In general, the additional structure implies \emph{possibly} better
uniqueness as well as stability to perturbations, however it is an
open question to show they are \emph{provably }better in certain cases.

The key observation is that the global model ${\cal M}$ imposes a
constraint on the allowed local supports.
\begin{definition}
Denote the set of allowed local supports by
\[
{\cal T}:=\left\{ \tau:\;\exists\left(s_{1},\dots,\tau,\dots,s_{P}\right)\in\Sigma_{{\cal M}}\right\} .
\]
\end{definition}
Recall the definition of the spark \eqref{eq:spark-def}. Clearly
$\sigma\left(D\right)$ can be equivalently rewritten as
\begin{equation}
\sigma\left(D\right)=\min\left\{ j:\;\exists s_{1},s_{2}\subset\left[1,\dots,m\right],\;\left|s_{1}\cup s_{2}\right|=j,\;\rank D_{s_{1}\cup s_{2}}<j\right\} .\label{eq:spark-rewrite}
\end{equation}

\begin{definition}
The \emph{globalized spark }$\sigma^{*}\left(D\right)$ is
\begin{equation}
\sigma^{*}\left(D\right):=\min\left\{ j:\;\exists s_{1},s_{2}\in{\cal T},\;\left|s_{1}\cup s_{2}\right|=j,\;\rank D_{s_{1}\cup s_{2}}<j\right\} .\label{eq:genspark-def}
\end{equation}
\end{definition}
The following proposition is immediate by comparing \eqref{eq:spark-rewrite}
with \eqref{eq:genspark-def}.
\begin{proposition}
$\sigma^{*}\left(D\right)\geqslant\sigma\left(D\right).$
\end{proposition}
The globalized spark provides a uniqueness result in the spirit of
\cite{donoho2003optimally}.
\begin{theorem}[Uniqueness]
Let $x\in{\cal M}\left(D,s,N,P\right)$. If $\exists\G\in\rho\left(x\right)$
for which $\|\G\|_{0,\infty}<\frac{1}{2}\sigma^{*}\left(D\right)$
(i.e. it is a sufficiently sparse solution of $P_{0,\infty}$), then
it is the unique solution (and so $\rho\left(x\right)=\left\{ \G\right\} $).
\end{theorem}
\begin{svmultproof2}
Suppose that $\exists\G_{0}\in\rho\left(x\right)$ which is different
from $\G$. Put $\G_{1}:=\G-\G_{0}$, then $\|\G_{1}\|_{0,\infty}<\sigma^{*}\left(D\right)$
, while $D_{G}\G_{1}=0$ and $M\G_{1}=0$. Denote $\beta_{j}:=\tilde{R_{j}}\G_{1}$.
By assumption, there exists an index $i$ for which $\beta_{i}\neq0$,
but we must have $D\beta_{j}=0$ for every $j$, and therefore $D_{\supp\beta_{i}}$
must be rank deficient \textendash{} contradicting the fact that $\|\beta_{i}\|<\sigma^{*}\left(D\right)$.
\end{svmultproof2}

In classical sparsity, we have the bound
\begin{equation}
\sigma\left(D\right)\geqslant\min\left\{ s:\;\mu_{1}\left(s-1\right)\geqslant1\right\} ,\label{eq:spark-coherence-classical}
\end{equation}
where $\mu_{1}$ is given by \prettyref{def:tropp-mu1}. In a similar
fashion, the globalized spark $\sigma^{*}$ can be bounded by an appropriate
analog of ``coherence'' \textendash{} however, computing this new
coherence appears to be in general intractable.
\begin{definition}
Given the model ${\cal M}$, we define the following globalized coherence
function
\[
\mu_{1}^{*}\left(s\right):=\max_{S\in{\cal T}\cup{\cal T},\left|S\right|=s}\max_{j\in S}\sum_{k\in S\setminus\left\{ j\right\} }\left|\langle d_{j},d_{k}\rangle\right|,
\]
where ${\cal T}\cup{\cal T}:=\left\{ s_{1}\cup s_{2}:\;s_{1},s_{2}\in{\cal T}\right\} .$
\end{definition}
\begin{theorem}
\label{thm:glob-spark-coh}The globalized spark $\sigma^{*}$ can
be bounded by the globalized coherence as follows\footnote{In general $\min\left\{ s:\;\mu_{1}^{*}\left(s-1\right)\geqslant1\right\} \neq\max\left\{ s:\;\mu_{1}^{*}\left(s\right)<1\right\} $
because the function $\mu_{1}^{*}$ need not be monotonic.}:
\[
\sigma^{*}\left(D\right)\geqslant\min\left\{ s:\;\mu_{1}^{*}\left(s\right)\geqslant1\right\} .
\]
\end{theorem}
\begin{svmultproof2}
Following closely the corresponding proof in \cite{donoho2003optimally},
assume by contradiction that
\[
\sigma^{*}\left(D\right)<\min\left\{ s:\;\mu_{1}^{*}\left(s\right)\geqslant1\right\} .
\]
Let $s^{*}\in{\cal T}\cup{\cal T}$ with $\left|s^{*}\right|=\sigma^{*}\left(D\right)$
for which $D_{s^{*}}$ is rank-deficient. Then the restricted Gram
matrix $G:=D_{s^{*}}^{T}D_{s^{*}}$ must be singular. On the other
hand, $\mu_{1}^{*}\left(\left|s^{*}\right|\right)<1$, and so in particular
\[
\max_{j\in s^{*}}\sum_{k\in s^{*}\setminus\left\{ j\right\} }\left|\langle d_{j},d_{k}\rangle\right|<1.
\]
But that means that $G$ is diagonally dominant and therefore $\det G\neq0$,
a contradiction.
\end{svmultproof2}

We see that  $\mu_{1}^{*}\left(s+1\right)\leqslant\mu_{1}\left(s\right)$
since the outer maximization is done on a smaller set. Therefore,
in general the bound of \prettyref{thm:glob-spark-coh} appears to
be sharper than \eqref{eq:spark-coherence-classical}. 

A notion of globalized RIP can also be defined as follows.
\begin{definition}
The globalized RIP constant of order $k$ associated to the model
${\cal M}$ is the smallest number $\delta_{k,{\cal M}}$ such that
\[
\left(1-\delta_{k,{\cal M}}\right)\|\alpha\|_{2}^{2}\leqslant\|D\alpha\|_{2}^{2}\leqslant\left(1+\delta_{k,{\cal M}}\right)\|\alpha\|_{2}^{2}
\]
for every $\alpha\in\RR^{m}$ with $\supp\alpha\in{\cal T}$.
\end{definition}
Immediately one can see the following (recall \prettyref{def:rip-classical}).
\begin{proposition}
The globalized RIP constant is upper bounded by the standard RIP constant:
\[
\delta_{k,{\cal M}}\leqslant\delta_{k}.
\]
\end{proposition}
\begin{minipage}[t]{1\columnwidth}%
\end{minipage}
\begin{definition}
The generalized RIP constant of order $k$ associated to signals of
length $N$ is the smallest number $\delta_{k}^{\left(N\right)}$
such that
\[
\left(1-\delta_{k}^{\left(N\right)}\right)\|\G\|_{2}^{2}\leqslant\|D_{G}\G\|_{2}^{2}\leqslant\left(1+\delta_{k}^{\left(N\right)}\right)\|\G\|_{2}^{2}
\]
for every $\G\in\RR^{mN}$ satisfying $M\G=0,\;\|\G\|_{0,\infty}\leqslant k$.
\end{definition}
\begin{proposition}
We have
\[
\delta_{k}^{\left(N\right)}\leqslant\frac{\delta_{k,{\cal M}}+\left(n-1\right)}{n}\leqslant\frac{\delta_{k}+\left(n-1\right)}{n}.
\]
\end{proposition}
\begin{svmultproof2}
Obviously it is enough to show only the leftmost inequality. If $\G=\left(\alpha_{i}\right)_{i=1}^{N}$
and $\|\G\|_{0,\infty}\leqslant k$, this gives $\|\alpha_{i}\|_{0}\leqslant k$
for all $i=1,\dots,P$. Further, setting $x:=D_{G}\G$ we clearly
have $\G\in\rho\left(x\right)$ and so $\supp\G\in\Sigma_{{\cal M}}$.
Thus $\supp\alpha_{i}\in{\cal T}$, and therefore
\[
\left(1-\delta_{k,{\cal M}}\right)\|\alpha_{i}\|_{2}^{2}\leqslant\|D\alpha_{i}\|_{2}^{2}\leqslant\left(1+\delta_{k,{\cal M}}\right)\|\alpha_{i}\|_{2}^{2}.
\]

By \prettyref{cor:equiv-objective} we know that for every $\G$ satisfying
$M\G=0$, we have
\[
\|D_{G}\G\|_{2}^{2}=\frac{1}{n}\sum_{i=1}^{N}\|D\alpha_{i}\|_{2}^{2}.
\]

Now for the lower bound,
\begin{align*}
\|D_{G}\G\|_{2}^{2} & \geqslant\frac{1-\delta_{k,{\cal M}}}{n}\sum_{i=1}^{N}\|\alpha_{i}\|_{2}^{2}=\left(1-1+\frac{1-\delta_{k,{\cal M}}}{n}\right)\|\G\|_{2}^{2}\\
 & =\left(1-\frac{\delta_{k,{\cal M}}+\left(n-1\right)}{n}\right)\|\G\|_{2}^{2}.
\end{align*}
For the upper bound,
\begin{align*}
\|D_{G}\G\|_{2}^{2} & \leqslant\frac{1+\delta_{k,{\cal M}}}{n}\sum_{i=1}^{N}\|\alpha_{i}\|_{2}^{2}<\left(1+\frac{\delta_{k,{\cal M}}+1}{n}\right)\|\G\|_{2}^{2}\\
 & \leqslant\left(1+\frac{\delta_{k,{\cal M}}+\left(n-1\right)}{n}\right)\|\G\|_{2}^{2}.
\end{align*}
\end{svmultproof2}

\begin{theorem}[Uniqueness and stability of $P_{0,\infty}$ via RIP]
Suppose that $\delta_{2s}^{\left(N\right)}<1$, and suppose further
that $x=D_{G}\G_{0}$ with $\|\G_{0}\|_{0,\infty}=s$ and $\|D_{G}\G_{0}-x\|_{2}\leqslant\varepsilon$.
Then every solution $\hat{\G}$ of the noise-constrained $P_{0,\infty}^{\varepsilon}$
problem
\[
\hat{\G}\leftarrow\arg\min_{\G}\|\G\|_{0,\infty}\;s.t.\;\|D_{G}\G-x\|\leqslant\varepsilon,\;M\G=0
\]
satisfies
\[
\|\hat{\G}-\G_{0}\|_{2}^{2}\leqslant\frac{4\varepsilon^{2}}{1-\delta_{2s}^{\left(N\right)}}.
\]
In particular, $\G_{0}$ is the unique solution of the noiseless $P_{0,\infty}$
problem.
\end{theorem}
\begin{svmultproof2}
Immediate using the definition of the globalized RIP:
\begin{eqnarray*}
\|\hat{\G}-\G_{0}\|_{2}^{2} & < & \frac{1}{1-\delta_{2s}^{\left(N\right)}}\|D_{G}\left(\hat{\G}-\G_{0}\right)\|_{2}^{2}\leqslant\frac{1}{1-\delta_{2s}^{\left(N\right)}}\left(\|D_{G}\hat{\G}-x\|_{2}+\|D_{G}\G_{0}-x\|_{2}\right)^{2}\\
 & \leqslant & \frac{4\varepsilon^{2}}{1-\delta_{2s}^{\left(N\right)}}.
\end{eqnarray*}
\end{svmultproof2}

\section{Pursuit algorithms\label{sec:Pursuit-algorithms}}

In this section we consider the problem of efficient projection onto
the model ${\cal M}$. First we treat the ``oracle'' setting, i.e.
when the supports of the local patches (and therefore of the global
vector $\G$) is known. We show that the patch averaging (LPA) method
is not a good projector, however repeated application of it does achieve
the desired result.

For the non-oracle setting, we consider ``local'' and ``globalized''
pursuits. The former type does not use any dependencies between the
patches, and tries to reconstruct the supports $\alpha_{i}$ completely
locally, using standard methods such as OMP \textendash{} and as we
demonstrate, it can be guaranteed to succeed in more cases than the
standard analysis would imply. However a possibly better alternative
exists, namely a ``globalized'' approach with the patch disagreements
as a major driving force. 

\subsection{Global (oracle) projection, local patch averaging (LPA) and the local-global
gap\label{sub:Global-(oracle)-projection}}

Here we briefly consider the question of efficient projection onto
the subspace $\ker A_{{\cal S}}$, given ${\cal S}$. 

As customary in the literature \cite{candes2006modernstatistical},
the projector onto $\ker A_{{\cal S}}$ can be called \emph{an oracle.
}In effect, we would like to compute
\begin{equation}
x_{G}\left(y,{\cal S}\right):=\arg\min_{x}\|y-x\|_{2}^{2}\qquad s.t.\;A_{{\cal S}}x=0,\label{eq:oracle-proj-def}
\end{equation}
given $y\in\RR^{N}$.

To make things concrete, let us assume the standard Gaussian noise
model:
\begin{equation}
y=x+{\cal N}\left(0,\sigma^{2}I\right).\label{eq:gaussian-noise-model}
\end{equation}

The following is well-known.
\begin{proposition}
\label{prop:oracle-perf}In the Gaussian noise model \eqref{eq:gaussian-noise-model},
the performance of the oracle estimator \eqref{eq:oracle-proj-def}
is
\[
MSE\left(x_{G}\right)=\left(\dim\ker A_{{\cal S}}\right)\sigma^{2}.
\]
\end{proposition}
Let us now turn to the LPA method. The (linear part of) LPA is the
solution to the minimization problem:
\[
\hat{x}=\arg\min_{x}\sum_{i=1}^{P}\left\Vert R_{i}x-P_{s_{i}}R_{i}y\right\Vert _{2}^{2},
\]
where $y$ is the noisy signal. This has a closed-form solution
\begin{align}
\hat{x}_{LPA} & =\left(\sum_{i}R_{i}^{T}R_{i}\right)^{-1}\left(\sum_{i}R_{i}^{T}P_{s_{i}}R_{i}\right)y=\underbrace{\left(\frac{1}{n}\sum_{i}R_{i}^{T}P_{s_{i}}R_{i}\right)}_{:=M_{A}}y.\label{eq:patch-averaging}
\end{align}

\begin{figure}
\centering{}\includegraphics[height=0.3\textheight]{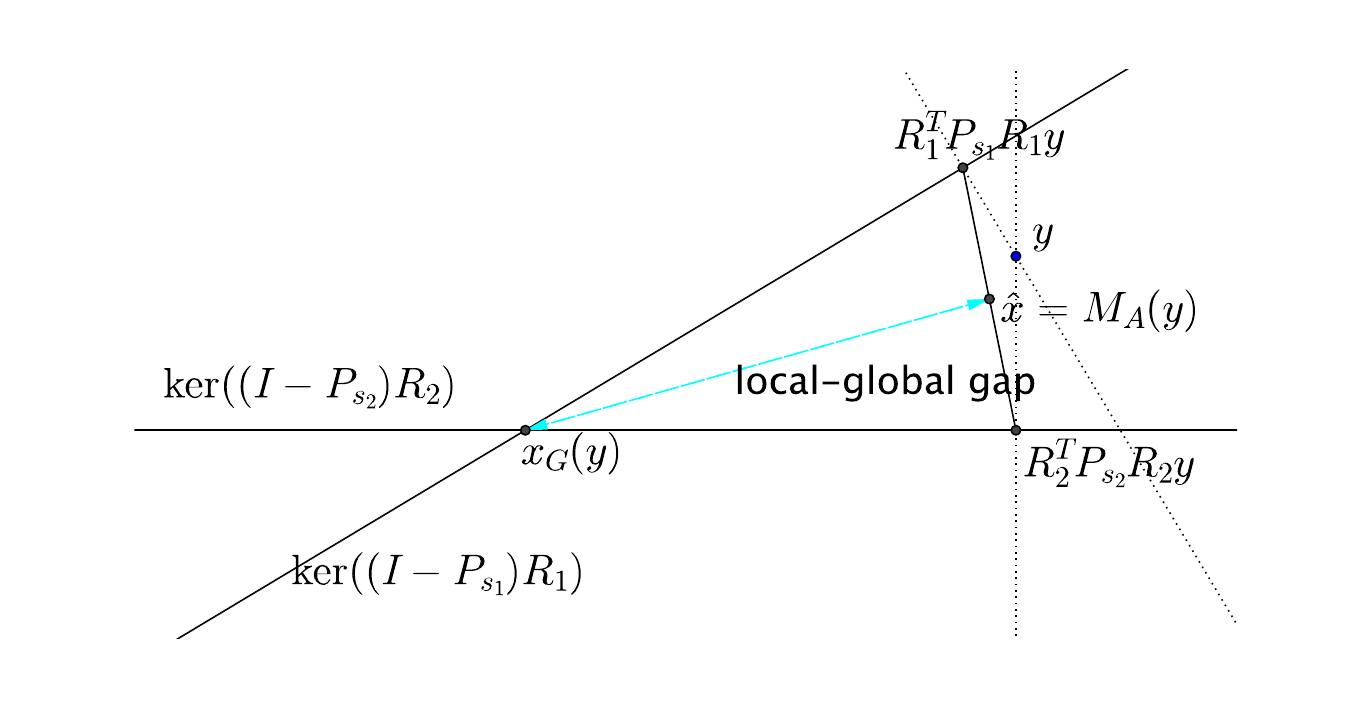}\caption{The local-global gap, oracle setting. Illustration for the case $P=2$. }
\label{fig:local-global-oracle-gap}
\end{figure}

Again, the following fact is well-established.
\begin{proposition}
In the Gaussian noise model \eqref{eq:gaussian-noise-model}, the
performance of the averaging estimator \eqref{eq:patch-averaging}
is
\[
MSE\left(\hat{x}_{LPA}\right)=\sigma^{2}\sum_{i=1}^{N}\lambda_{i},
\]
where $\left\{ \lambda_{1},\dots,\lambda_{N}\right\} $ are the eigenvalues
of $M_{A}M_{A}^{T}$.
\end{proposition}
Thus, there exists a \emph{local-global gap} in the oracle setting,
illustrated in \prettyref{fig:local-global-oracle-gap}. In \prettyref{sub:Piecewise-constant-signals}
we estimate this gap for a specific case of piecewise-constant signals.

The following result is proved in \nameref{app:proof-of-lpa-iters}.
\begin{theorem}
\label{thm:lpa-iters}Iterations of \eqref{eq:patch-averaging} converge
to $x_{G}$.
\end{theorem}
To conclude, \emph{the iterated LPA algorithm provides an efficient
method for computing the global oracle projection $x_{G}$.}

\subsection{Local pursuit guarantees}

Now we turn to the question of projection onto the model ${\cal M}$
when the support of $\G$ is not known. 

Here we show that the local OMP \cite{chen1989orthogonal,pati1993orthogonal}
in fact succeeds in more cases than can be predicted by the classical
unconstrained sparse model for each patch. We use the modified coherence
function (which is unfortunately intractable to compute)
\[
\eta_{1}^{*}\left(s\right):=\max_{S\in{\cal T}}\left(\max_{j\in S}\sum_{k\in S\setminus\left\{ j\right\} }\left|\langle d_{k},d_{j}\rangle\right|+\max_{j\notin S}\sum_{k\in S}\left|\langle d_{k},d_{j}\rangle\right|\right).
\]

The proof of the following theorem is very similar to proving the
guarantee for the standard OMP via the Babel function (\prettyref{def:tropp-mu1}),
see e.g. \cite[Theorem 5.14]{foucart2013amathematical} \textendash{}
and therefore we do not reproduce it here.
\begin{theorem}
\label{thm:local-omp-better-guarantee}If $\eta_{1}^{*}\left(s\right)<1$
then the local OMP will recover the supports of any $x\in{\cal M}$.
\end{theorem}
Since the modified coherence function takes the allowed local supports
into consideration, one can readily conclude that

\[
\eta_{1}^{*}\left(s\right)\leqslant\mu_{1}\left(s\right)+\mu_{1}\left(s-1\right),
\]

and therefore \prettyref{thm:local-omp-better-guarantee} gives in
general a better guarantee than the one based on $\mu_{1}$. 

\subsection{Globalized pursuits}

We now turn to consider several pursuit algorithms, aiming at solving
the $P_{0,\infty}/P_{0,\infty}^{\varepsilon}$ problems, in the globalized
model. The main question is how to project the patch supports onto
the nonconvex set $\Sigma_{{\cal M}}$.

The core idea is to relax the constraint $M_{*}\G=0,\;\|\G\|_{0,\infty}\leqslant s$,
and allow for some patch disagreements, so that the term $\|M_{*}\G_{k}\|$
is not exactly zero. Intuitive explanation is as follows: the disagreement
term ``drives'' the pursuit, and the probability of success is higher
because we only need to ``jump-start'' it with the first patch,
and then by strengthening the weight of the penalty related to this
constraint the supports will ``align'' themselves correctly. Justifying
this intuition, at least in some cases, is a future research goal. 

\subsubsection{Q-OMP}

Given $\beta>0$, we define
\[
Q_{\beta}:=\begin{bmatrix}D_{G}\\
\beta M_{*}
\end{bmatrix}.
\]

The main idea of the Q-OMP algorithm is to substitute the matrix $Q_{\beta}$
as a proxy for the constraint $M_{*}\G=0$, by plugging it as a dictionary
to the OMP algorithm. Then, given the obtained support ${\cal S}$,
as a way to ensure that this constraint is met, one can construct
the matrix $A_{{\cal S}}$ and project the signal onto the subspace
$\ker A_{{\cal S}}$ (in \prettyref{sub:Global-(oracle)-projection}
we show how such a projection can be done efficiently). The Q-OMP
algorithm is detailed in \prettyref{alg:qomp}. Let us re-emphasize
the point that various values of $\beta$ correspond to different
weightings of the model constraint $M_{*}\G=0$, and this might possibly
become useful when considering relaxed models (see \prettyref{sec:Discussion}).

\begin{algorithm}
Given: noisy signal $y$, dictionary $D$, local sparsity $s$, parameter
$\beta>0$.
\begin{enumerate}
\item Construct the matrix $Q_{\beta}$.
\item Run the OMP algorithm on the vector $Y:=\begin{bmatrix}y\\
\vec{0}
\end{bmatrix}$, with the dictionary $Q_{\beta}$ and sparsity $sN$. Obtain the
global support vector $\hat{\G}$ with $\supp\hat{\G}=\hat{{\cal S}}$.
\item Construct the matrix $A_{{\cal \hat{S}}}$ and project $y$ onto $\ker A_{\hat{{\cal S}}}$.
\end{enumerate}
\caption{The Q-OMP algorithm \textendash{} A globalized pursuit}
\label{alg:qomp}
\end{algorithm}

\subsubsection{ADMM-based approach}

In what follows we extend the above idea and develop an ADMM-type
pursuit \cite{boyd2011distributed}.

We start with the following global objective:

\[
\hat{x}\leftarrow\arg\min_{x}\|y-x\|_{2}^{2}\quad\text{s.t.}\;x=D_{G}\G,M_{*}\G=0,\;\|\G\|_{0,\infty}<K.
\]
Clearly, it is equivalent to $\hat{x}=D_{G}\hat{\G}$, where
\begin{eqnarray}
\hat{\G}\leftarrow\arg\min_{\G}\|y-D_{G}\G\|_{2}^{2}\quad s.t.\;M_{*}\G & = & 0,\;\|\G\|_{0,\infty}<K.\label{eq:noisy-pursuit-constrained-sparsity}
\end{eqnarray}

Applying \prettyref{cor:equiv-objective}, we have the following result.
\begin{proposition}
The following problem is equivalent to \eqref{eq:noisy-pursuit-constrained-sparsity}:

\begin{equation}
\begin{split}\hat{\G}\leftarrow\arg\min_{\left\{ \alpha_{i}\right\} }\sum_{i=1}^{P}\|R_{i}y-D\alpha_{i}\|_{2}^{2}\\
\textrm{s.t.}\;S_{B}D\alpha_{i}=S_{T}D\alpha_{i+1}\textrm{ and }\|\alpha_{i}\|_{0}<K\text{ \textrm{for }} & i=1,\dots,P.
\end{split}
\label{eq:local-problem}
\end{equation}
\end{proposition}
We propose to approximate solution of the nonconvex problem \eqref{eq:local-problem}
as follows. Define new variables $z_{i}$ (which we would like to
be equal to $\alpha_{i}$ eventually), and rewrite the problem in
ADMM form (here $Z$ is the concatenation of all the $z_{i}$'s):
\[
\left\{ \hat{\G},\hat{Z}\right\} \leftarrow\arg\min_{\G,Z}\sum_{i=1}^{P}\|R_{i}y-D\alpha_{i}\|_{2}^{2}\quad s.t.\;S_{B}D\alpha_{i}=S_{T}Dz_{i+1},\;\alpha_{i}=z_{i},\;\|\alpha_{i}\|_{0}<K.
\]
The constraints can be written in concise form
\[
\underbrace{\begin{bmatrix}I\\
S_{B}D
\end{bmatrix}}_{:=A}\alpha_{i}=\underbrace{\begin{bmatrix}I & 0\\
0 & S_{T}D
\end{bmatrix}}_{:=B}\begin{pmatrix}z_{i}\\
z_{i+1}
\end{pmatrix},
\]
and so globally we would have the following structure (for $N=3)$
\[
\underbrace{\begin{bmatrix}A\\
 & A\\
 &  & A
\end{bmatrix}}_{:=\tilde{A}}\begin{pmatrix}\alpha_{1}\\
\alpha_{2}\\
\alpha_{3}
\end{pmatrix}=\underbrace{\begin{bmatrix}I\\
 & S_{T}D\\
 & I\\
 &  & S_{T}D\\
 &  & I\\
S_{T}D
\end{bmatrix}}_{:=\tilde{B}}\begin{pmatrix}z_{1}\\
z_{2}\\
z_{3}
\end{pmatrix}
\]

Our ADMM-based method is defined in \prettyref{alg:ADMM}.

\begin{algorithm}
Given: noisy signal $y$, dictionary $D$, local sparsity $s$, parameter
$\rho>0$. The augmented Lagrangian is
\[
L_{\rho}\left(\left\{ \alpha_{i}\right\} ,\left\{ z_{i}\right\} ,\left\{ u_{i}\right\} \right)=\sum_{i=1}^{P}\|R_{i}y-D\alpha_{i}\|_{2}^{2}+\frac{\rho}{2}\sum_{i=1}^{P}\|A\alpha_{i}-B\begin{pmatrix}z_{i}\\
z_{i+1}
\end{pmatrix}+u_{i}\|_{2}^{2}.
\]

\begin{enumerate}
\item Repeat until convergence:

\begin{enumerate}
\item Minimization wrt $\left\{ \alpha_{i}\right\} $ is a batch-OMP:
\begin{eqnarray*}
\alpha_{i}^{k+1} & \leftarrow & \arg\min_{\alpha_{i}}\|R_{i}y-D\alpha_{i}\|_{2}^{2}+\frac{\rho}{2}\|A\alpha_{i}-B\begin{pmatrix}z_{i}^{k}\\
z_{i+1}^{k}
\end{pmatrix}+u_{i}^{k}\|_{2}^{2},\quad s.t.\|\alpha_{i}\|_{0}<K\\
\alpha_{i}^{k+1} & \leftarrow & OMP\left(\tilde{D}=\begin{bmatrix}D\\
\sqrt{\frac{\rho}{2}}A
\end{bmatrix},\;\tilde{y}_{i}^{k}=\begin{pmatrix}R_{i}y\\
\sqrt{\frac{\rho}{2}}\left(B\begin{pmatrix}z_{i}^{k}\\
z_{i+1}^{k}
\end{pmatrix}-u_{i}^{k}\right)
\end{pmatrix},K\right).
\end{eqnarray*}
\item Minimization wrt $z$ is a least squares problem with a sparse matrix,
which can be implemented efficiently:
\[
Z^{k+1}\leftarrow\arg\min_{Z}\|\tilde{A}\G^{k+1}+U^{k}-\tilde{B}Z\|_{2}^{2}
\]
\item Dual update:
\[
U^{k+1}\leftarrow\tilde{A}\G^{k+1}-\tilde{B}Z+U^{k}.
\]
\end{enumerate}
\item Compute $\hat{y}:=D_{G}\hat{\G}$.
\end{enumerate}
\caption{The ADMM-based pursuit for $P_{0,\infty}$}
\label{alg:ADMM}

\end{algorithm}

\section{\label{sec:Examples}Examples }

We now turn to present several classes of signals that belong to the
proposed globalized model, where each of these is obtained by imposing
a special structure on the local dictionary. Then, we demonstrate
how one can sample from ${\cal M}$ and generate such signals.

\subsection{\label{sub:Piecewise-constant-signals}Piecewise constant signals}

\begin{figure}
\begin{centering}
\includegraphics[width=0.3\textwidth]{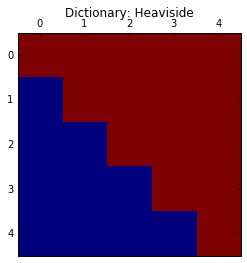}
\par\end{centering}
\caption{Heaviside dictionary $H_{4}$. Red is 1, blue is 0.}
\label{fig:heaviside4}

\end{figure}

The (unnormalized) Heaviside $n\times n$ dictionary $H_{n}$ is the
upper triangular matrix with 1's in the upper part (see \prettyref{fig:heaviside4}).
Formally, each local atom $d_{i}$ of length $n$ is expressed as
a step function, given by $d_{i}^{T}=[\mathbf{1_{i}}\ ,\ \mathbf{0_{n-i}}]^{T}$,
$1\leq i\leq n$, where $\mathbf{1_{i}}$ is a vector of ones of length
$i$. Similarly, $\mathbb{\mathbf{0}}_{n-i}$ is a zero-vector of
length $n-i$. The following property is verified by noticing that
$H_{n}^{-1}$ is the discrete difference operator.

\begin{proposition}
If a patch $x_{i}\in\RR^{n}$ has $L-1$ steps, then its (unique)
representation in the Heaviside dictionary $H_{n}$ has at most $L$
nonzeros.
\end{proposition}
\begin{corollary}
Let $x\in\RR^{N}$ be a piecewise-constant signal with at most $L-1$
steps per each segment of length $n$(in the periodic sense). Then
\[
x\in{\cal M}\left(H_{n},L,N,P=N\right).
\]
\end{corollary}
\begin{remark}
The model ${\cal M}\left(H_{n},L,N,P=N\right)$ contains also some
signals having exactly $L$ steps in a particular patch, but those
patches must have their last segment with zero height.
\end{remark}
As an example, one might synthesize signals with sparsity $\|\G\|_{0,\infty}\leq2$
according to the following scheme:
\begin{enumerate}
\item Draw at random the support of $\Gamma$ with the requirement that
the distance between the jumps within the signal will be at least
the length of a patch (this allows at most 2 non-zeros per patch,
one for the step and the second for the bias/DC).
\item Multiply each step by a random number.
\end{enumerate}
The global subspace $A_{{\cal S}}$ and the corresponding global oracle
denoiser $x_{G}$ \eqref{eq:oracle-proj-def} in the PWC model can
be explicitly described.
\begin{proposition}
\label{prop:pwc-oracle}Let $x\in\RR^{N}$ consist of $s$ constant
segments with lengths $\ell_{r}$, $r=1,\dots,s$, and let $\G$ be
the (unique) global representation of $x$ in ${\cal M}$ (i.e. $\rho\left(x\right)=\left\{ \G\right\} $).
Then
\begin{equation}
\ker A_{\supp\G}=\ker\left(I_{N}-\diag\left(B_{r}\right)_{r=1}^{s}\right),\label{eq:pwc-kerA-char}
\end{equation}
where $B_{r}=\frac{1}{\ell_{r}}\mathbf{1}_{\ell_{r}\times\ell_{r}}$.
Therefore, $\dim\ker A_{\supp\G}=s$ and $MSE\left(\hat{x}_{G}\right)=s\sigma^{2}$
under the Gaussian noise model \eqref{eq:gaussian-noise-model}.
\end{proposition}
\begin{svmultproof2}
Every signal $y\in\ker A_{\supp\G}$ has the same ``local jump pattern''
as $x$, and therefore it also has the same \emph{global} jump pattern.
That is, every such $y$ consists of $s$ constant segments with lengths
$\ell_{r}$. It is an easy observation that such signals satisfy
\[
y=\diag\left(B_{r}\right)_{r=1}^{s}y.
\]
This proves \eqref{eq:pwc-kerA-char}. It is easy to see that $\dim\ker\left(I_{\ell_{r}}-B_{r}\right)=1$,
and therefore
\[
\ker\left(I_{N}-\diag\left(B_{r}\right)_{r=1}^{s}\right)=s.
\]
The proof is finished by invoking \prettyref{prop:oracle-perf}.
\end{svmultproof2}

In other words, the global oracle is the averaging operator within
the constant segments of the signal, which is quite intuitive.

It turns out that the LPA performance (and the local-global gap) can
be accurately described by the following result. We provide an outline
of proof in \ref{sec:lpa-pwc-perf-proof}.
\begin{theorem}
\label{thm:pwc-lpa-performance}Let $x\in\RR^{N}$ consist of $s$
constant segments with lengths $\ell_{r}$, $r=1,\dots,s$, and assume
the Gaussian noise model \eqref{eq:gaussian-noise-model}. Then
\begin{enumerate}
\item There exists a function $R\left(n,\alpha\right):\NN\times\NN\to\RR^{+}$,
with $R\left(n,\alpha\right)>1$, such that
\[
MSE\left(\hat{x}_{LPA}\right)=\sigma^{2}\sum_{r=1}^{s}R\left(n,\ell_{r}\right).
\]
\item The function $R\left(n,\alpha\right)$ satisfies:
\begin{enumerate}
\item $R\left(n,\alpha\right)=1+\frac{\alpha\left(2\alpha H_{\alpha}^{\left(2\right)}-3\alpha+2\right)-1}{n^{2}}$
if $n\geqslant\alpha$, where $H_{\alpha}^{\left(2\right)}=\sum_{k=1}^{\alpha}\frac{1}{k^{2}}$;
\item $R\left(n,\alpha\right)=\frac{11}{18}+\frac{2\alpha}{3n}+\frac{6\alpha-11}{18n^{2}}$
if $n\leqslant\frac{\alpha}{2}$.
\end{enumerate}
\end{enumerate}
\end{theorem}
\begin{corollary}
The function $R\left(n,\alpha\right)$ is monotonically increasing
in $\alpha$ (with $n$ fixed) and monotonically decreasing in $n$
(with $\alpha$ fixed). Furthermore,
\begin{enumerate}
\item $\lim_{n\to\infty}R\left(n,n\right)=\frac{\pi^{2}}{3}-2\approx1.28968$;
\item $\lim_{n\to\infty}R\left(n,2n\right)=\frac{35}{18}\approx1.9444.$
\end{enumerate}
\end{corollary}
Thus, for reasonable choices of the patch size, the local-global gap
is roughly a constant multiple of the number of segments, reflecting
the global complexity of the signal.

For numerical examples of reconstructing the PWC signals using our
local-global framework, see \prettyref{sub:pwc-numerical}.

\subsection{\label{sub:Signature-Dictionary}Signature-type dictionaries}

Another type of signals that comply with our model are those represented
via a signature dictionary, which has been shown to be effective for
image restoration \cite{aharon2008sparse}. This dictionary is constructed
from a small signal, $x\in\RR^{m}$, such that its every patch (in
varying location, extracted in a cyclic fashion), $R_{i}x\in\RR^{n}$,
is a possible atom in the representation, namely $d_{i}=R_{i}x$.
As such, every consecutive pair of atoms $(i,i+1)$ is essentially
a pair of overlapping patches that satisfy $S_{B}d_{i}=S_{T}d_{i+1}$
(before normalization). The complete algorithm is presented for convenience
in \prettyref{alg:signature-dict}.

\begin{algorithm}
\begin{enumerate}
\item Choose the base signal $x\in\RR^{m}$.
\item Compute $D(x)=[R_{1}x,R_{2}x,...,R_{m}x]$, where $R_{i}$ extracts
the $i$-th patch of size $n$ in a cyclic fashion. 
\item Normalization: $\widetilde{D}(x)=[d_{1},\ldots,d_{m}]$, where $d_{i}=\frac{R_{i}x}{\left\Vert R_{i}x\right\Vert _{2}}$. 
\end{enumerate}
\caption{Constructing the signature dictionary}
\label{alg:signature-dict}
\end{algorithm}

Given $D$ as above, one can generate signals $y\in\RR^{N}$, where
$N$ is an integer multiple of $m$, with $s$ non-zeros per patch,
by the easy procedure outlined below.
\begin{enumerate}
\item Init: Construct a base signal $b\in\RR^{N}$ by replicating $x\in\RR^{m}$
$N/m$ times (note that $b$ is therefore periodic). Set $y=0$.
\item Repeat for $j=1,\dots,s$:
\begin{enumerate}
\item Shift: Circularly shift the base signal by $t_{j}$ positions, denoted
by $\mbox{shift}(b,t_{j})$, for some $t_{j}=0,1,\dots,m-1$ (drawn
at random).
\item Aggregate: $y=y+\omega_{j}\mbox{\ensuremath{\cdot}shift}(b,t_{j})$,
where $\omega$ is an arbitrary random scalar.
\end{enumerate}
\end{enumerate}
Notice that a signal constructed in this way must be periodic, as
it is easily seen that 
\[
\ker A_{{\cal S}}=\spn\left\{ \mbox{shift}\left(b,t_{i}\right)\right\} _{i=1}^{s},
\]
while the support sequence ${\cal S}$ is
\[
{\cal S}=\left(\left[t_{1},t_{2},\dots,t_{s}\right],\left[t_{1},t_{2},\dots,t_{s}\right]+1,\dots,\left[t_{1},t_{2},\dots,t_{s}\right]+N\right)\;\left(\mod m\right).
\]

Assuming that there are no additional relations between the single
atoms of $D$ except those from the above construction, the dependency
graph ${\cal G}_{D,1}$ of the resulting dictionary is easily seen
to be \emph{cyclic, }and all ${\cal S}\in\Sigma_{{\cal M}}$ are of
the above form. 

\begin{figure}
\centering\subfloat[The dictionary matrix $D$]{\includegraphics[height=0.25\paperwidth]{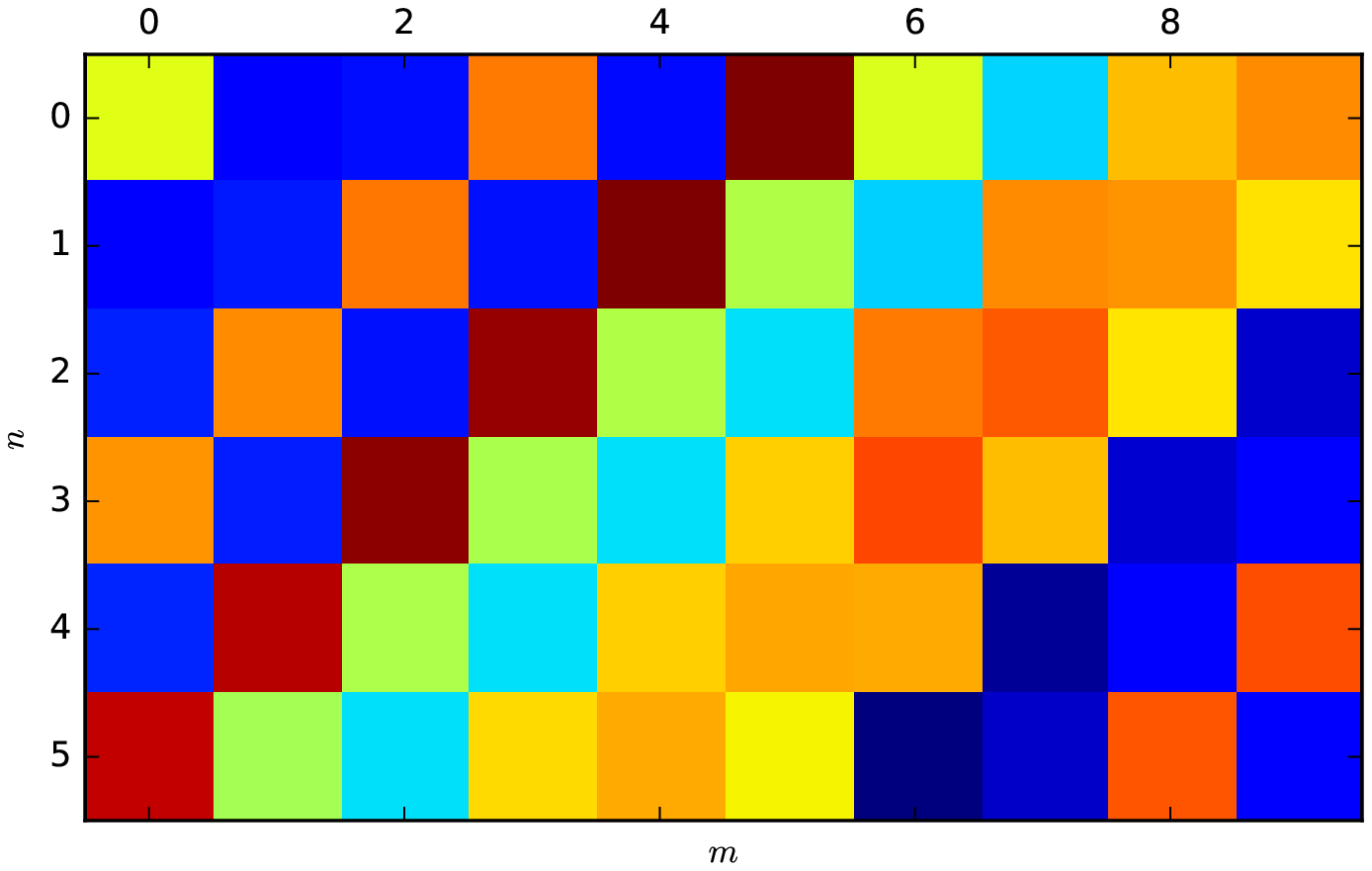}

}

\centering\subfloat[The dependency graph ${\cal G}_{D,1}$. The numerical values above
the edges are the transfer matrices (scalars) $C_{i,j}$ , satisfying
$S_{B}d_{i}=C_{i,j}S_{T}d_{j}$ (see \prettyref{sub:arbitrary-graphs}).]{\includegraphics[width=0.7\paperwidth]{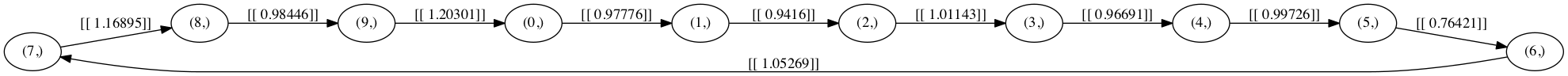}

}

\centering\subfloat[The signal $x\in\ker A_{{\cal S}}$ for ${\cal S}$ generated by $t_{1}=6$
and $s=1$, with $P=N=30$.]{\includegraphics[width=0.4\paperwidth]{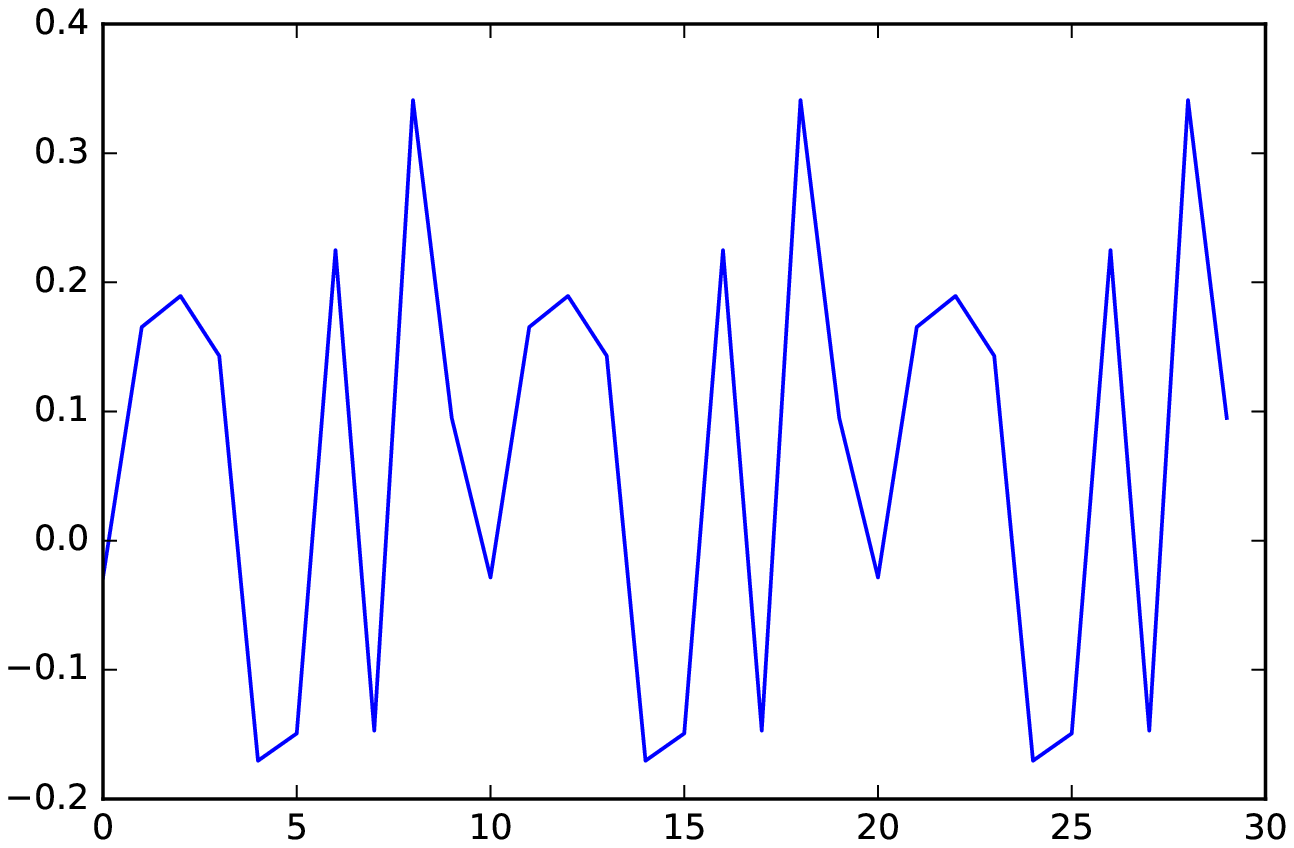}

}

\centering\subfloat[The coefficient matrix $\protect\G$ corresponding to the signal $x$
in $\left(c\right)$]{\includegraphics[width=0.4\paperwidth]{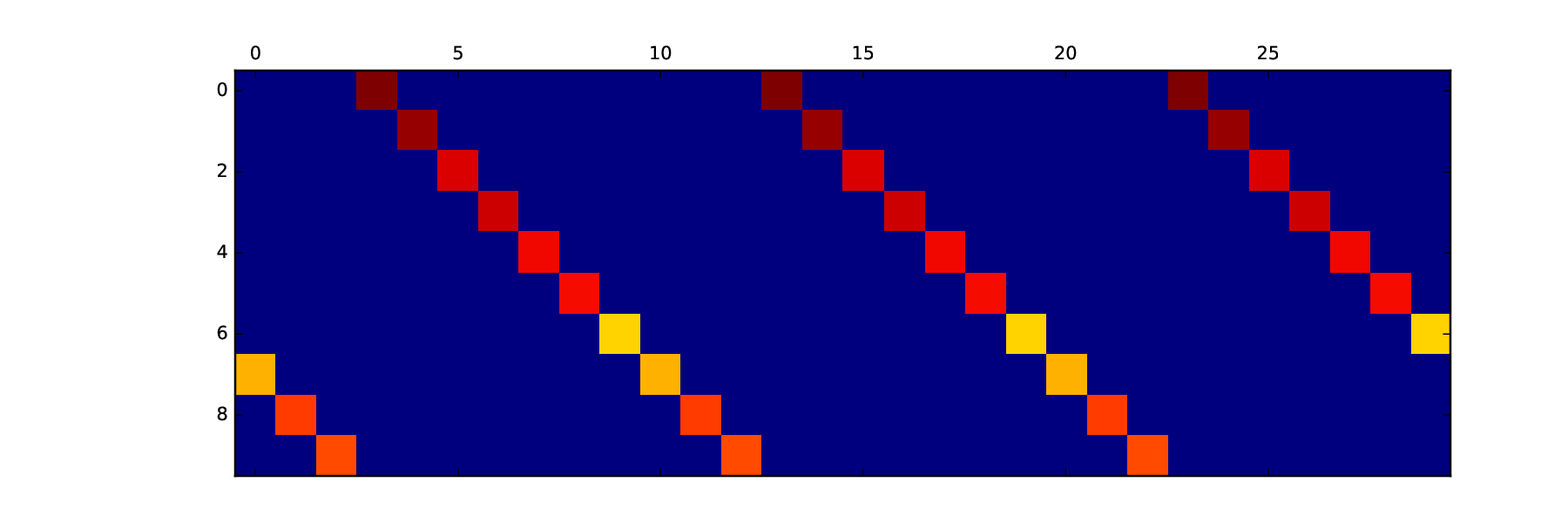}

}

\caption{An example of the signature dictionary with $n=6,\;m=10$. See \prettyref{rem:hankel-dict}.}
\label{fig:singature-dict}

\end{figure}

In \prettyref{fig:singature-dict} we give an example of a signature-type
dictionary $D$ for $\left(n,m\right)=\left(6,10\right)$, its dependency
graph ${\cal G}_{D}$, and a signal $x$ with $N=P=30$ together with
its corresponding sparse representation $\G$. 
\begin{remark}
\label{rem:hankel-dict}It might seem that every $n\times m$ Hankel
matrix such as the one shown in \prettyref{fig:singature-dict} produces
a signature-type dictionary with a nonempty signal space ${\cal M}$.
However this is not the case, because such a dictionary will usually
fail to generate signals of length larger than $n+m-1$, as its dependency
graph will not be cyclic (but rather consist of a single chain of
nodes).
\end{remark}

\subsubsection{Multi-signature dictionaries}

One can generalize the construction of \prettyref{sub:Signature-Dictionary}
and consider $s$-tuples of initial base signals $x_{i},\dots,x_{s}$,
instead of a single $x$. The desired dictionary $D$ will consist
of corresponding $s$-tuples of atoms, which are constructed from
those base signals. In order to avoid ending up with the same structure
as the case $s=1$, we also require a ``mixing'' of the atoms. The
complete procedure is outlined in \prettyref{alg:multi-sig-dict}.

\begin{algorithm}
\begin{enumerate}
\item Input: $n,m,s$ such that $s$ divides $m$. Put $r:=\frac{m}{s}$.
\item Select a signal basis matrix $X\in\RR^{r\times s}$ and $r$ nonsingular
transfer matrices $M_{i}\in\RR^{s\times s},\;i=1,\dots,r$.
\item Repeat for $i=1,\dots,r$:
\begin{enumerate}
\item Let $Y_{i}=\left[y_{i,1},\dots,y_{i,s}\right]\in\RR^{n\times s}$,
where each $y_{i,j}$ is the $i$-th patch (of length $n$) of the
signal $x_{j}$.
\item Put the $s$-tuple $\left[d_{i,1},\dots,d_{i,s}\right]=Y_{i}\times M_{i}$
as the next $s$ atoms in $D$.
\end{enumerate}
\end{enumerate}
\caption{Constructing the multi-signature dictionary}

\label{alg:multi-sig-dict}
\end{algorithm}

In order to generate a signal of length $N$ from ${\cal M}$, one
can follow these steps (again we assume that $m$ divides $N$ ):
\begin{enumerate}
\item Create a base signal matrix $X^{G}\in\RR^{N\times s}$ by stacking
$s\frac{N}{m}$ copies of the original basis matrix $X$. Set $y=0$.
\item Repeat for $j=1,\dots,k$: 
\begin{enumerate}
\item Select a base signal $b_{j}\in\ran{X^{G}}$ and shift it (in a circular
fashion) by some $t_{j}=0,1,\dots,R-1$.
\item Aggregate: $y=y+\mbox{shift}(b_{j},t_{j})$ (note that here we do
not need to multiply by a random scalar).
\end{enumerate}
\end{enumerate}
This procedure will produce a signal $y$ of local sparsity $k\cdot s$.
The corresponding support sequence can be written as
\[
{\cal S}=\left(s_{1},s_{2},\dots,s_{N}\right),
\]
where $s_{i}=s_{1}+i\;\left(\mod m\right)$ and 
\[
s_{1}=\left[\left(t_{1},1\right),\left(t_{1},2\right),\dots,\left(t_{1},s\right),\dots,\left(t_{k},1\right),\left(t_{k},2\right),\dots,\left(t_{k},s\right)\right].
\]
Here $\left(t_{j},i\right)$ denotes the atom $d_{t_{j},i}$ in the
notation of \prettyref{alg:multi-sig-dict}. The corresponding signal
space is
\[
\ker A_{{\cal S}}=\spn\left\{ \mbox{shift}\left(X^{G},t_{j}\right)\right\} _{j=1}^{k},
\]
and it is of dimension $k\cdot s$.

\begin{figure}
\centering\subfloat[The dictionary $D$]{\includegraphics[width=0.25\paperwidth]{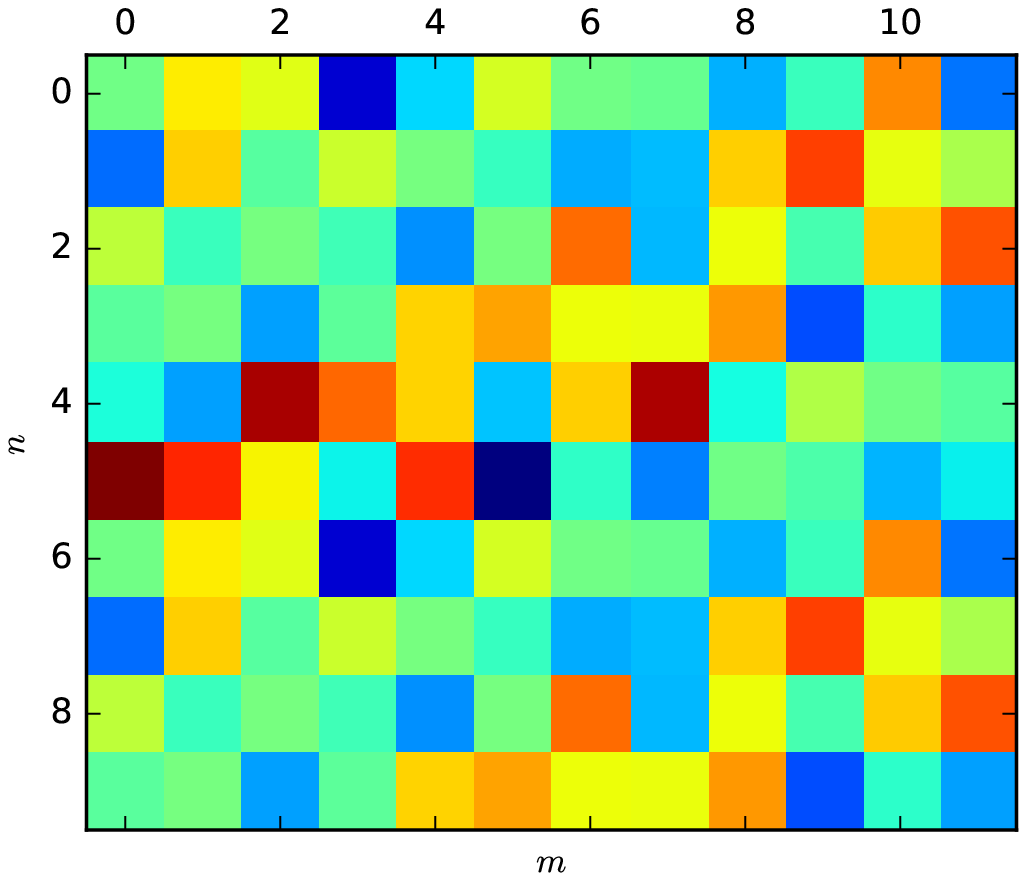}

}

\centering\subfloat[The dependency graph ${\cal G}_{D,2}$. The numerical values above
the edges are the transfer matrices $C_{i,j}$, explained below in
\prettyref{sub:arbitrary-graphs}.]{\includegraphics[width=0.7\paperwidth]{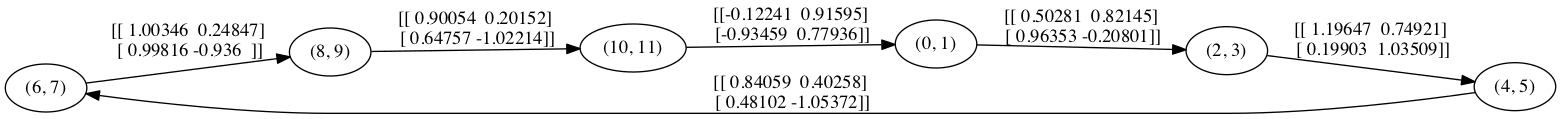}

}

\centering\subfloat[The first signal and its sparse representation in $\ker A_{{\cal S}}$
with $N=24,$ $k=1$ and $t_{1}=5$. ]{\hskip-5em\includegraphics[width=0.4\paperwidth]{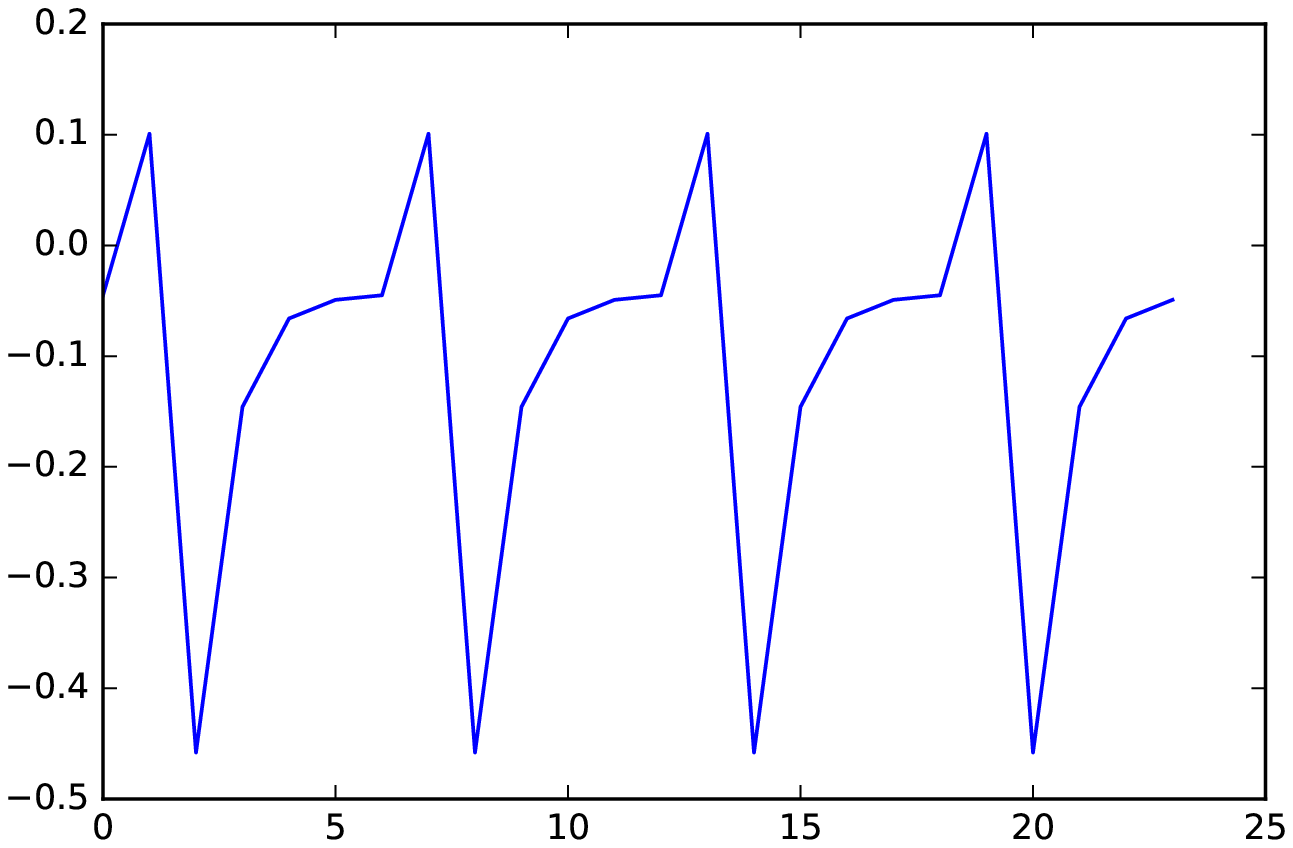}\includegraphics[width=0.4\paperwidth]{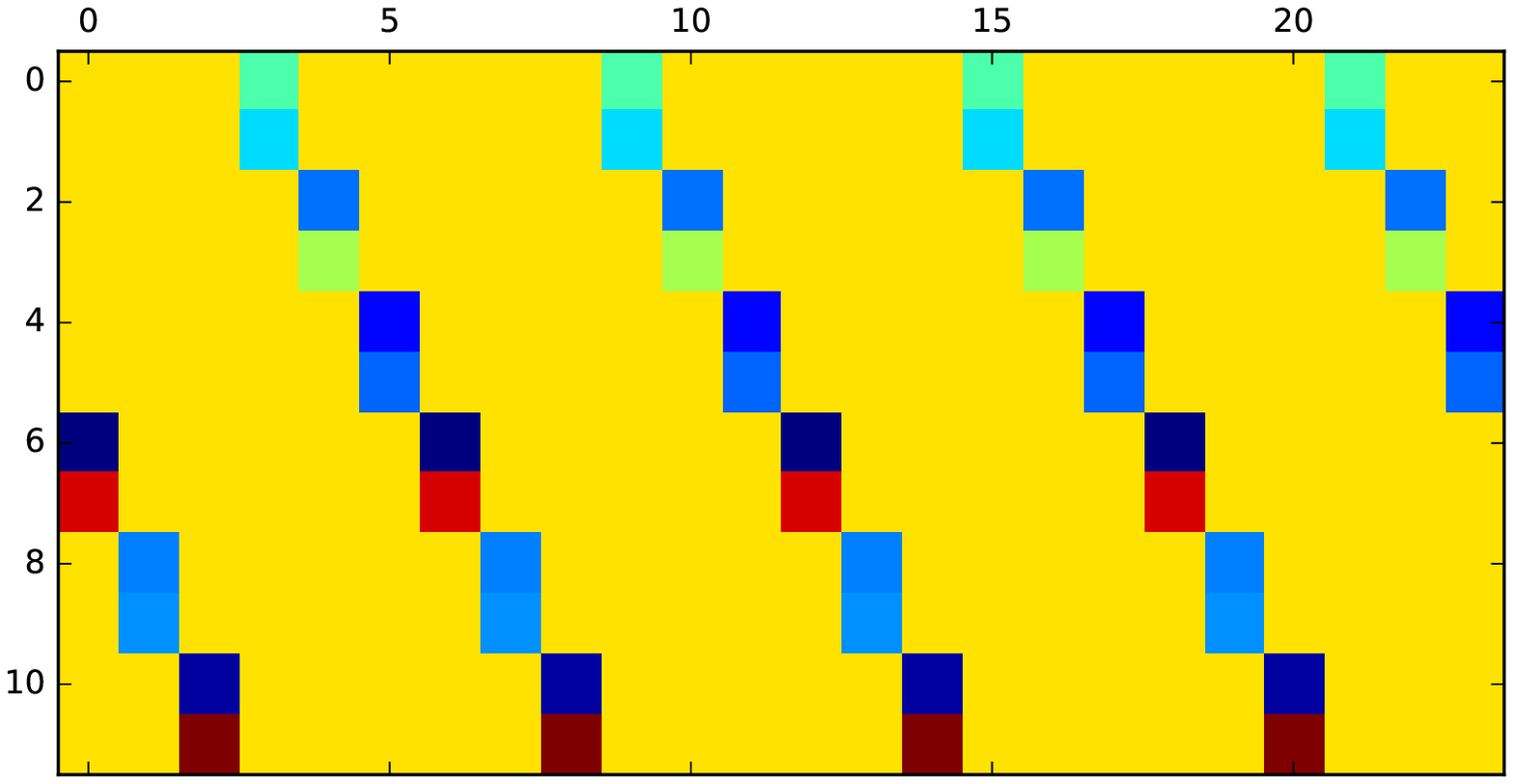}

}

\centering\subfloat[The second signal and its sparse representation in $\ker A_{{\cal S}}$.]{\hskip-5em\includegraphics[width=0.4\paperwidth]{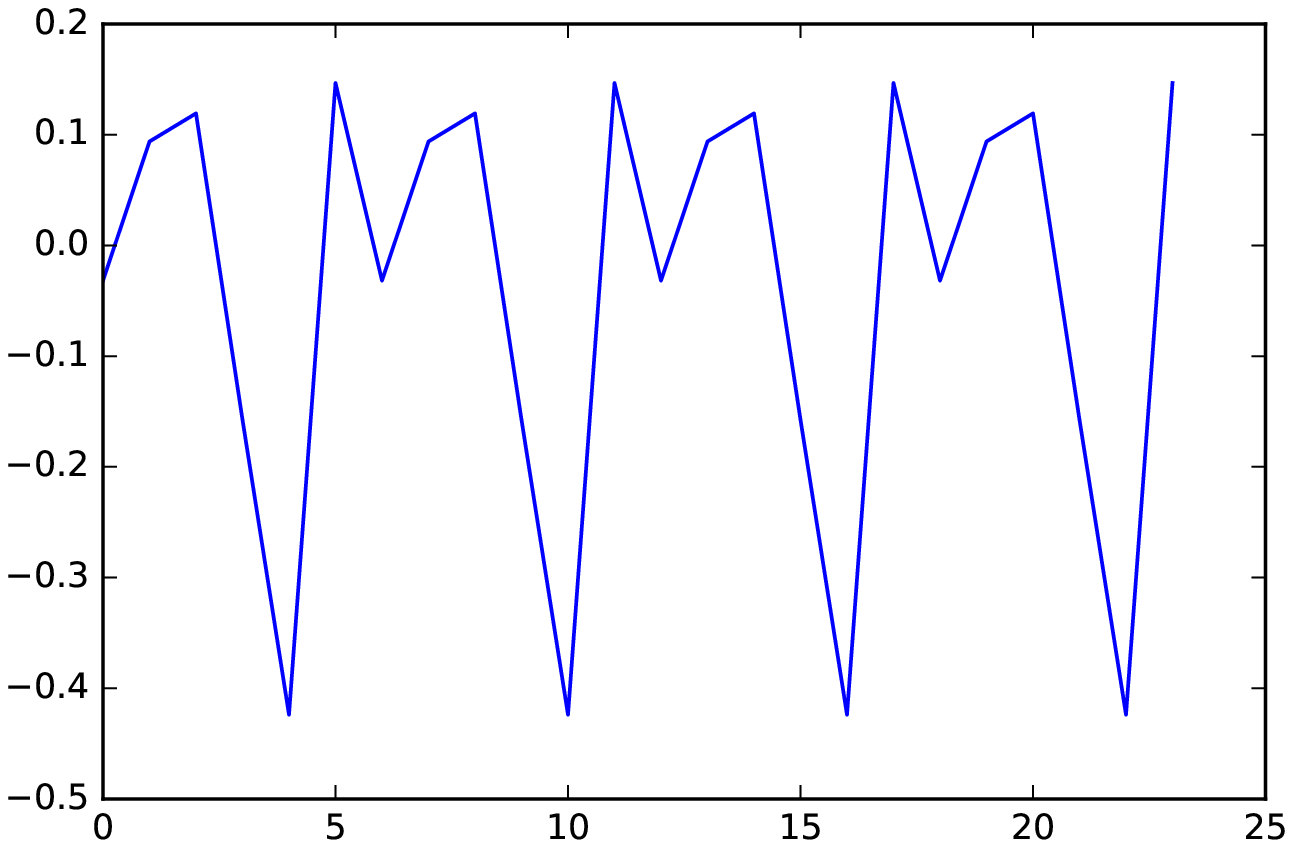}\includegraphics[width=0.4\paperwidth]{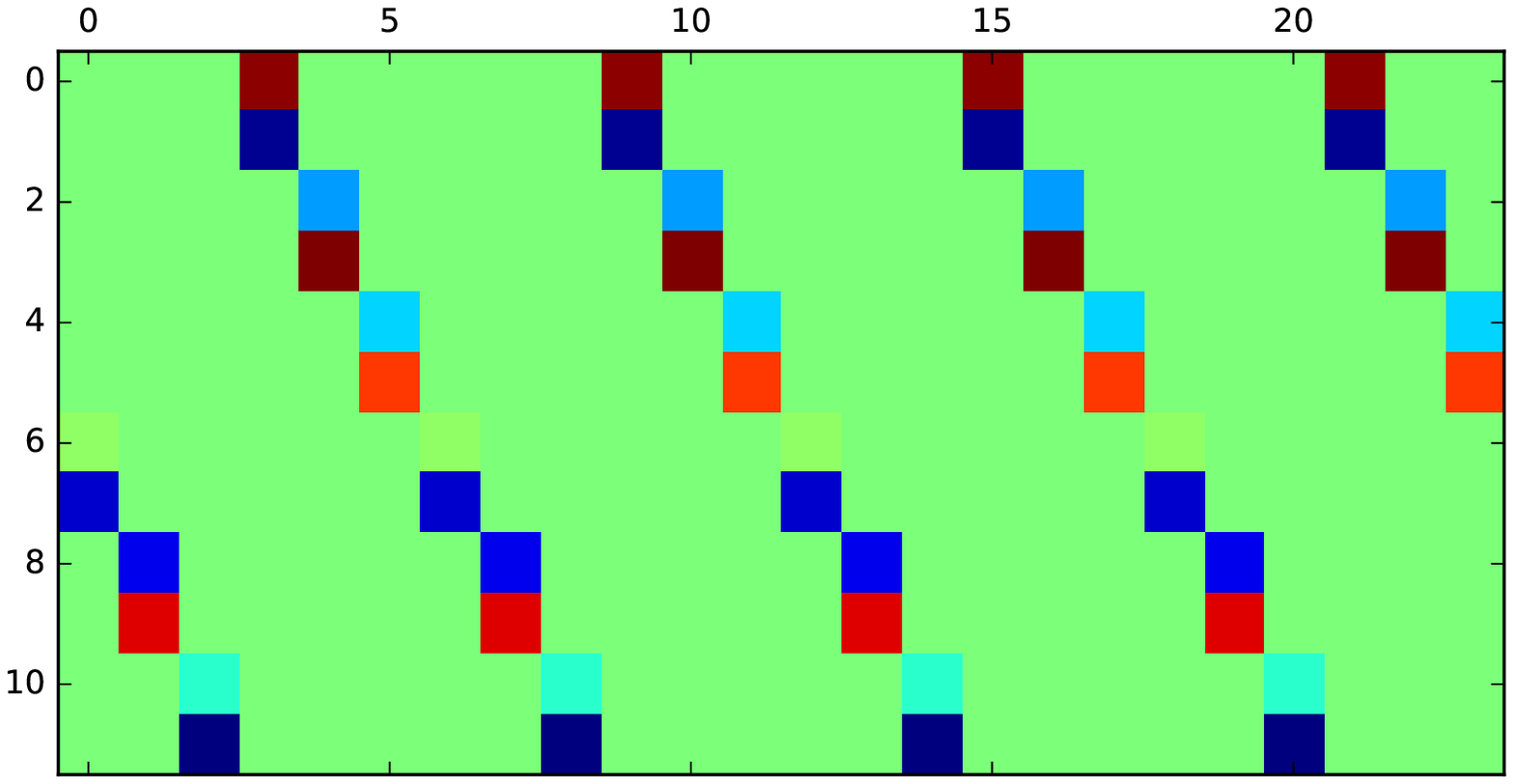}}

\caption{Example of multi-signature dictionary with $n=10,\;m=12,\;s=2$.}

\label{fig:multi-sig-dict}
\end{figure}

An example of a multi-signature dictionary and corresponding signals
may be seen in \prettyref{fig:multi-sig-dict}.

\subsection{\label{sub:csc}Convolutional dictionaries}

An important class of signals is the \emph{sparse convolution model,}
where each signal $x\in\RR^{N}$ can be written as a linear combination
of shifted ``waveforms'' $\vec{d_{i}}\in\RR^{n}$, each $\vec{d_{i}}$
being a column in the local dictionary $D'\in\RR^{n\times m}$. More
conveniently, any such $x$ can be represented as a circular convolution
of $\vec{d_{i}}$ with a (sparse) ``feature map'' $\vec{\psi_{i}}\in\RR^{N}$:
\begin{equation}
x=\sum_{i=1}^{m}\vec{d_{i}}*_{N}\vec{\psi_{i}}.\label{eq:csc-model}
\end{equation}
Such signals arise in various applications, such as audio classification
\cite{blumensath_sparse_2006,grosse_shift-invariance_2012,smith_efficient_2005},
neural coding \cite{ekanadham_unified_2014,quiroga_spike_2007}, mid-level
image representation and denoising \cite{kavukcuoglu_learning_2010,zeiler_deconvolutional_2010,zeiler_adaptive_2011}.

Formally, the convolutional class can be re-cast into the patch-sparse
model of this paper as follows. First, we can rewrite \eqref{eq:csc-model}
as
\[
x=\underbrace{\begin{bmatrix}\mat C_{1} & \mat C_{2} & \dots & \mat C_{m}\end{bmatrix}}_{:=\mat E}\vec{\Psi},
\]
where each $\mat C_{i}\in\RR^{N\times N}$ is a banded circulant matrix
with its first column being equal to $\vec{d_{i}}$, and $\vec{\Psi}\in\RR^{Nm}$
is the concatenation of the $\vec{\psi_{i}}$'s. It is easy to see
that by permuting the columns of $\mat E$ one obtains precisely the
global convolutional dictionary $nD_{G}$ based on the local dictionary
$D'$ (recall \eqref{eq:DG-def}). Therefore we obtain
\begin{equation}
x=\underbrace{D_{G}\left(D'\right)}_{:=D_{G}'}\vec{\G'}.\label{eq:conv-repr-csc}
\end{equation}

While it is tempting to conclude from comparing \eqref{eq:conv-repr-csc}
and \eqref{eq:global-representation-conv} that the convolutional
model is equivalent to the patch-sparse model, an essential ingredient
is missing, namely the requirement of equality on overlaps, $M\vec{\G'}=0.$
Indeed, nothing in the definition of the convolutional model restricts
the representation $\vec{\Psi}$ (and therefore $\vec{\G'}),$ therefore
in principle the number of degrees of freedom remains $Nm$, as compared
to $N\left(m-n+1\right)$ from \prettyref{prop:dimkerA}. 

To fix this, following \textcolor{red}{\cite{papyan2016working_1,papyan2016working_2}}
we apply $R_{i}$ to \eqref{eq:conv-repr-csc} and obtain $R_{i}x=R_{i}D_{G}'\vec{\G'}$.
The ``stripe'' $\Omega_{i}'=R_{i}D_{G}'$ has only $\left(2n-1\right)m$
nonzero consecutive columns, and in fact the nonzero portion of $\Omega_{i}'$
is equal for all $i$. This implies that every $x_{i}$ has a representation
$x_{i}=\Theta\vec{\gamma_{i}}$ in the ``pseudo-local'' dictionary
\[
\Theta\left(D'\right):=\begin{bmatrix}Z_{B}^{\left(n-1\right)}D' & \dots & D' & \dots & Z_{T}^{\left(n-1\right)}D'\end{bmatrix}\in\RR^{n\times\left(2n-1\right)m},
\]
where the operators $Z_{B}^{\left(k\right)}$ and $Z_{T}^{\left(k\right)}$
are given by \prettyref{def:stsb-def} in \nameref{app:proof1}. If
we now assume that our convolutional signals satisfy
\[
\|\vec{\gamma_{i}}\|_{0}\leqslant s\qquad\forall i,
\]
then we have shown that they belong to ${\cal M}\left(\Theta\left(D'\right),s,P,N\right)$,
and thus can be formally treated by the framework we have developed.

It turns out that this direct approach is quite naive, as the dictionary
$\Theta\left(D'\right)$ is extremely bad-equipped for sparse reconstruction
(for example it has repeated atoms, and therefore $\mu\left(\Theta\left(D'\right)\right)=1$).
To tackle this problem, a convolutional sparse coding framework was
recently developed in \textcolor{red}{\cite{papyan2016working_1,papyan2016working_2}},
where the explicit dependencies between the sparse representation
vectors $\vec{\gamma_{i}}$ (and therefore the special structure of
the corresponding constraint $M\left(D'\right)\vec{\G'}=0$) were
exploited quite extensively, resulting in efficient recovery algorithms
and nontrivial theoretical guarantees. We refer the reader to \textcolor{red}{\cite{papyan2016working_1,papyan2016working_2}
}for further details and examples.

\subsection{\label{sub:arbitrary-graphs}Arbitrary dependency graphs}

The examples considered in the previous sections are somewhat special.
For the most general case, one can define an abstract graph ${\cal G}$
with some desirable properties, and subsequently look for a nontrivial
realization $D$ of the graph, so that in addition ${\cal R}_{{\cal G}}\neq\emptyset$.
Let us therefore discuss each one of those steps, along with a specific
example.

\subsubsection{Defining an abstract ${\cal G}$ with desirable properties}

\begin{figure}
\hskip-5em\includegraphics[width=0.8\paperwidth]{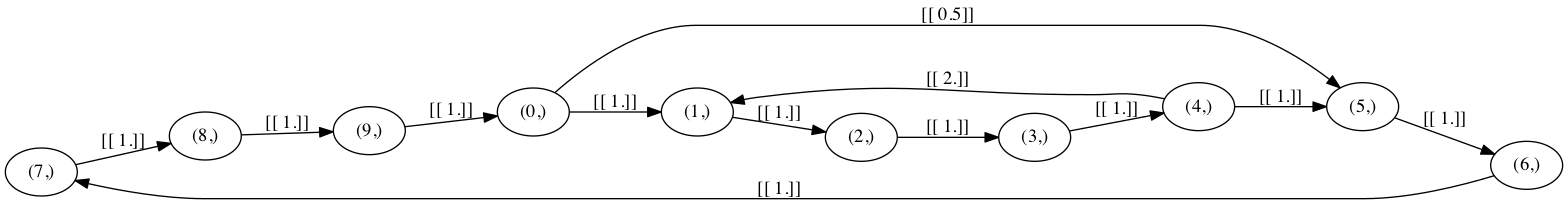}\caption{A possible dependency graph ${\cal G}$ with $m=10$. In this example,
$\left|{\cal C_{G}}\left(70\right)\right|=37614$.}

\label{fig:possible-graph}
\end{figure}

In this context, we would want ${\cal G}$ to contain \emph{sufficiently
many different long cycles}, which would correspond to long signals
and a rich resulting model ${\cal M}$. In contrast with the models
from \prettyref{sub:Signature-Dictionary}, one therefore should allow
for some branching mechanism. An example of a possible ${\cal G}$
is given in \prettyref{fig:possible-graph}. It differs only slightly
from the example in \prettyref{fig:singature-dict}. Notice that due
to the structure of ${\cal G}$ there are many possible paths in ${\cal C_{G}}\left(P\right)$.
In fact, a direct search algorithm yields $\left|{\cal C_{G}}\left(70\right)\right|=37614$.

\subsubsection{Finding a dictionary $D$ which has ${\cal G}$ as its dependency
graph}

Every edge in ${\cal G}$ corresponds to a conditions of the form
\eqref{eq:basic-rank-inequality} imposed on the entries of $D$.
As discussed in \prettyref{thm:measure-zero-for-dicts}, this in turn
translates to a set of algebraic equations. So the natural idea would
be to write out the large system of such equations and look for a
solution over the field $\RR$ by well-known algorithms in numerical
algebraic geometry \cite{basu2006algorithms}. However, this approach
is highly impractical because these algorithms have (single or double)
exponential running time. We consequently propose a simplified, more
direct approach to the problem.

\begin{figure}
\centering\includegraphics[width=0.3\paperwidth]{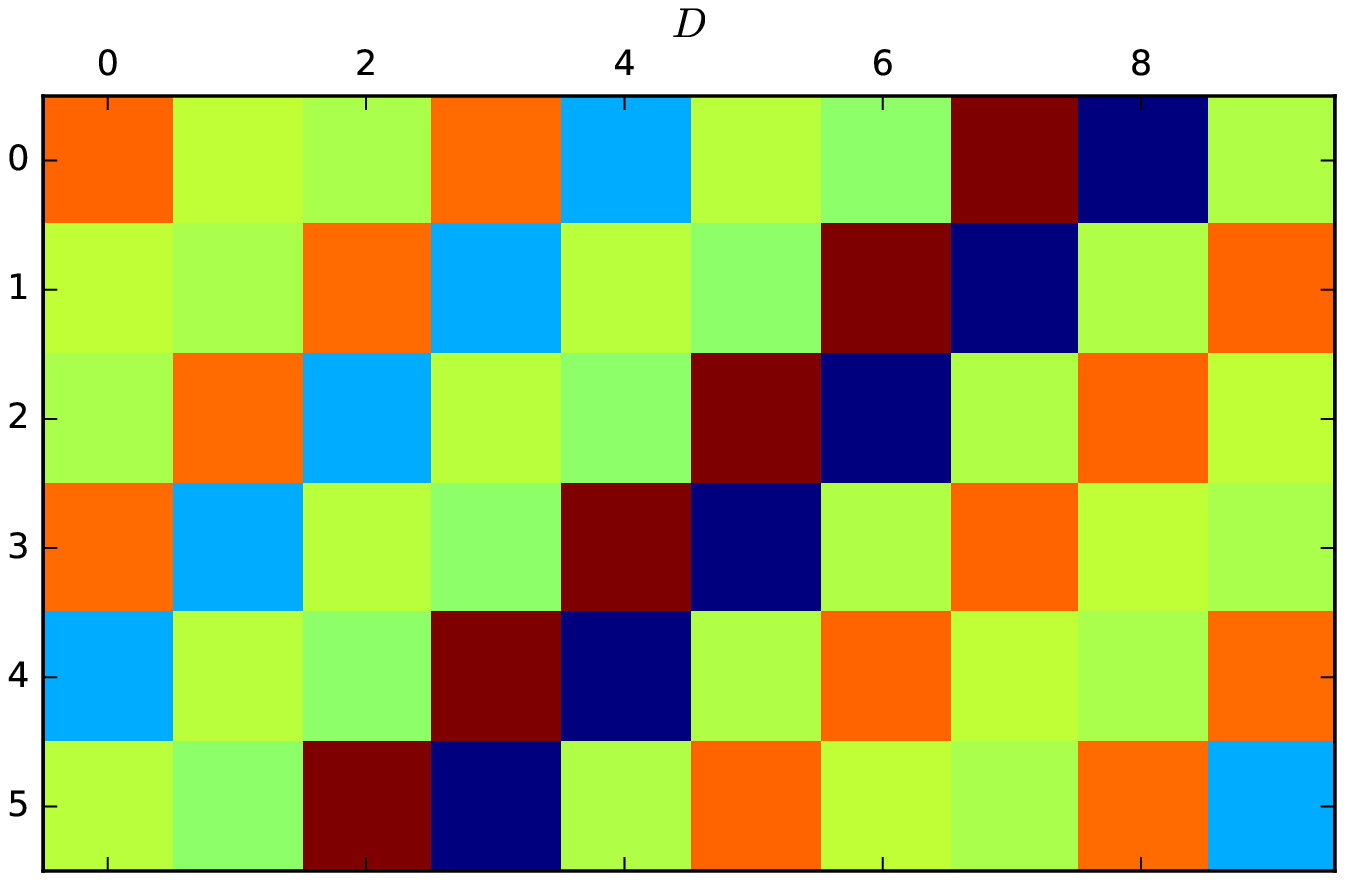}

\caption{A realization $D\in\RR^{6\times10}$ of ${\cal G}$ from \prettyref{fig:possible-graph}.}
\label{fig:realization-D}
\end{figure}

In detail, we replace the low-rank conditions \eqref{eq:basic-rank-inequality}
with more explicit and restrictive ones below.
\begin{description}
\item [{Assumptions({*})}] For each $\left(s_{i},s_{j}\right)\in{\cal G}$
we have $\left|s_{i}\right|=\left|s_{j}\right|=k$. We require that
$\spn S_{B}D_{s_{i}}=\spn S_{T}D_{s_{j}}=\Lambda_{i,j}$ with $\dim\Lambda_{i,j}=k$.
Thus there exists a non-singular transfer matrix $C_{i,j}\in\RR^{k\times k}$
such that
\begin{equation}
S_{B}D_{s_{i}}=C_{i,j}S_{T}D_{s_{j}}.\label{eq:explicit-linear-equation-for-edge}
\end{equation}
\end{description}
In other words, every column in $S_{B}D_{s_{i}}$ must be a specific
linear combination of the columns in $S_{T}D_{s_{j}}$. This is much
more restrictive than the low-rank condition, but on the other hand,
given the matrix $C_{i,j}$, it defines a set of linear constraints
on $D$. To summarize, the final algorithm is presented in \prettyref{alg:find-realization-of-graph}.
In general, nothing guarantees that for a particular choice of ${\cal G}$
and the transfer matrices, there is a nontrivial solution $D$, however
in practice we do find such solutions. For example, taking the graph
from \prettyref{fig:possible-graph} and augmenting it with the matrices
$C_{i,j}$ (scalars in this case), we obtain a solution over $\RR^{6}$
which is shown in \prettyref{fig:realization-D}. Notice that while
the resulting dictionary has a Hankel-type structure similar to what
we have seen previously, the additional dependencies between the atoms
produce a rich signal space structure, as we shall demonstrate in
the following section.

\begin{algorithm}
\begin{enumerate}
\item Input: a graph ${\cal G}$ satisfying the \textbf{Assumptions({*})}
above, and the dimension $n$ of the realization space $\RR^{n}$.
\item Augment the edges of ${\cal G}$ with arbitrary nonsingular transfer
matrices $C_{i,j}$.
\item Construct the system of linear equations given by \eqref{eq:explicit-linear-equation-for-edge}.
\item Find a nonzero $D$ solving the system above over $\RR^{n}$.
\end{enumerate}
\caption{Finding a realization $D$ of the graph ${\cal G}$}
\label{alg:find-realization-of-graph}

\end{algorithm}

\subsubsection{The resulting signal space}

\begin{figure}
\centering\includegraphics[width=0.4\paperwidth]{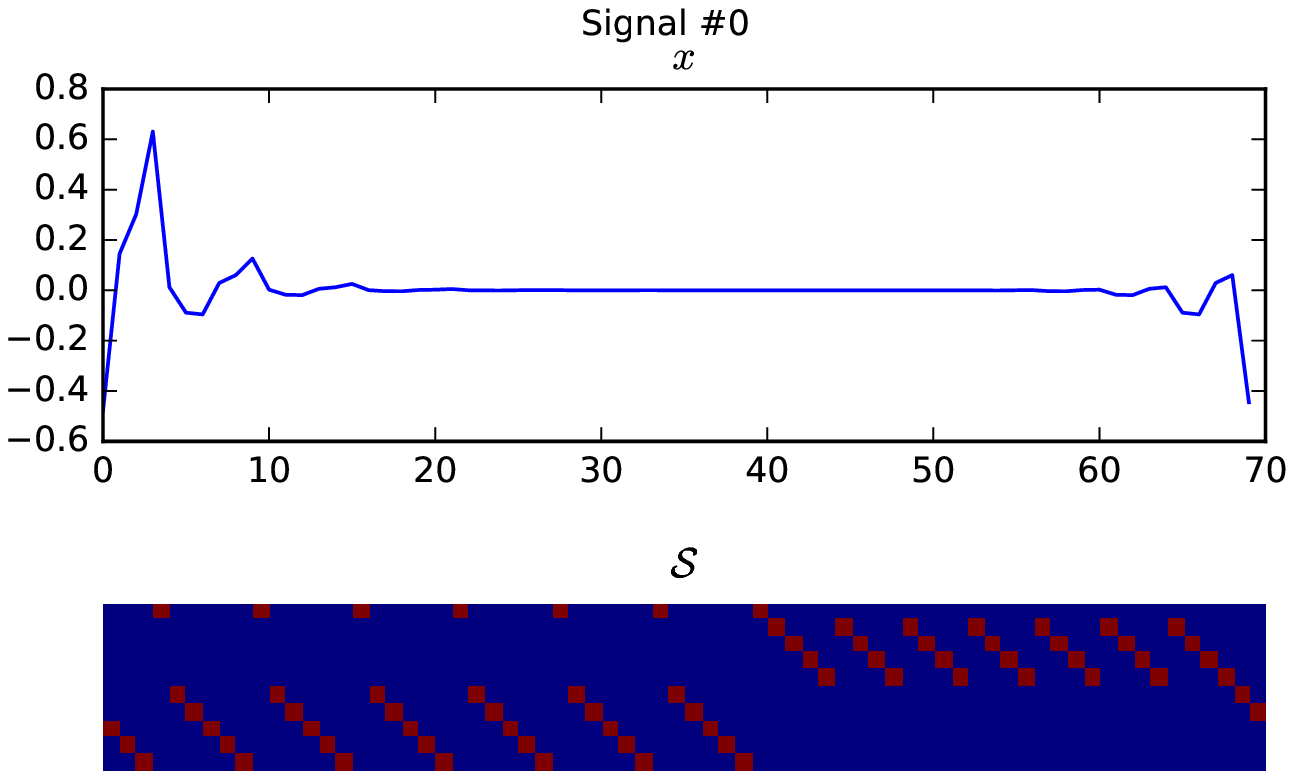}

\includegraphics[width=0.4\paperwidth]{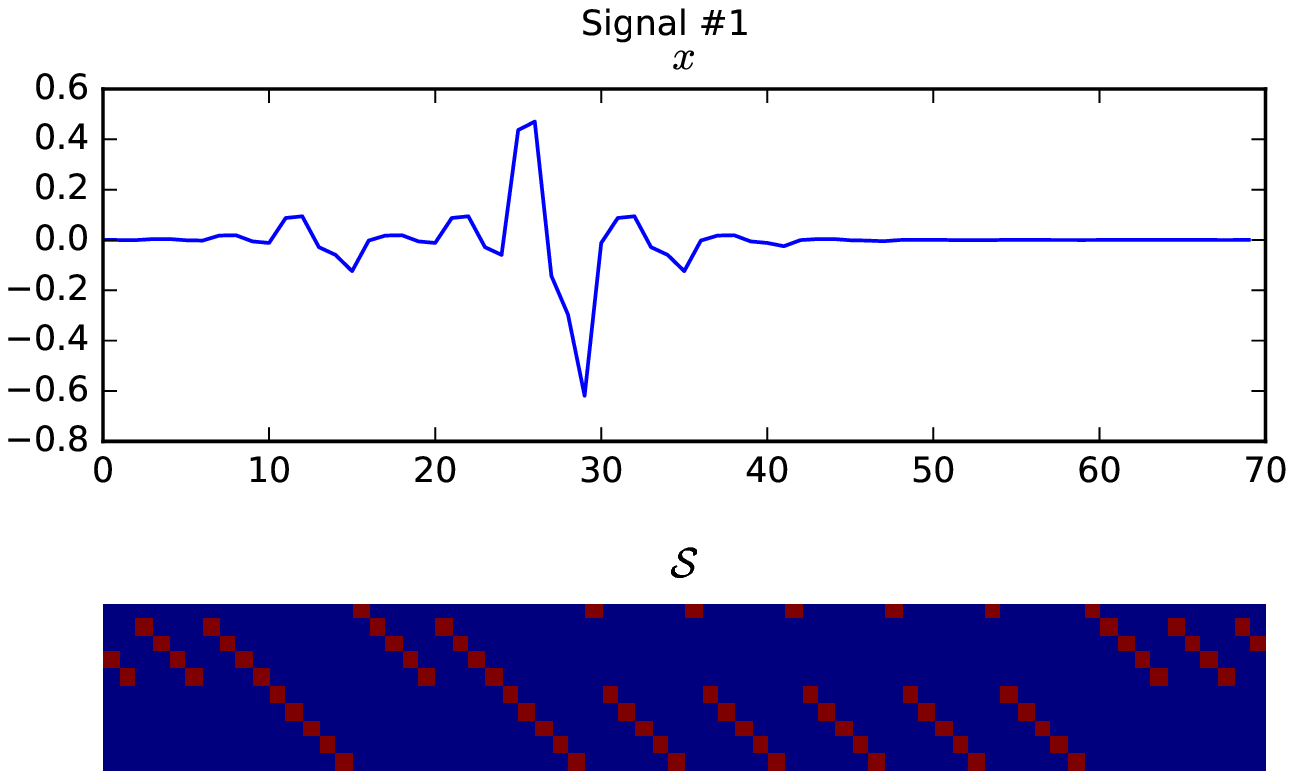}

\includegraphics[width=0.4\paperwidth]{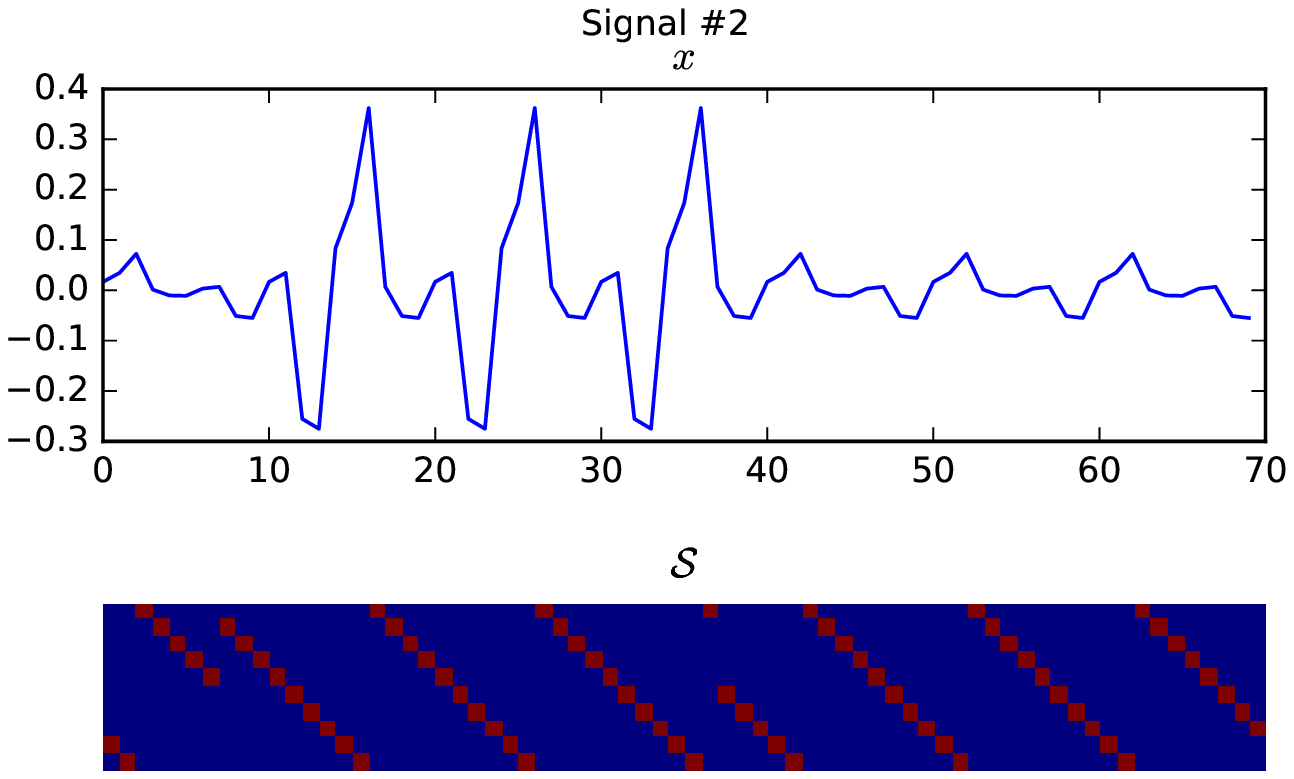}

\caption{Examples of signals from ${\cal M}$ and the corresponding supports
 ${\cal S}$}
\label{fig:signals-arbitrary-model}
\end{figure}

Given ${\cal G}_{D}$ and the signal length $N=P$, the signals $x$
can be generated according to \prettyref{alg:gen-sig-model} \vpageref{alg:gen-sig-model}.
Not all paths in ${\cal C_{G}}$ are realizable, but it turns out
that in our example we have $\left|{\cal R_{G}}\left(70\right)\right|=17160$.
Three different signals and their supports ${\cal S}$ are shown in
\prettyref{fig:signals-arbitrary-model}. As can be seen from these
examples, the resulting model ${\cal M}$ is indeed much richer than
the signature-type construction from \prettyref{sub:Signature-Dictionary}.

Using the restricted construction of this section, the following estimate
can be easily shown.
\begin{proposition}
\label{prop:dimkerAs-bound-arbitrary-graph}Assume that the model
satisfies \textbf{Assumptions({*}) }above. Then for every ${\cal S}\in{\cal R_{G}}\left(P\right)$
\[
\dim\ker A_{{\cal S}}\leqslant k.
\]
\end{proposition}
\begin{svmultproof2}
The idea is to construct a spanning set for $\ker M_{*}^{\left({\cal S}\right)}$
and invoke \prettyref{prop:dimkerA}. Let us relabel the nodes along
${\cal S}$ to be $1,2,\dots,P$. Starting from an arbitrary $\alpha_{1}$
with support $\left|s_{1}\right|=k$, we use \eqref{eq:explicit-linear-equation-for-edge}
to obtain, for $i=1,2,\dots,P-1$, a formula for the next portion
of the global representation vector $\Gamma$
\begin{equation}
\alpha_{i+1}=C_{i,i+1}^{-1}\alpha_{i}.\label{eq:building-Gamma}
\end{equation}
This gives a set $\Delta$ consisting of overall $k$ linearly independent
vectors $\Gamma_{i}$ with $\supp\G_{i}={\cal S}$. It may happen
that equation \eqref{eq:building-Gamma} is not satisfied for $i=P$.
However, every $\G$ with $\supp\G={\cal S}$ and $M_{*}^{\left({\cal S}\right)}\Gamma_{{\cal S}}=0$
must belong to $\spn\Delta$, and therefore
\[
\dim\ker M_{*}^{\left({\cal S}\right)}\leqslant\dim\spn\Delta=k.
\]
\end{svmultproof2}

We believe that \prettyref{prop:dimkerAs-bound-arbitrary-graph} can
be extended to more general graphs, not necessarily satisfying \textbf{Assumptions({*})}.\textbf{
}In particular, the following estimate appears to hold for a general
model ${\cal M}$ and ${\cal S}\in{\cal R_{G}}\left(P\right)$:
\[
\dim\ker A_{{\cal S}}\leqslant\left|s_{1}\right|+\sum_{i}\left(\left|s_{i+1}\right|-\rank\left[S_{B}D_{s_{i}}\;S_{T}D_{s_{i+1}}\right]\right).
\]
We leave the rigorous proof of this result to a future work.

\subsubsection{Further remarks}

While the presented model is the hardest to analyze theoretically,
even in the restricted case of \textbf{Assumptions({*})} (when does
a nontrivial realization of a given ${\cal G}$ exist? how does the
answer depend on $n$? When ${\cal R_{G}}\left(P\right)\neq\emptyset$?
etc?), we hope that this construction will be most useful in applications
such as denoising of natural signals.

\section{\label{sec:Numerical-experiments}Numerical experiments}

In this section, we test the effectiveness of the globalized model
for recovering the signals from Section \ref{sec:Examples}, both
in the noiseless and noisy cases. These results are also compared
to the classical LPA approach.

\subsection{Signature-type signals}

In this section we investigate the performance of the pursuit algorithms
on signals complying with the signature dictionary model elaborated
in \prettyref{sub:Signature-Dictionary}, constructed from one base
signal and allowing for varying values of $s$. 

\subsubsection{Constructing the dictionary}

In the context of the LPA algorithm, the condition for its success
in recovering the representation is a function of the mutual coherence
of the local dictionary \textendash{} the smaller this measure the
larger the number of non-zeros that are guaranteed to be recovered.
Leveraging this, we aim at constructing $D\in\RR^{n\times m}$ of
a signature type that has a small coherence. This can be cast as an
optimization problem
\[
x_{0}=\arg\min_{x\in\RR^{m}}\mu\left(\tilde{D}\left(x\right)\right),\;D=D\left(x_{0}\right),
\]
where $\tilde{D}\left(x\right)$ is computed by \prettyref{alg:signature-dict}
and $\mu$ is the (normalized) coherence function.

In our experiments, we choose $n=15$ and $m=20$, and minimize the
above loss function via gradient descent, resulting in $\mu(D(x))=0.26$.
We used the TensorFlow open source package \cite{tensorflow2015-whitepaper}.
As a comparison, the coherence of a random signature dictionary is
about $0.5.$

\subsubsection{Noiseless case}

\begin{figure}
\centering\includegraphics[clip,width=0.5\paperwidth]{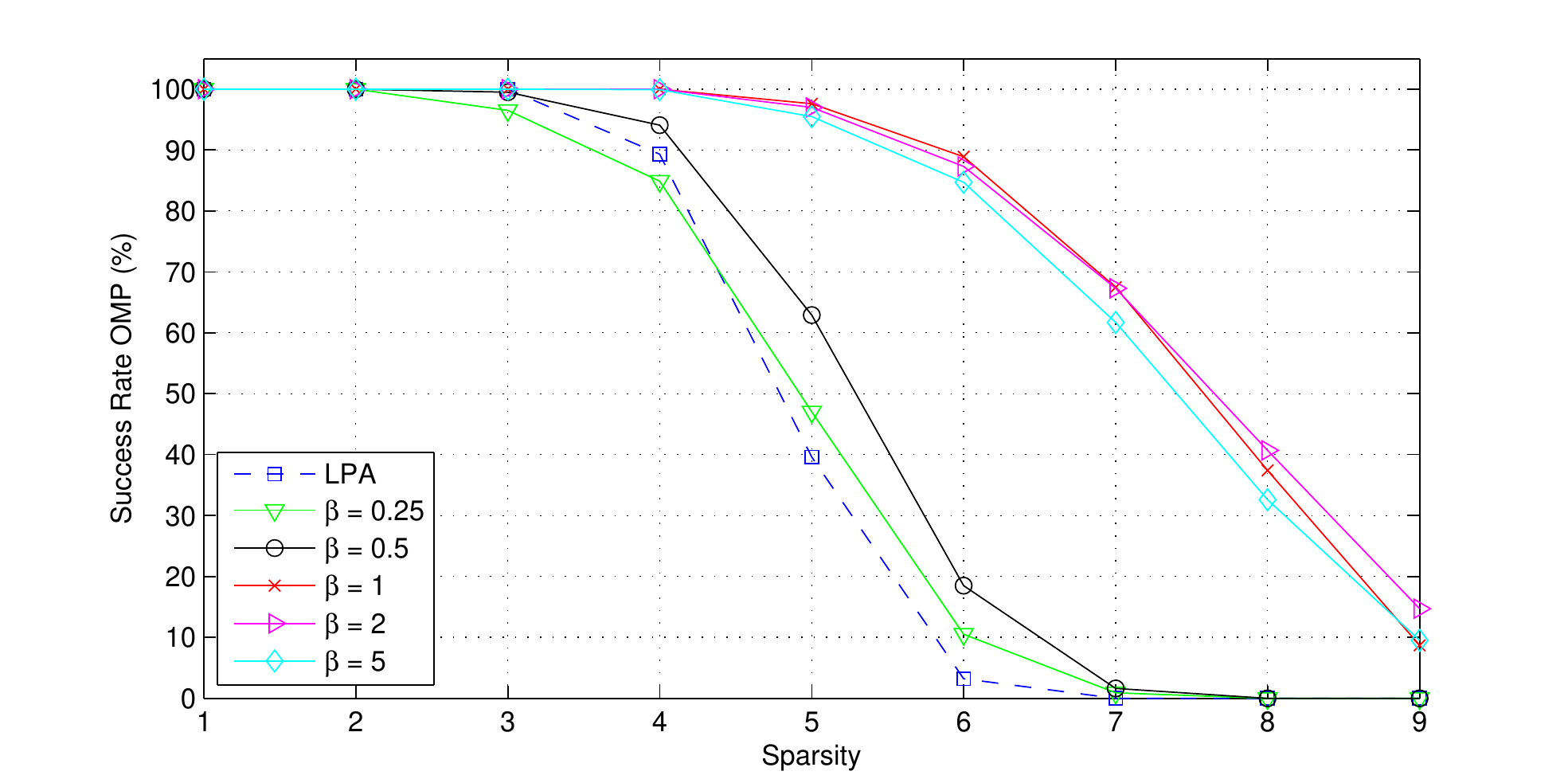}

\caption{Probability of the success (\%) of the globalized OMP (as a function
of $\beta$) and the LPA algorithms to perfectly recover the sparse
representations of test signals from the signature dictionary model,
averaged over $10^{3}$ realizations.\label{fig:ProbOfSuccess}}
\end{figure}

In this setting, we test the ability of the globalized OMP to perfectly
recover the sparse representation of clean signature-type signals.
Figure \ref{fig:ProbOfSuccess} compares the proposed algorithm (for
different choices of $\beta\in\left\{ 0.25,0.5,1,2,5\right\} $) with
the LPA one by providing their probability of success in recovering
the true sparse vectors, averaged over $10^{3}$ randomly generated
signals of length $N=100$. 

From a theoretical perspective, since $\mu(D)=0.26$, the LPA algorithm
is guaranteed to recover the representation when $\|\G\|_{0,\infty}\leq2$,
while as can be seen in practice it successfully recover these for
$\|\G\|_{0,\infty}\leq3$. Comparing the LPA approach to the globalized
OMP, one can observe that for $\beta\geq1$ the latter consistently
outperforms the former, having a perfect recovery for $\|\G\|_{0,\infty}\leq4$.
Another interesting insight of this experiment is the effect of $\beta$
on the performance; roughly speaking, a relatively large value of
this parameter results in a better success-rate than the very small
ones, thereby emphasizing importance of the constraint $M_{*}\G=0$.
On the other hand, $\beta$ should not be too large since then the
importance of the signal is reduced compared to the constraint, which
might lead to deterioration in the success-rate (see the curve that
corresponds to$\beta=5$ in Figure \ref{fig:ProbOfSuccess}). 

\subsubsection{Noisy case}

In what follows, the stability of the proposed globalized OMP and
the ADMM-pursuit are tested and compared to the traditional LPA algorithm.
In addition, we provide the projected versions of these algorithms,
which by definition satisfy the constraint of our model. More specifically,
given the estimated support $\hat{S}$ of each pursuit, we ensure
that the constraint $M_{*}\hat{\G}=0$ is met by constructing the
matrix $A_{\hat{S}}$ and then projecting the signal onto the subspace
$\ker A_{\hat{{\cal S}}}$. In addition to the above, we provide the
restoration performance of the oracle estimator, serving as an indication
for the best possible denoising that can be achieved. In this case,
the oracle projection matrix $A_{S}$ is constructed according to
the ground-truth support $S$.

Per each local cardinality $1\leq\|\G\|_{0,\infty}\leq4$ we generate
10 random signature-type signals, where each of these is corrupted
by white additive Gaussian noise with standard deviation $\sigma$,
ranging from 0.05 up-to 0.5. The global number of non-zeros is injected
to the globalized OMP, and the information regarding the local sparsity
is utilized both by the LPA algorithm and our ADMM-pursuit (which
is based on local sparse recovery operations). Following Figure \ref{fig:denoising_k1},
which plots the Mean Squared Error (MSE) of the estimation as a function
of the noise level, the ADMM-pursuit achieves the best denoising performance,
having similar results to the oracle estimator for all noise-levels
and sparsity factors. The source of superiority of the ADMM pursuit
might be its inherent ability to obtain an estimation that perfectly
fits to the globalized model, i.e., a reconstruction that is identical
to its projected version. The second best algorithm being the globalized
OMP; for relatively small noise levels, its projected version performs
as good as the oracle one, indicating that it successfully recovers
the true supports. For large noise levels, however, this algorithm
tends to err and results in local estimations that do not ``agree''
with each other on the overlaps. Yet, the denoising performance of
the globalized OMP is better than the one of the LPA algorithm. Notice
that the latter performs similarly to the oracle estimator only for
very low noise levels and relatively small sparsity factors. This
sheds light on the difficulty of finding the true supports, the non
trivial solution of this problem, and the great advantage of the proposed
globalized model.

\begin{figure}
\leftskip-3cm

\subfloat[$\|\protect\G\|_{0,\infty}\leq1$]{\includegraphics[width=0.7\textwidth]{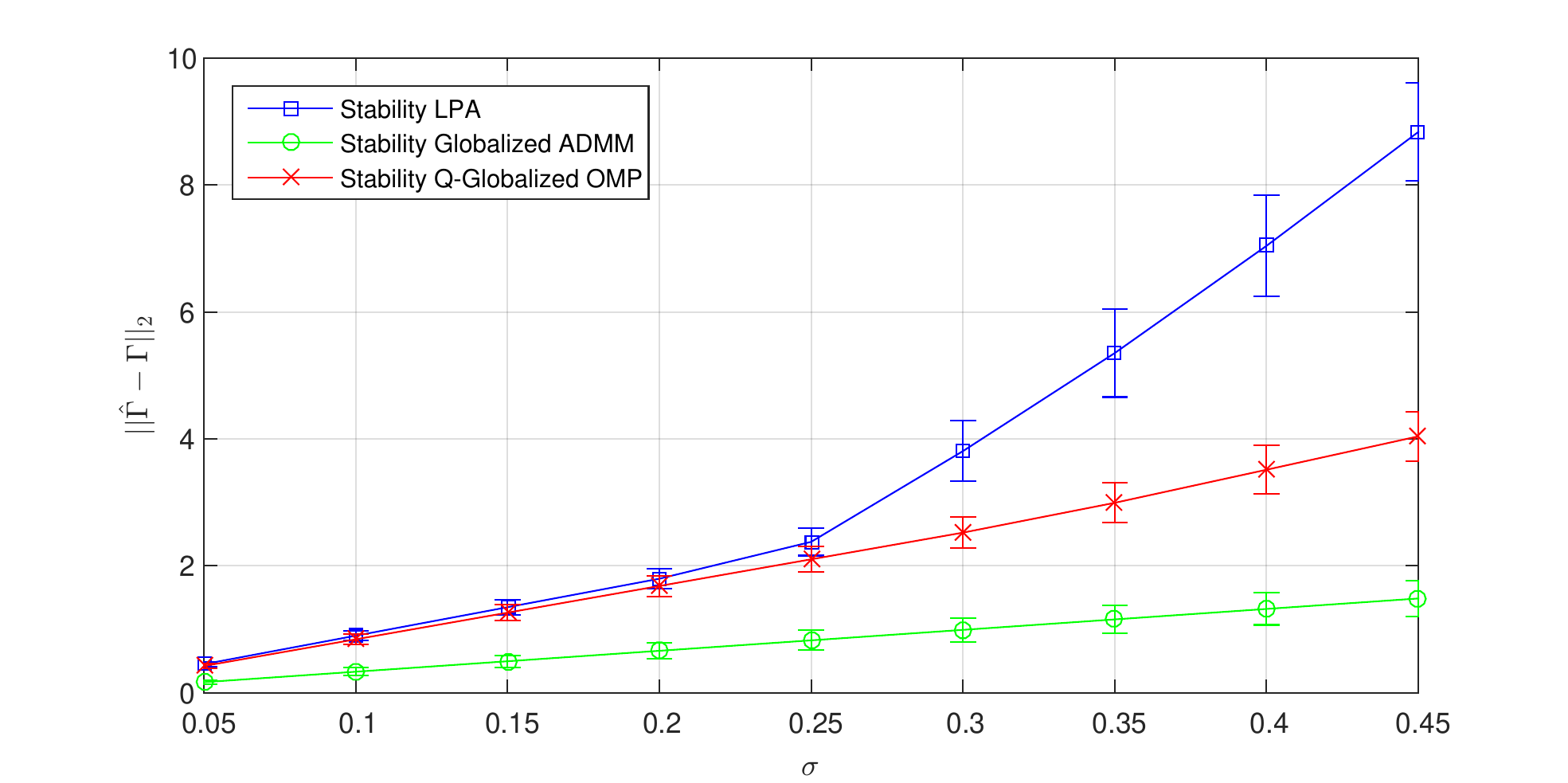}

}\subfloat[$\|\protect\G\|_{0,\infty}\leq2$]{\includegraphics[clip,width=0.7\textwidth]{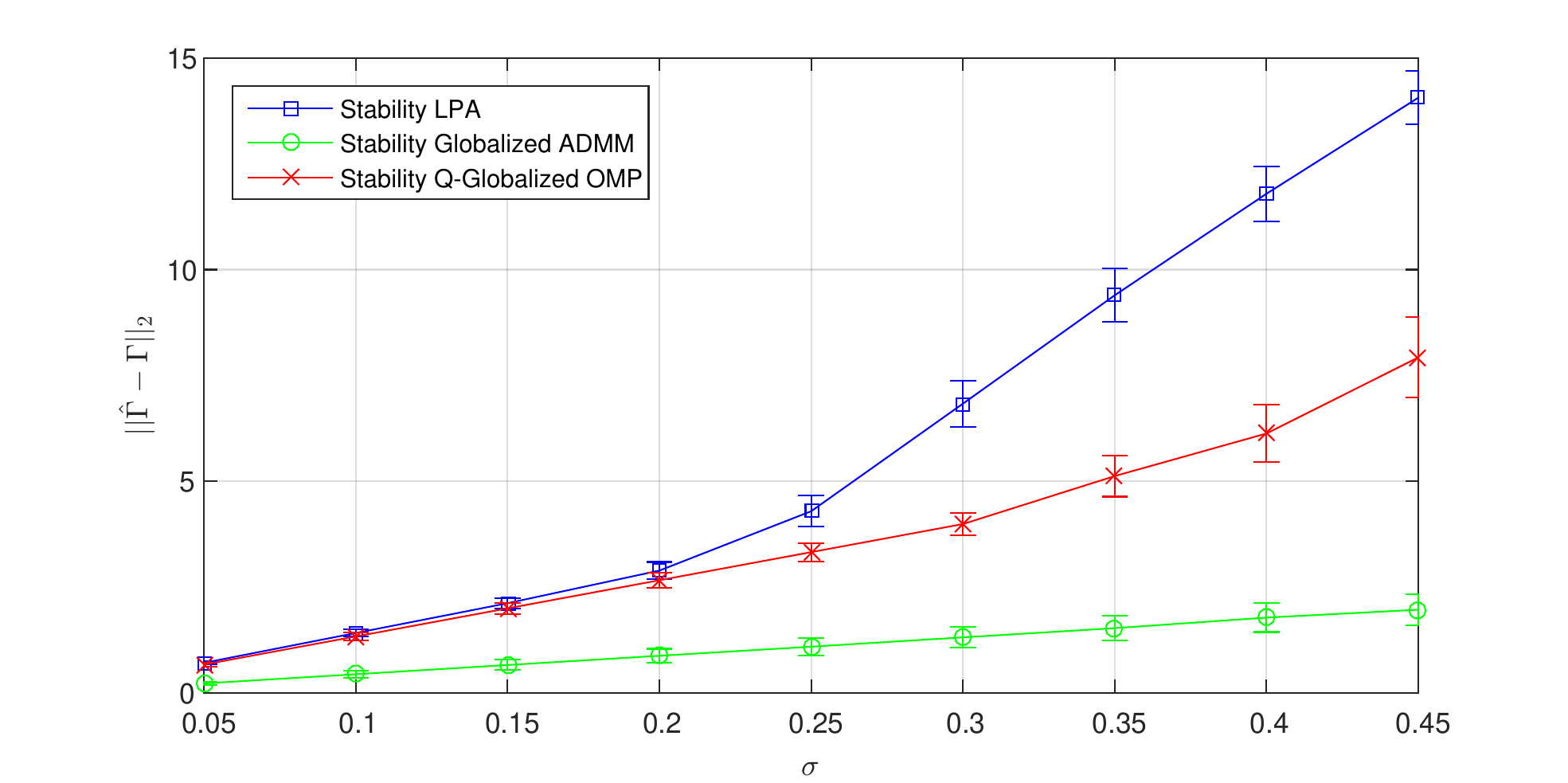}

}

\subfloat[$\|\protect\G\|_{0,\infty}\leq3$]{\includegraphics[clip,width=0.7\textwidth]{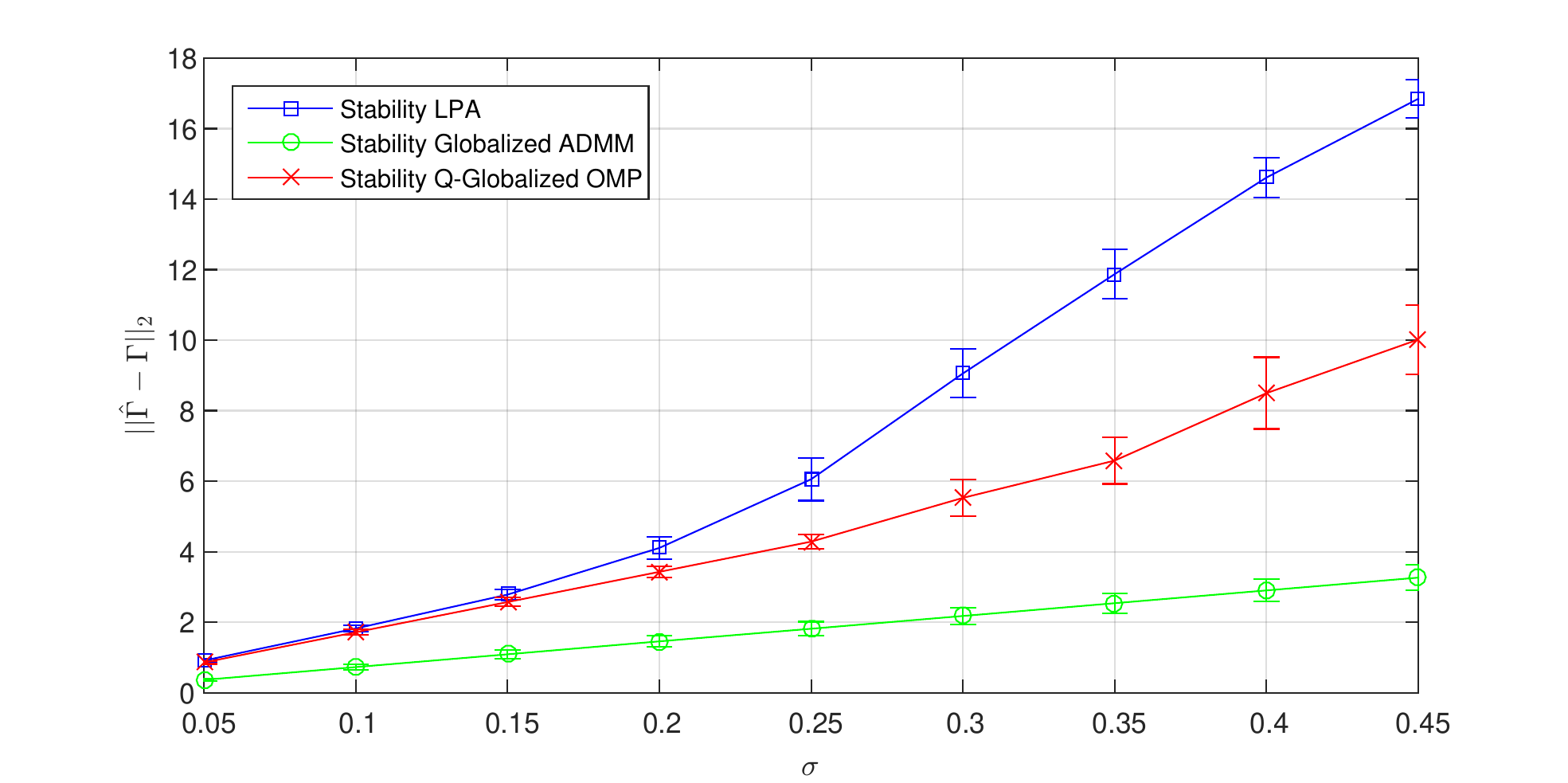}

}\subfloat[$\|\protect\G\|_{0,\infty}\leq4$]{\includegraphics[clip,width=0.7\textwidth]{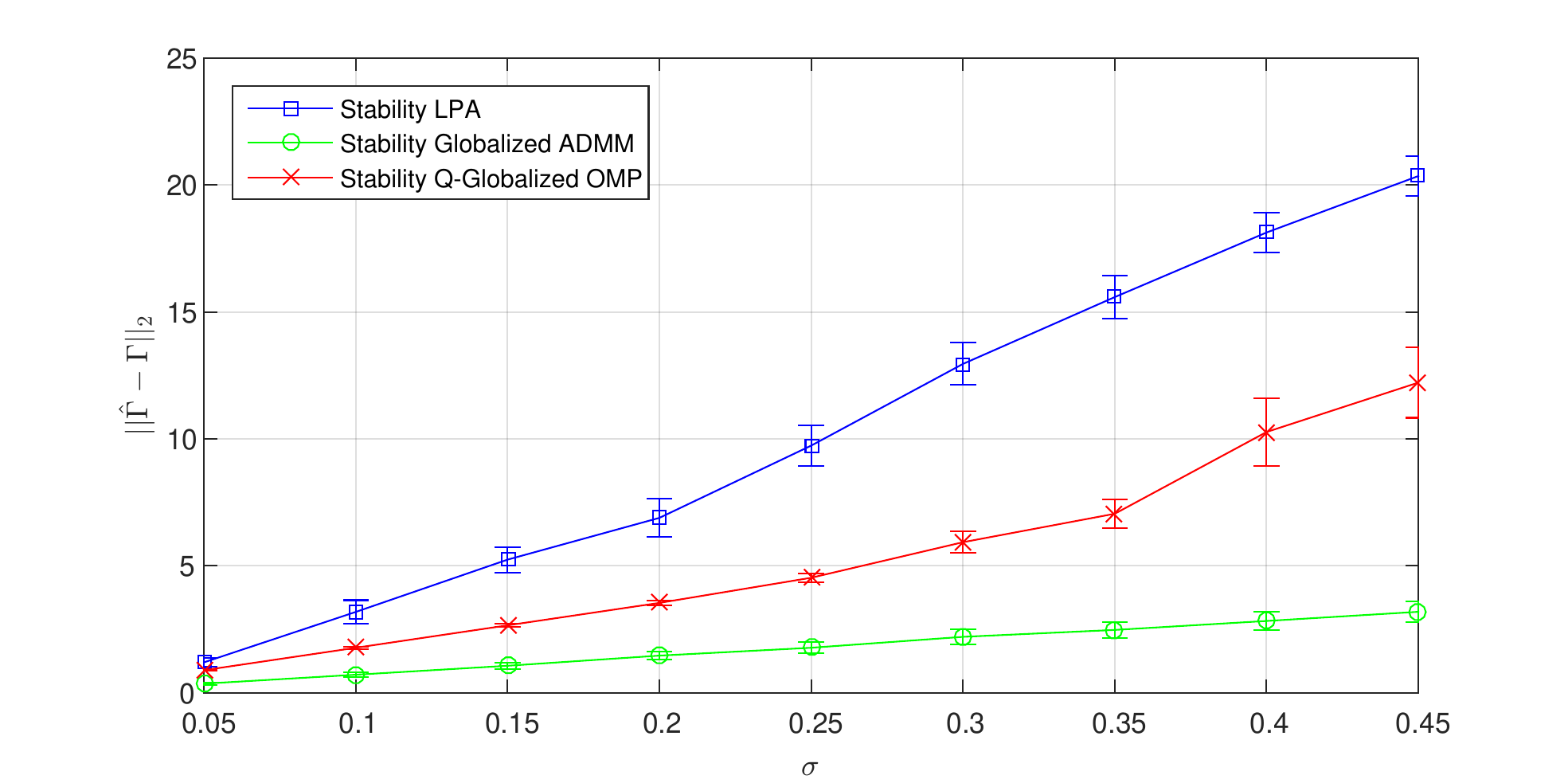}

}

\caption{Stability of the globalized OMP, ADMM-pursuit and LPA algorithm for
various noise levels and sparsity factors. The signals are drawn from
the signature dictionary model.}
\label{fig:stability_k1}
\end{figure}

Similar conclusion holds for the stable recovery of the sparse representations.
Per each pursuit algorithm, Figure \ref{fig:stability_k1} illustrates
the $\ell_{2}$ distance between the original sparse vector $\G$
and its estimation $\hat{\G}$, averaged over the different noise
realizations. As can be seen, the ADMM-pursuit achieves the most stable
recovery, the globalized OMP is slightly behind it, and both of them
outperform the LPA algorithm especially in the challenging cases of
high noise levels and/or large sparsity factors. 

\begin{figure}
\leftskip-3cm

\subfloat[$\|\protect\G\|_{0,\infty}\leq1$]{\includegraphics[width=0.7\textwidth]{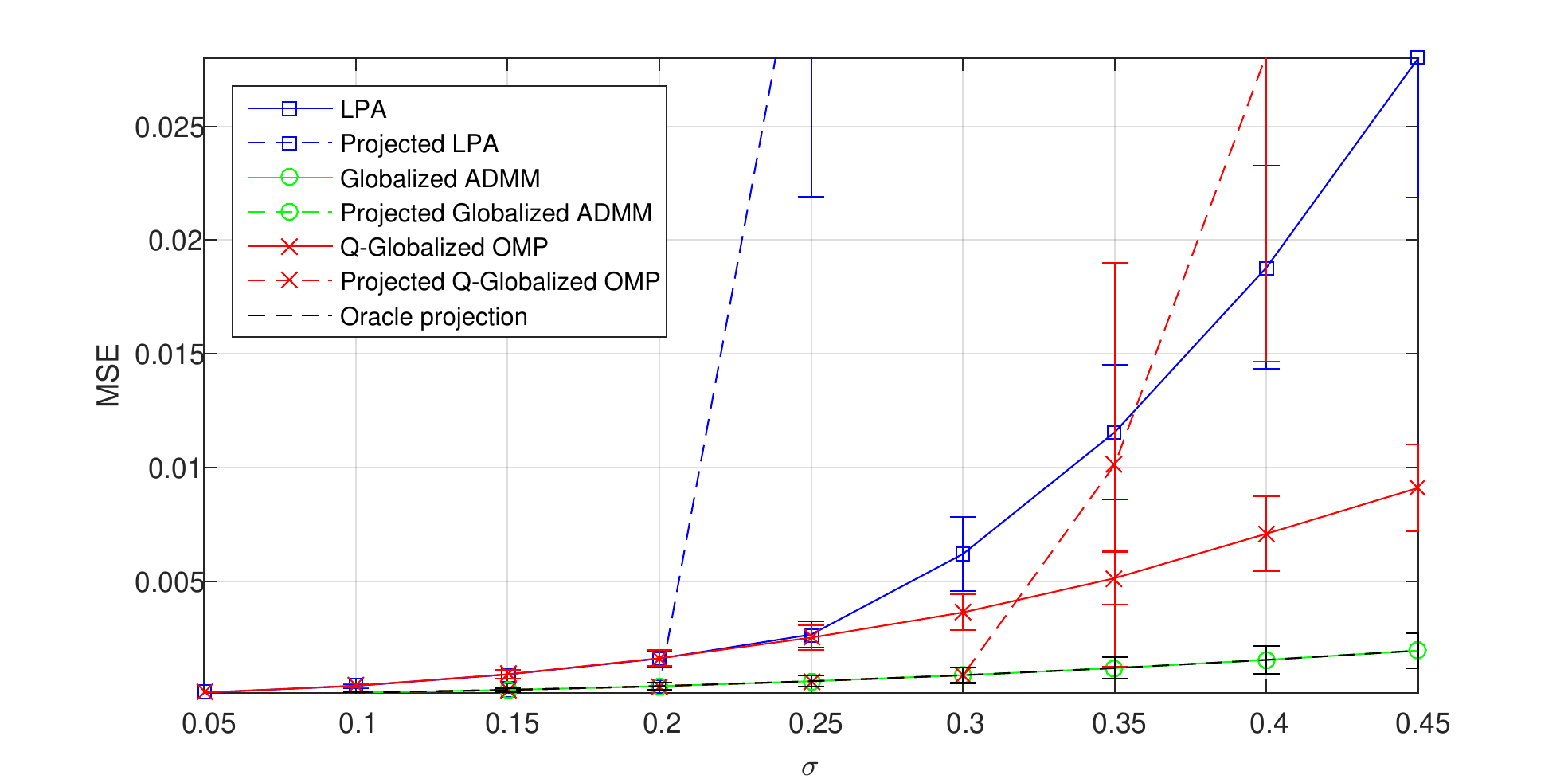}

}\subfloat[$\|\protect\G\|_{0,\infty}\leq2$]{\includegraphics[clip,width=0.7\textwidth]{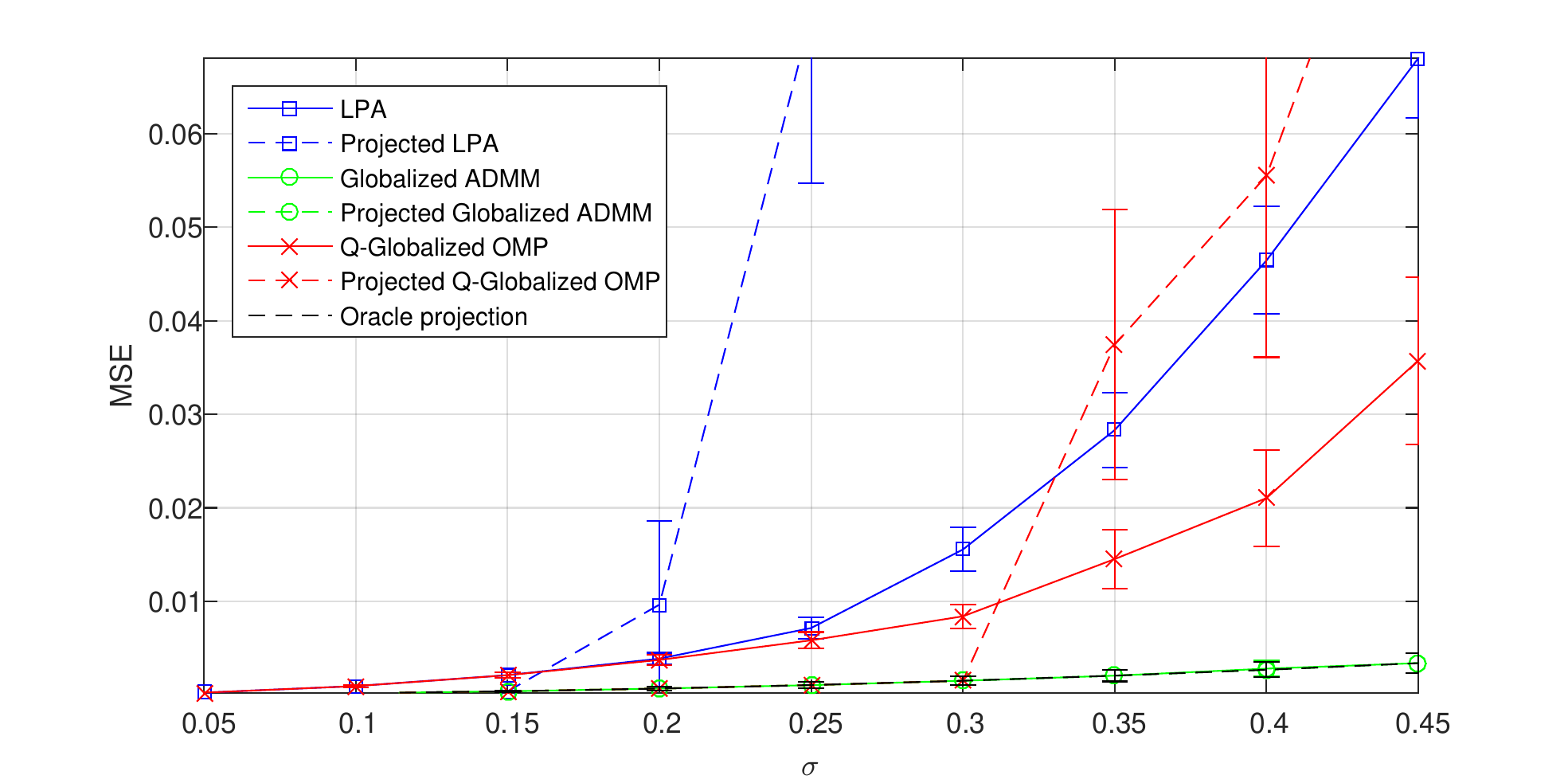}

}

\subfloat[$\|\protect\G\|_{0,\infty}\leq3$]{\includegraphics[clip,width=0.7\textwidth]{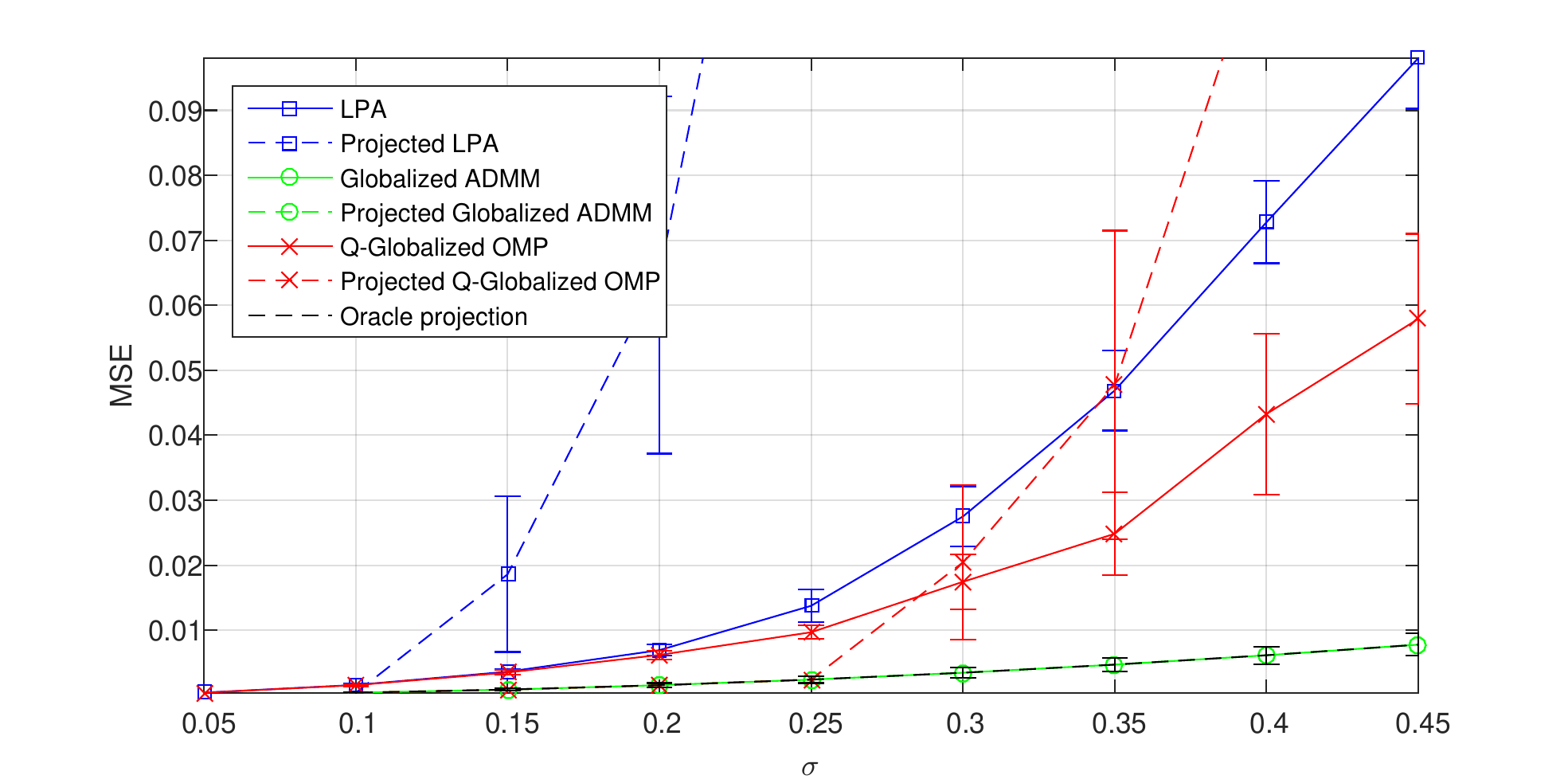}

}\subfloat[$\|\protect\G\|_{0,\infty}\leq4$]{\includegraphics[clip,width=0.7\textwidth]{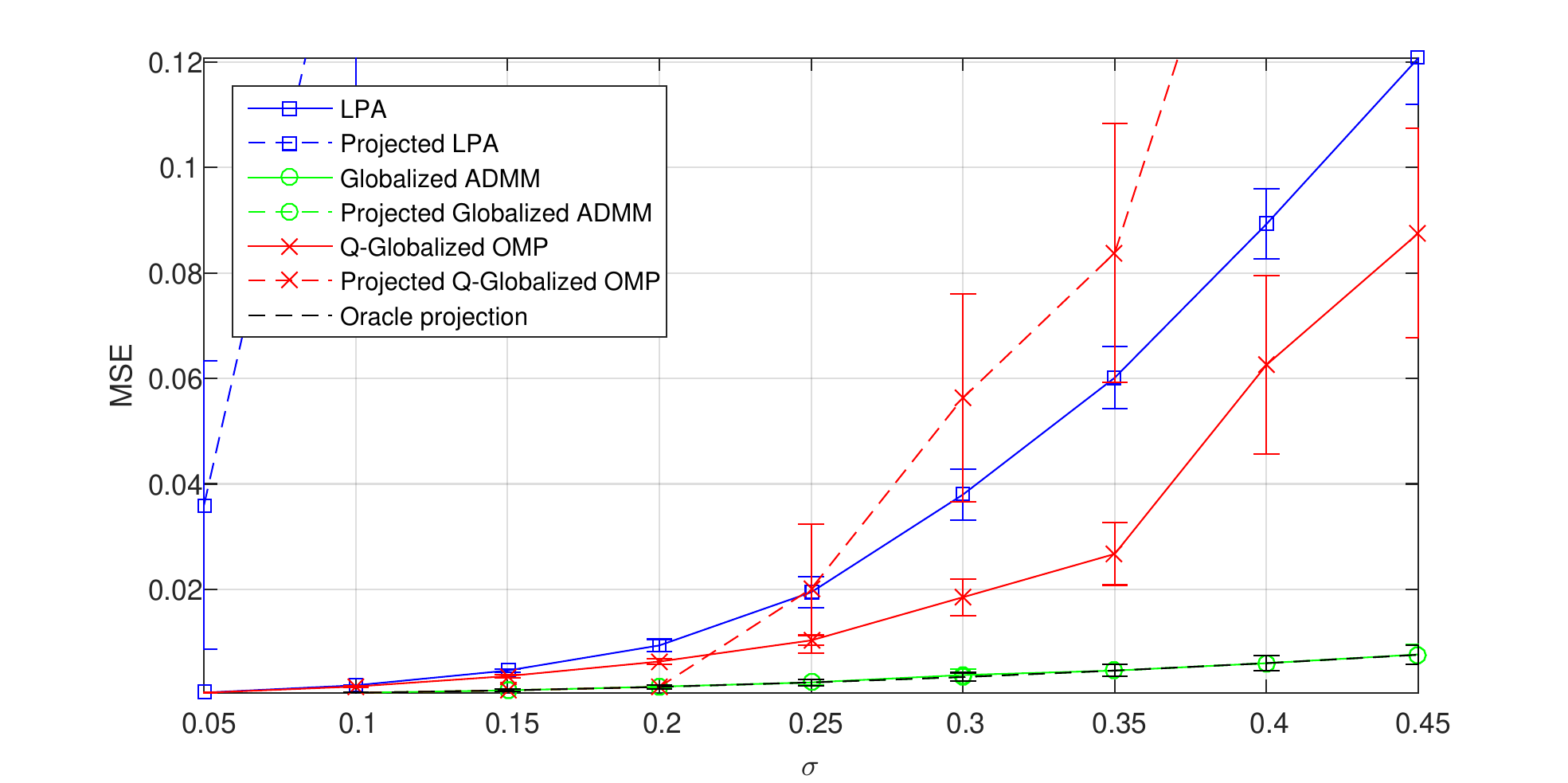}

}

\caption{Denoising performance of the globalized OMP, ADMM-pursuit and LPA
algorithm along with their projected versions for various noise levels
and sparsity factors. The projected version of the oracle estimator
is provided as well, demonstrating the best possible restoration that
can be achieved. The signals are drawn from the signature dictionary
model.}
\label{fig:denoising_k1}
\end{figure}

\subsection{\label{sub:pwc-numerical}Denoising PWC Signals}

In this scenario, we test the ability of the globalized ADMM-pursuit
to restore corrupted PWC signals, and compare these to the outcome
of the LPA algorithm. Similarly to the previous subsection, the projected
versions of the two pursuit algorithms are provided along with the
one of the oracle estimator. Following the description in Section
\ref{sub:Piecewise-constant-signals}, we generate a signal of length
$N=200$, composed of patches of size $n=m=20$ with a local sparsity
of at most $2$ non-zeros in the $\ell_{0,\infty}$-sense. These signals
are then contaminated by a white additive Gaussian noise with $\sigma$
in the range of $0.1$ to $0.9$.

\begin{figure}
\centering

\includegraphics[width=0.8\textwidth]{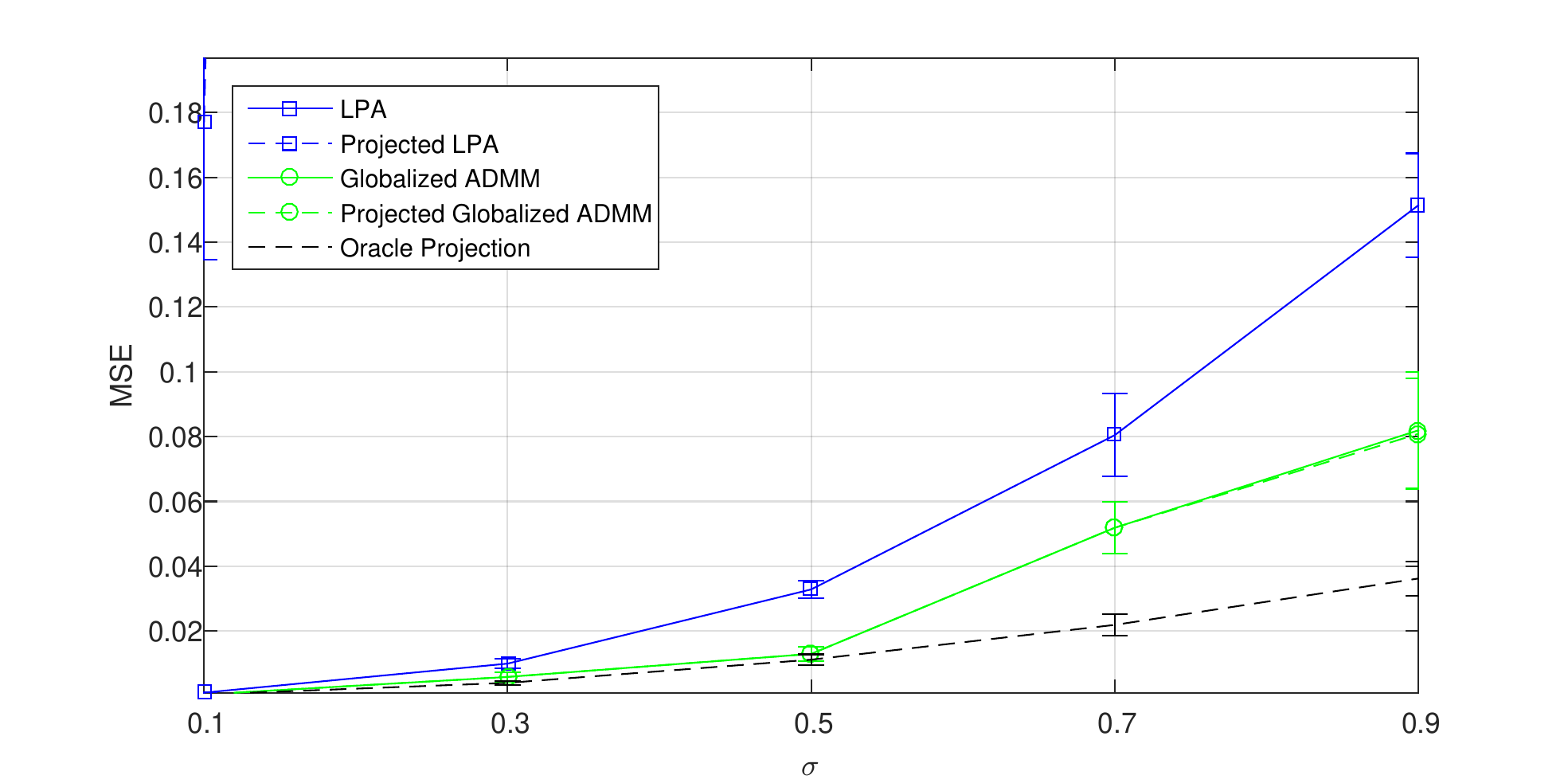}

\caption{Denoising performance of the ADMM-pursuit and the LPA algorithm for
various noise levels, tested for signals from the piecewise-constant
model with $\|\protect\G\|_{0,\infty}\leq2$.}

\label{fig:denoising_pwc}
\end{figure}

\begin{figure}
\centering

\includegraphics[width=0.8\textwidth]{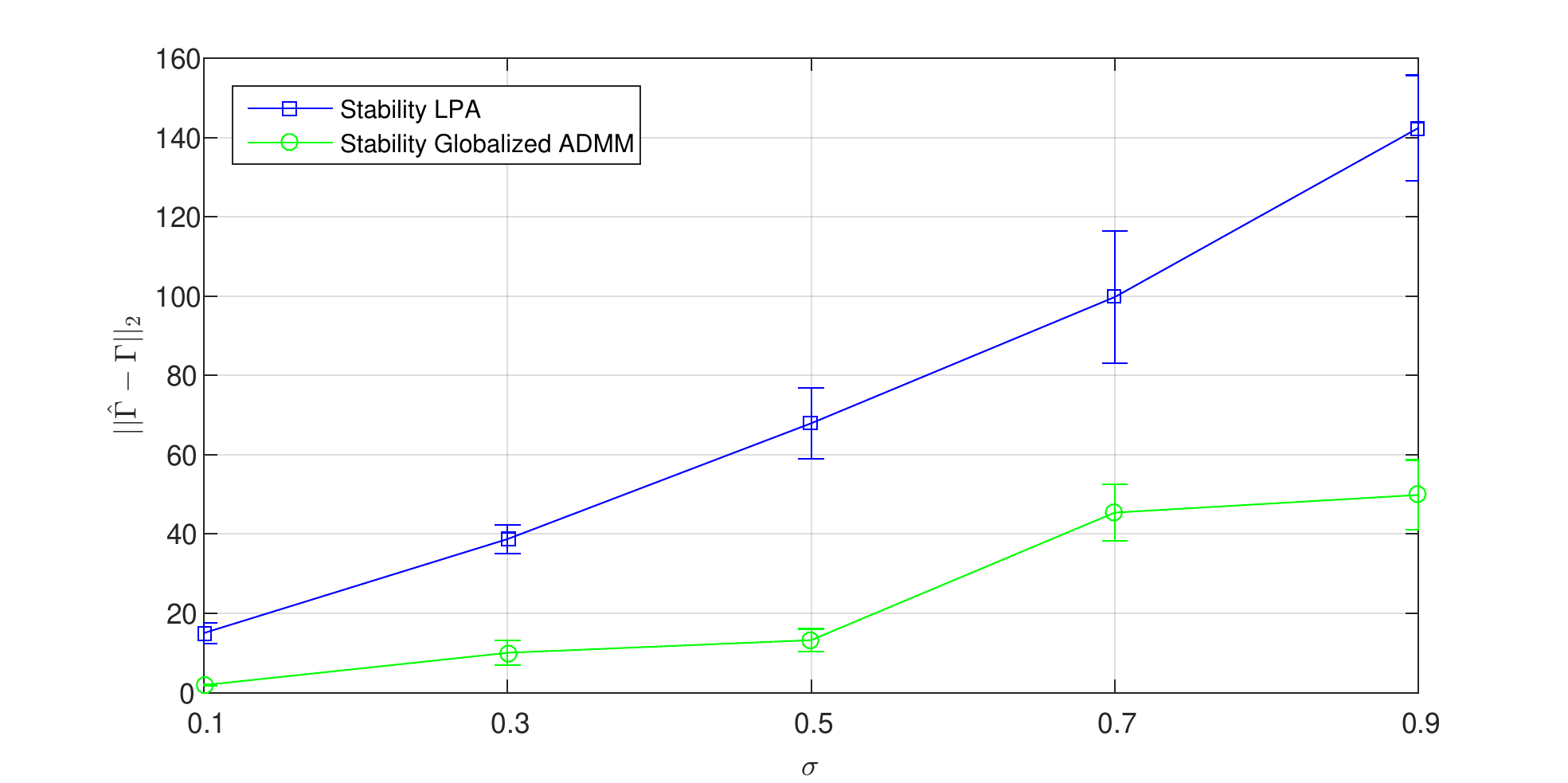}

\caption{Stability of the ADMM-pursuit and the LPA algorithm for various noise
levels, tested for signals from the piecewise-constant model with
$\|\protect\G\|_{0,\infty}\leq2$.}

\label{fig:stability_pwc}
\end{figure}

\begin{figure}
\centering

\includegraphics[width=0.8\textwidth]{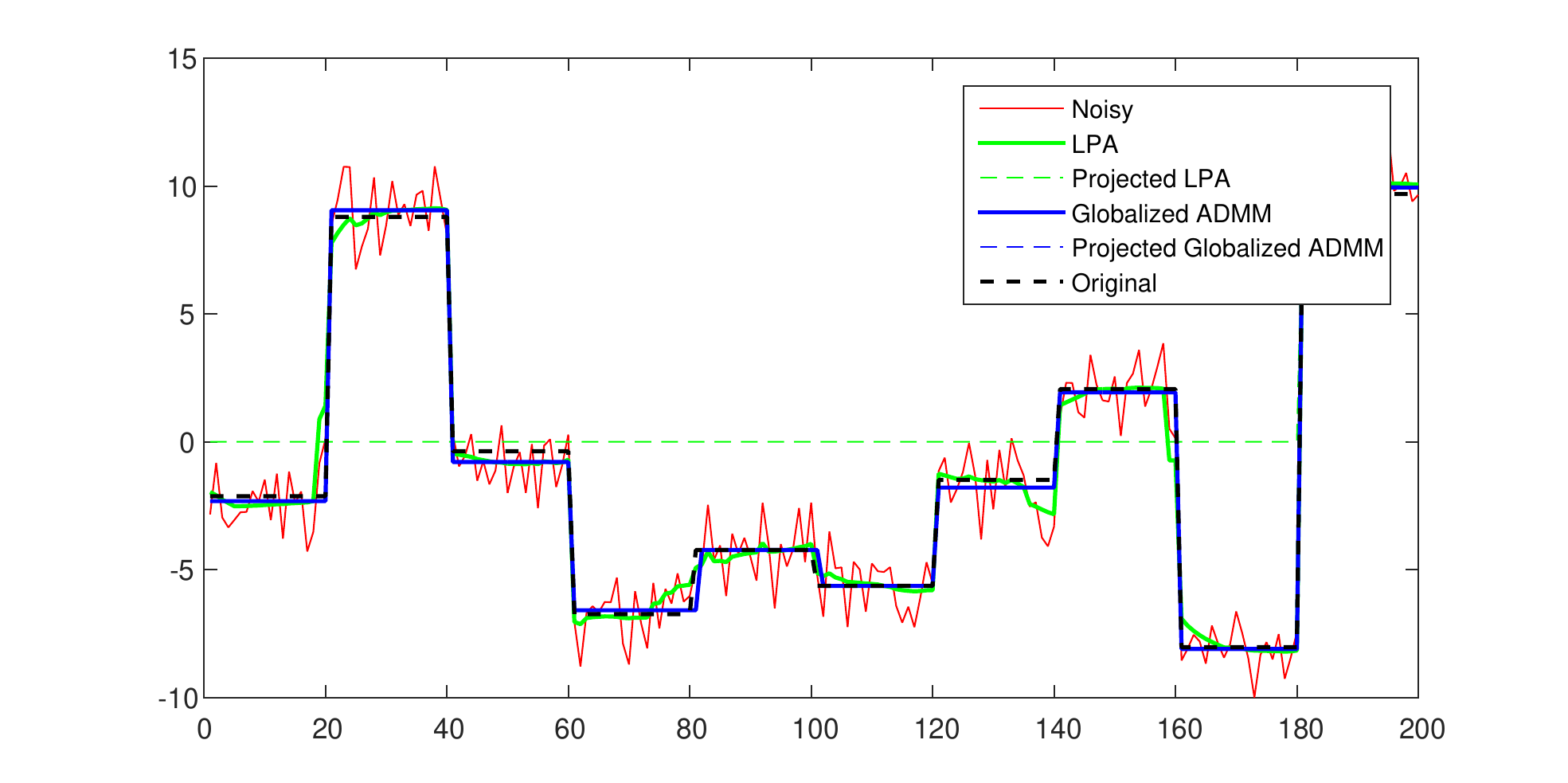}

\caption{Denoising of a PWC signal contaminated with additive Gaussian noise
($\sigma=1.1)$ via several pursuit algorithms: Input noisy signal
(MSE = 1.173), LPA algorithm (MSE = 0.328), projected LPA (MSE = 24.672),
ADMM-pursuit (MSE = 0.086), projected ADMM-pursuit (MSE = 0.086),
and oracle estimator (MSE = 0.047). }

\label{fig:vis_pwc}
\end{figure}

The restoration performance (in terms of MSE) of the above-mentioned
algorithms is illustrated in Figure \ref{fig:denoising_pwc} and the
stability of the estimates is demonstrated in Figure \ref{fig:stability_pwc},
where the results are averaged over 10 noise realizations. As can
be seen, the globalized approach significantly outperforms the LPA
algorithm for all noise levels. Furthermore, when $\sigma\leq0.5$,
the ADMM-pursuit performs similarly to the oracle estimator. One can
also notice that, similarly to the previous subsection, the ADMM-pursuit
and its projected version resulting in the very same estimation, i.e.
this algorithm forces the signal to conform with the patch-sparse
model globally. On the other hand, following the visual illustration
given in Figure \ref{fig:vis_pwc}, the projected version of the LPA
algorithm has zero segments, which are the consequence of a complete
disagreement in the support (local inconsistency). This is also reflected
in Figure \ref{fig:denoising_pwc}, illustrating that even for a very
small noise level ($\sigma=0.1$) the projected version of the LPA
algorithm has a very large estimation error ($\text{{MSE}}\approx0.18$)
compared to the one of the ADMM-pursuit ($\text{{MSE}}\approx0.0004$),
indicating that the former fails in obtaining a consistent representation
of the signal.

\section{\label{sec:Discussion}Discussion}

In this work we have presented an extension of the classical theory
of sparse representations to signals which are locally sparse, together
with novel pursuit algorithms. We envision several promising research
directions which might emerge from this work.

\subsection{Relation to other models}

Viewed globally, the resulting signal model can be considered a sort
of ``structured sparse'' model, however, in contrast to other such
constructions (\cite{tropp2006algorithms,huang2010benefit,huang2011learning,kyrillidis2015structured}
and others), our model incorporates both structure in the representation
coefficients and a structured dictionary. 

The recently developed framework of Convolutional Sparse Coding (CSC)
\cite{papyan2016working_1,papyan2016working_2,vardan2016convolutional}
bears some similarities to our work, in that it, too, has a convolutional
representation of the signal via a dictionary identical in structure
to $D_{G}$. However, the underlying local sparsity assumptions are
drastically different in the two models, resulting in very different
guarantees and algorithms. That said, we believe that it would be
important to provide precise connections between the results, possibly
leading to their deeper understanding. First steps in this direction
are outlined in \prettyref{sub:csc}.

\subsection{Further extensions}

The decomposition of the global signal $x\in\RR^{N}$ into its patches:
\begin{equation}
x\mapsto\left(R_{i}x\right)_{i=1}^{P},\label{eq:our-decomp}
\end{equation}
is a special case of a more general decomposition, namely
\begin{equation}
x\mapsto\left(w_{i}{\cal P}_{i}x\right)_{i=1}^{P},\label{eq:fusion-frames-decomp}
\end{equation}
where ${\cal P}_{i}$ is the (orthogonal) projection onto a subspace
$W_{i}$ of $\RR^{N}$, and $w_{i}$ are some weights. This observation
naturally places our theory, at least partially, into the framework
of \emph{fusion frames}, a topic which is generating much interest
recently in the applied harmonic analysis community \cite[Chapter 13]{finiteframes}.
In fusion frame theory, which is motivated by applications such as
distributed sensor networks, the starting point is precisely the decomposition
\prettyref{eq:fusion-frames-decomp}. Instead of the reconstruction
formula $x=\sum_{i}\frac{1}{n}R_{i}^{T}R_{i}x$, in fusion frame theory
we have
\[
x=\sum_{i}w_{i}^{2}S_{{\cal W}}^{-1}\left({\cal P}_{i}x\right),
\]
where $S_{{\cal W}}$ is the associated \emph{fusion frame operator}.
The natural extension of our work to this setting would seek to enforce
some sparsity of the projections ${\cal P}_{i}x$. Perhaps the most
immediate variant of \eqref{eq:our-decomp} in this respect would
be to drop the periodicity requirement, resulting in a slightly modified
$R_{i}$ operators near the endpoints of the signal. We would like
to mention some recent works which investigate different notions of
fusion frame sparsity \cite{boufounos2011sparserecovery,aceska2015localsparsity,ayaz2016uniform}.

Another intriguing possible extension of our work is to relax the
complete overlap requirement between patches and consider an ``approximate
patch sparsity'' model, where the patch disagreement vector $M\G$
is not zero but ``small''. In some sense, one can imagine a full
``spectrum'' of such models, ranging from a complete agreement (this
work) to an arbitrary disagreement (such as in the CSC framework mentioned
above). 

\subsection{Learning models from data}

The last point above brings us to the question of how to obtain ``good''
models, reflecting the structure of the signals at hand (such as speech/images
etc.) We hope that one might use the ideas presented here in order
to create novel learning algorithms. In this regard, the main difficulty
is how to parametrize the space of allowed models in an efficient
way. While we presented some initial ideas in \prettyref{sec:Examples},
in the most general case (incorporating the approximate sparsity direction
above) the problem remains widely open.
\begin{acknowledgement}
The research leading to these results has received funding from the
European Research Council under European Union\textquoteright s Seventh
Framework Programme, ERC Grant agreement no. 320649. The authors would
also like to thank Jeremias Sulam, Vardan Papyan, Raja Giryes and
Gitta Kutinyok for inspiring discussions.
\end{acknowledgement}

\section*{\label{app:proof-dimkerM}Appendix A: Proof of \prettyref{lem:dim-ker-M}}

\addcontentsline{toc}{section}{Appendix A}
\begin{svmultproof2}
Denote $Z:=\ker M$ and consider the linear map $A:Z\to\RR^{N}$ given
by the restriction of the ``averaging map'' $D_{G}:\RR^{mP}\to\RR^{N}$
to $Z$. 

\begin{enumerate}
\item Let us see first that $im\left(A\right)=\RR^{N}$. Indeed, for every
$x\in\RR^{N}$ consider its patches $x_{i}=R_{i}x$. Since $D$ is
full rank, there exist $\left\{ \alpha_{i}\right\} $ for which $D\alpha_{i}=x_{i}$.
Then setting $\G:=\left(\alpha_{1},\dots,\alpha_{P}\right)$ we have
both $D_{G}\G=x$ and $M\G=0$ (by construction, see \prettyref{sec:Local-global-sparsity}),
i.e. $\G\in Z$ and the claim follows.
\item Define
\[
J:=\ker D\times\ker D\times\dots\ker D\subset\RR^{mP}.
\]
We claim that $J=\ker A$.

\begin{enumerate}
\item In one direction, let $\G=\left(\alpha_{1},\dots,\alpha_{P}\right)\in\ker A$,
i.e. $M\G=0$ and $D_{G}\G=0$. Immediately we see that $\frac{1}{n}D\alpha_{i}=0$
for all $i$, and therefore $\alpha_{i}\in\ker D$ for all $i$, thus
$\G\in J$.
\item In the other direction, let $\G=\left(\alpha_{1},\dots,\alpha_{P}\right)\in J$,
i.e. $D\alpha_{i}=0$. Then the local representations agree, i.e.
$M\G=0$, thus $\G\in Z$. Furthermore, $D_{G}\G=0$ and therefore
$\G\in\ker A$.
\end{enumerate}
\item By the fundamental theorem of linear algebra we conclude
\begin{eqnarray*}
\dim Z & = & \dim im\left(A\right)+\dim\ker A=N+\dim J\\
 & = & N+\left(m-n\right)N=N\left(m-n+1\right).
\end{eqnarray*}
\end{enumerate}
\end{svmultproof2}

\section*{\label{app:proof1}Appendix B: Proof of \prettyref{lem:equiv-constraint}}

\addcontentsline{toc}{section}{Appendix B}

We start with an easy observation.
\begin{proposition}
\label{prop:f1}For any vector $\rho\in\RR^{N}$, we have
\[
\|\rho\|_{2}^{2}=\frac{1}{n}\sum_{j=1}^{N}\|R_{j}\rho\|_{2}^{2}.
\]
\end{proposition}
\begin{svmultproof2}
Since
\[
\|\rho\|_{2}^{2}=\sum_{j=1}^{N}\rho_{j}^{2}=\frac{1}{n}\sum_{j=1}^{N}n\rho_{j}^{2}=\frac{1}{n}\sum_{j=1}^{N}\sum_{k=1}^{n}\rho_{j}^{2},
\]
we can rearrange the sum and get
\begin{eqnarray*}
\|\rho\|_{2}^{2} & = & \frac{1}{n}\sum_{k=1}^{n}\sum_{j=1}^{N}\rho_{j}^{2}=\frac{1}{n}\sum_{k=1}^{n}\sum_{j=1}^{N}\rho_{\left(j+k\right)\mod N}^{2}=\frac{1}{n}\sum_{j=1}^{N}\sum_{k=1}^{n}\rho_{\left(j+k\right)\mod N}^{2}\\
 & = & \frac{1}{n}\sum_{j=1}^{N}\|R_{j}\rho\|_{2}^{2}.
\end{eqnarray*}
\end{svmultproof2}

\begin{corollary}
\label{cor:equiv-objective}Given $M\G=0$, we have
\[
\|y-D_{G}\G\|_{2}^{2}=\frac{1}{n}\sum_{j=1}^{N}\|R_{j}y-D\alpha_{j}\|_{2}^{2}.
\]
\end{corollary}
\begin{svmultproof2}
Using \prettyref{prop:f1}, we get
\begin{eqnarray*}
\|y-D_{G}\G\|_{2}^{2} & = & \frac{1}{n}\sum_{j=1}^{N}\|R_{j}y-R_{j}D_{G}\G\|_{2}^{2}=\frac{1}{n}\sum_{j=1}^{N}\|R_{j}y-\Omega_{j}\G\|_{2}^{2}.
\end{eqnarray*}
Now since $M\G=0$, then by definition of $M$ we have $\Omega_{j}\G=D\alpha_{j}$
(see \eqref{eq:q-omega}), and this completes the proof.
\end{svmultproof2}

Recall \prettyref{def:stsb-def} and \eqref{eq:ri-def}. Multiplying
the corresponding matrices gives
\begin{proposition}
We have the following equality for all $i=1,\dots P$:
\begin{equation}
S_{B}R_{i}=S_{T}R_{i+1}.\label{eq:str-eq}
\end{equation}
\end{proposition}
To facilitate the proof, we introduce extension of \prettyref{def:stsb-def}
to multiple shifts as follows.
\begin{definition}
\label{def:sbst-ext}Let $n$ be fixed. For $k=0,\dots,n-1$ let
\begin{enumerate}
\item $S_{T}^{\left(k\right)}:=\begin{bmatrix}I_{n-k} & \vec{0}\end{bmatrix}$
and $S_{B}^{\left(k\right)}:=\begin{bmatrix}\vec{0} & I_{n-k}\end{bmatrix}$
denote the operators extracting the top (resp. bottom) $n-k$ entries
from a vector of length $n$; the matrices have dimension $\left(n-k\right)\times n$.
\item $Z_{B}^{\left(k\right)}:=\begin{bmatrix}S_{B}^{\left(k\right)}\\
\vec{0}_{k\times n}
\end{bmatrix}$ and $Z_{T}^{\left(k\right)}:=\begin{bmatrix}\vec{0}_{k\times n}\\
S_{T}^{\left(k\right)}
\end{bmatrix}$ .
\item $W_{B}^{\left(k\right)}:=\begin{bmatrix}\vec{0}_{k\times n}\\
S_{B}^{\left(k\right)}
\end{bmatrix}$ and $W_{T}^{\left(k\right)}:=\begin{bmatrix}S_{T}^{\left(k\right)}\\
\vec{0}_{k\times n}
\end{bmatrix}$ .
\end{enumerate}
Note that $S_{B}=S_{B}^{\left(1\right)}$ and $S_{T}=S_{T}^{\left(1\right)}$.
We have several useful consequences of the above definitions. The
proofs are carried out via elementary matrix identities and are left
to the reader.
\end{definition}
\begin{proposition}
\label{prop:decomposition-single}For any $n\in\NN$ the following
hold:
\begin{enumerate}
\item \label{enu:sbst-z-fold}$Z_{T}^{\left(k\right)}=\left(Z_{T}^{\left(1\right)}\right)^{k}$
and $Z_{B}^{\left(k\right)}=\left(Z_{B}^{\left(1\right)}\right)^{k}$
for $k=0,\dots,n-1$;
\item $W_{T}^{\left(k\right)}W_{T}^{\left(k\right)}=W_{T}^{\left(k\right)}$
and $W_{B}^{\left(k\right)}W_{B}^{\left(k\right)}=W_{B}^{\left(k\right)}$
for $k=0,\dots,n-1$;
\item \label{enu:sbst-wtwb-commute}$W_{T}^{\left(k\right)}W_{B}^{\left(j\right)}=W_{B}^{\left(j\right)}W_{T}^{\left(k\right)}$
for $j,k=0,\dots,n-1$;
\item \label{enu:sbst-fold-right}$Z_{B}^{\left(k\right)}=Z_{B}^{\left(k\right)}W_{B}^{\left(k\right)}$
and $Z_{T}^{\left(k\right)}=Z_{T}^{\left(k\right)}W_{T}^{\left(k\right)}$
for $k=0,\dots,n-1$;
\item \label{enu:sbst-triple-identity}$W_{B}^{\left(k\right)}=Z_{T}^{\left(1\right)}W_{B}^{\left(k-1\right)}Z_{B}^{\left(1\right)}$
and $W_{T}^{\left(k\right)}=Z_{B}^{\left(1\right)}W_{T}^{\left(k-1\right)}Z_{T}$
for $k=1,\dots,n-1$;
\item \label{enu:sbst-zfold-w}$Z_{B}^{\left(k\right)}Z_{T}^{\left(k\right)}=W_{T}^{\left(k\right)}$
and $Z_{T}^{\left(k\right)}Z_{B}^{\left(k\right)}=W_{B}^{\left(k\right)}$
for $k=0,\dots,n-1$;
\item \label{enu:sbst-resolution-identity}$\left(n-1\right)I_{n\times n}=\sum_{k=1}^{n-1}\left(W_{B}^{\left(k\right)}+W_{T}^{\left(k\right)}\right).$
\end{enumerate}
\end{proposition}
\begin{minipage}[t]{1\columnwidth}%
\end{minipage}
\begin{proposition}
\label{prop:stb}If the vectors $u_{1},\dots,u_{N}\in\RR^{n}$ satisfy
pairwise
\[
S_{B}u_{i}=S_{T}u_{i+1},
\]
 then they also satisfy for each $k=0,\dots,n-1$ the following:
\begin{align}
W_{B}^{\left(k\right)}u_{i} & =Z_{T}^{\left(k\right)}u_{i+k},\label{eq:stb-first-claim}\\
Z_{B}^{\left(k\right)}u_{i} & =W_{T}^{\left(k\right)}u_{i+k}.\label{eq:stb-second-claim}
\end{align}
\end{proposition}
\begin{svmultproof2}
It is easy to see that the condition $S_{B}u_{i}=S_{T}u_{i+1}$ directly
implies
\begin{align}
Z_{B}^{\left(1\right)}u_{i} & =W_{T}^{\left(1\right)}u_{i+1},\quad W_{B}^{\left(1\right)}u_{i}=Z_{T}^{\left(1\right)}u_{i+1}\quad\forall i.\label{eq:sbst-given-rewrite}
\end{align}

Let us first prove \eqref{eq:stb-first-claim} by induction on $k$.
The base case $k=1$ is precisely \eqref{eq:sbst-given-rewrite}.
Assuming validity for $k-1$ and $\forall i$, we have
\begin{eqnarray*}
\begin{split}W_{B}^{\left(k\right)}u_{i}= & Z_{T}^{\left(1\right)}W_{B}^{\left(k-1\right)}Z_{B}^{\left(1\right)}u_{i} & \left(\text{by Proposition \ref{prop:decomposition-single}, item \ref{enu:sbst-triple-identity}}\right)\\
= & Z_{T}^{\left(1\right)}W_{B}^{\left(k-1\right)}W_{T}^{\left(1\right)}u_{i+1} & \left(\text{by \eqref{eq:sbst-given-rewrite}}\right)\\
= & Z_{T}^{\left(1\right)}W_{T}^{\left(1\right)}W_{B}^{\left(k-1\right)}u_{i+1} & \left(\text{by Proposition \ref{prop:decomposition-single}, item \ref{enu:sbst-wtwb-commute}}\right)\\
= & Z_{T}^{\left(1\right)}W_{T}^{\left(1\right)}Z_{T}^{\left(k-1\right)}u_{i+k} & \left(\text{by the induction hypothesis}\right)\\
= & Z_{T}^{\left(1\right)}Z_{T}^{\left(k-1\right)}u_{i+k} & \left(\text{by Proposition \ref{prop:decomposition-single}, item \ref{enu:sbst-fold-right}}\right)\\
= & Z_{T}^{\left(k\right)}u_{i+k}. & \left(\text{by Proposition \ref{prop:decomposition-single}, item \ref{enu:sbst-z-fold}}\right)
\end{split}
\end{eqnarray*}

To prove \eqref{eq:stb-second-claim} we proceed as follows:
\begin{eqnarray*}
\begin{split}Z_{B}^{\left(k\right)}u_{i} & =Z_{B}^{\left(k\right)}W_{B}^{\left(k\right)}u_{i} & \left(\text{by Proposition \ref{prop:decomposition-single}, item \ref{enu:sbst-fold-right}}\right)\\
 & =Z_{B}^{\left(k\right)}Z_{T}^{\left(k\right)}u_{i+k} & \left(\text{by \eqref{eq:stb-first-claim} which is already proved}\right)\\
 & =W_{T}^{\left(k\right)}u_{i+k}. & \left(\text{by Proposition \ref{prop:decomposition-single}, item \ref{enu:sbst-zfold-w}}\right)
\end{split}
\end{eqnarray*}
This finishes the proof of \prettyref{prop:stb}.

\end{svmultproof2}

\begin{figure}
\begin{centering}
\def\svgwidth{3.5cm}
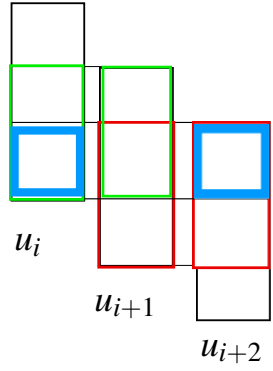
\par\end{centering}
\caption{Illustration to the proof of \prettyref{prop:stb}. The green pair
is equal, as well as the red pair. It follows that the blue elements
are equal as well.}
\label{fig:proof-of-overlaps}
\end{figure}

\begin{example}
To help the reader understand the claim of \prettyref{prop:stb},
consider the case $k=2$, and take some three vectors $u_{i},u_{i+1},u_{i+2}.$
We have $S_{B}u_{i}=S_{T}u_{i+1}$ and also $S_{B}u_{i+1}=S_{T}u_{i+2}$.
Then clearly $S_{B}^{\left(2\right)}u_{i}=S_{T}^{\left(2\right)}u_{i+2}$
(see \prettyref{fig:proof-of-overlaps}) and therefore $W_{B}^{\left(2\right)}u_{i}=Z_{T}^{\left(2\right)}u_{i+2}$. 
\end{example}

Let us now present the proof of \prettyref{lem:equiv-constraint}.
\begin{svmultproof2}
We show equivalence in two directions.

\begin{itemize}
\item $\left(1\right)\Longrightarrow\left(2\right)$: Let $M\G=0$. Define
$x:=D_{G}\G$, and then further denote $x_{i}:=R_{i}x$. Then on the
one hand:
\begin{eqnarray*}
\begin{split}x_{i} & =R_{i}D_{G}\G\\
 & =\Omega_{i}\G & \text{(definition of }\Omega_{i})\\
 & =D\alpha_{i}. & \left(M\G=0\right)
\end{split}
\end{eqnarray*}
On the other hand, because of \eqref{eq:str-eq} we have $S_{B}R_{i}x=S_{T}R_{i+1}x$,
and by combining the two we conclude that $S_{B}D\alpha_{i}=S_{T}D\alpha_{i+1}$.
\item $\left(2\right)\Longrightarrow\left(1\right)$: In the other direction,
suppose that $S_{B}D\alpha_{i}=S_{T}D\alpha_{i+1}$. Denote $u_{i}:=D\alpha_{i}$.
Now consider the product $\Omega_{i}\G$ where $\Omega_{i}=R_{i}D_{G}$.
One can easily be convinced  that in fact 
\[
\Omega_{i}\G=\frac{1}{n}\left(\sum_{k=1}^{n-1}\left(Z_{B}^{\left(k\right)}u_{i-k}+Z_{T}^{\left(k\right)}u_{i+k}\right)+u_{i}\right).
\]
Therefore
\begin{eqnarray*}
\begin{split}\left(\Omega_{i}-Q_{i}\right)\G & =\frac{1}{n}\left(u_{i}+\sum_{k=1}^{n-1}\left(Z_{B}^{\left(k\right)}u_{i-k}+Z_{T}^{\left(k\right)}u_{i+k}\right)\right)-u_{i}\\
 & =\frac{1}{n}\left(\sum_{k=1}^{n-1}\left(W_{T}^{\left(k\right)}u_{i}+W_{B}^{k}u_{i}\right)-\left(n-1\right)u_{i}\right) & \left(\text{by Proposition \ref{prop:stb}}\right)\\
 & =0. & \left(\text{by Proposition \ref{prop:decomposition-single}, item \ref{enu:sbst-resolution-identity}}\right)
\end{split}
\end{eqnarray*}
Since this holds for all $i$, we have shown that $M\G=0$. 
\end{itemize}
\end{svmultproof2}

\section*{\label{app:proof-of-lpa-iters}Appendix C: Proof of \prettyref{thm:lpa-iters}}

\addcontentsline{toc}{section}{Appendix C}

Recall that $M_{A}=\frac{1}{n}\sum_{i}R_{i}^{T}P_{s_{i}}R_{i}$. We
first show that $M_{A}$ is a contraction.
\begin{proposition}
\label{prop:ma-contraction}$\left\Vert M_{A}\right\Vert _{2}\leqslant1$.
\end{proposition}
\begin{svmultproof2}
Closely following a similar proof in \cite{romano2015boosting}, divide
the index set $\left\{ 1,\dots,N\right\} $ into $n$ groups representing
\emph{non-overlapping }patches: for $i=1,\dots,n$ let
\[
K\left(i\right):=\left\{ i,i+n,\dots,i+\left(\left\lfloor \frac{N}{n}\right\rfloor -1\right)n\right\} \;\mod N.
\]
Now
\begin{align*}
\left\Vert M_{A}x\right\Vert _{2} & =\frac{1}{n}\left\Vert \sum_{i=1}^{N}R_{i}^{T}P_{s_{i}}R_{i}x\right\Vert _{2}\\
 & =\frac{1}{n}\left\Vert \sum_{i=1}^{n}\sum_{j\in K\left(i\right)}R_{j}^{T}P_{s_{j}}R_{j}x\right\Vert _{2}\\
 & \leqslant\frac{1}{n}\sum_{i=1}^{n}\left\Vert \sum_{j\in K\left(i\right)}R_{j}^{T}P_{j}R_{j}x\right\Vert _{2}.
\end{align*}
By construction, $R_{j}R_{k}^{T}=\boldsymbol{0}_{n\times n}$ for
$j,k\in K\left(i\right)$ and $j\neq k$. Therefore for all $i=1,\dots,n$
we have
\begin{align*}
\left\Vert \sum_{j\in K\left(i\right)}R_{j}^{T}P_{s_{j}}R_{j}x\right\Vert _{2}^{2} & =\sum_{j\in K\left(i\right)}\left\Vert R_{j}^{T}P_{s_{j}}R_{j}x\right\Vert _{2}^{2}\\
 & \leqslant\sum_{j\in K\left(i\right)}\left\Vert R_{j}x\right\Vert _{2}^{2}\leqslant\left\Vert x\right\Vert _{2}^{2}.
\end{align*}
Substituting in back into the preceding inequality finally gives
\[
\left\Vert M_{A}x\right\Vert _{2}\leqslant\frac{1}{n}\sum_{i=1}^{n}\left\Vert x\right\Vert _{2}=\left\Vert x\right\Vert _{2}.
\]
\end{svmultproof2}

\begin{minipage}[t]{1\columnwidth}%
\end{minipage}

Now let us move on to prove \prettyref{thm:lpa-iters}.
\begin{svmultproof2}
Define
\[
\hat{P}_{i}:=\left(I-P_{s_{i}}\right)R_{i}.
\]
It is easy to see that
\[
\sum_{i}\hat{P}_{i}^{T}\hat{P}_{i}=A_{{\cal S}}^{T}A_{{\cal S}}.
\]
Let the SVD of $A_{{\cal S}}$ be
\[
A_{{\cal S}}=U\Sigma V^{T}.
\]
Now
\begin{eqnarray*}
V\Sigma^{2}V^{T}=A_{{\cal S}}^{T}A_{{\cal S}}=\sum_{i}\hat{P}_{i}^{T}\hat{P}_{i} & = & \sum_{i}R_{i}^{T}R_{i}-\underbrace{\sum_{i}R_{i}^{T}P_{s_{i}}R_{i}}_{:=T}\\
 & = & nI-T.
\end{eqnarray*}
 Therefore $T=nI-V\Sigma V^{T}$, and
\begin{eqnarray*}
M_{A} & = & \frac{1}{n}T=I-\frac{1}{n}V\Sigma^{2}V^{T}=V\left(I-\frac{\Sigma^{2}}{n}\right)V^{T}.
\end{eqnarray*}
This shows that the eigenvalues of $M_{A}$ are $\tau_{i}=1-\frac{\sigma_{i}^{2}}{n}$
where $\left\{ \sigma_{i}\right\} $ are the singular values of $A_{{\cal S}}$.
Thus we obtain
\begin{eqnarray*}
M_{A}^{k} & = & V\diag\left\{ \tau_{i}^{k}\right\} V^{T}.
\end{eqnarray*}
If $\sigma_{i}=0$ then $\tau_{i}=1$, and in any case, by \prettyref{prop:ma-contraction}
we have $\left|\tau_{i}\right|\leqslant1$. Let the columns of the
matrix $W$ consist of the singular vectors of $A_{{\cal S}}$ corresponding
to $\sigma_{i}=0$ (and so $\spn W=\nullsp{A_{{\cal S}}}$), then
\[
\lim_{k\to\infty}M_{A}^{k}=WW^{T}.
\]
Thus, as $k\to\infty$,  $M_{A}^{k}$ tends to the orthogonal projector
onto $\nullsp{A_{{\cal S}}}$. 
\end{svmultproof2}

\section*{\label{sec:lpa-pwc-perf-proof}Appendix D: Proof of \prettyref{thm:pwc-lpa-performance}}

\addcontentsline{toc}{section}{Appendix D}

Recall that the signal consists of $s$ constant segments of corresponding
lengths $\ell_{1},\dots,\ell_{s}$. We would like to compute the MSE
for every pixel within every such segment of length $\alpha:=\ell_{r}$.
For each patch, the oracle provides the locations of the jump points
within the patch.

\begin{figure}
\centering\definecolor{qqttzz}{rgb}{0.,0.2,0.6}
\definecolor{uuuuuu}{rgb}{0.26666666666666666,0.26666666666666666,0.26666666666666666}
\definecolor{ffwwqq}{rgb}{1.,0.4,0.}
\definecolor{qqqqff}{rgb}{0.,0.,1.}
\begin{tikzpicture}[line cap=round,line join=round,>=triangle 45,x=1.0cm,y=1.0cm]
\clip(-4.,-3.) rectangle (4.,3.);
\draw [line width=2.4pt] (-2.,0.)-- (2.010124146417792,0.);
\draw [line width=2.4pt] (-0.03691261665051161,0.09921290330405406) -- (-0.03691261665051161,-0.09921290330405406);
\draw [line width=2.4pt] (0.047036763068303504,0.09921290330405406) -- (0.047036763068303504,-0.09921290330405406);
\draw [line width=1.6pt,dash pattern=on 1pt off 1pt on 3pt off 4pt,color=ffwwqq] (-3.042102160296354,2.0110532917649464)-- (1.,2.);
\draw [line width=0.4pt,dash pattern=on 3pt off 3pt,color=uuuuuu] (1.,-3.) -- (1.,3.);
\draw [line width=0.4pt,dash pattern=on 3pt off 3pt,color=uuuuuu] (-2.,-3.) -- (-2.,3.);
\draw [line width=1.6pt,color=qqttzz] (-2.,0.)-- (-2.0054690399568864,-1.9999850446891443);
\draw [line width=1.6pt,color=qqttzz] (-2.0054690399568864,-1.9999850446891443)-- (-4.,-2.);
\draw [line width=1.6pt,color=qqttzz] (2.010124146417792,0.)-- (2.,1.);
\draw [line width=1.6pt,color=qqttzz] (2.,1.)-- (4.,1.);
\begin{scriptsize}
\draw [fill=qqqqff] (-2.,0.) circle (2.5pt);
\draw[color=qqqqff] (-2.4315612168867893,-0.4463740054585484) node {$-k$};
\draw [fill=qqqqff] (2.010124146417792,0.) circle (2.5pt);
\draw[color=qqqqff] (1.994860622832553,-0.6142727648961783) node {$\alpha-k-1$};
\draw [fill=qqqqff] (0.,0.) circle (2.5pt);
\draw[color=qqqqff] (-0.0657150611747272,0.3168021738034059) node {$O$};
\draw [fill=qqqqff] (-3.042102160296354,2.0110532917649464) circle (2.5pt);
\draw [fill=qqqqff] (1.,2.) circle (2.5pt);
\draw [fill=uuuuuu] (-1.99450852707683,2.0081886046244746) circle (1.5pt);
\draw[color=uuuuuu] (-2.2941895046196374,1.6447287257192067) node {$a_j$};
\draw [fill=uuuuuu] (1.,0.) circle (1.5pt);
\draw[color=uuuuuu] (1.2316844435705974,0.2557480794624496) node {$b_j$};
\end{scriptsize}
\end{tikzpicture}

\caption{The oracle estimator for the pixel $O$ in the segment (black). The
orange line is patch number $j=1,\dots,n$, and the relevant pixels
are between $a_{j}$ and $b_{j}$. The signal itself is shown to extend
beyond the segment (blue line).}

\label{fig:pwc-illustration}
\end{figure}
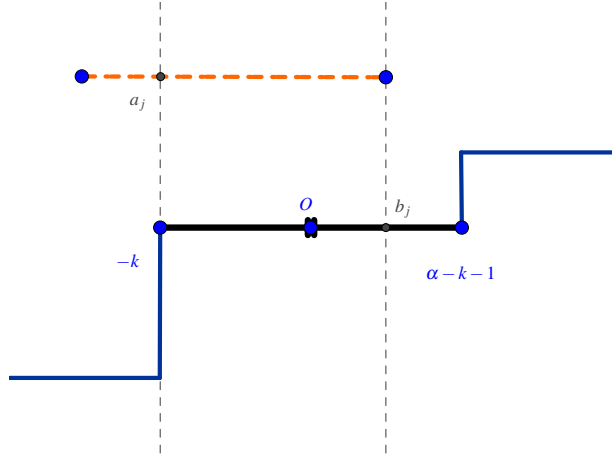

Let us calculate the MSE for pixel with index $0$ inside a constant
(\textbf{nonzero}) segment $\left[-k,\alpha-k-1\right]$ with value
$v$ (\prettyref{fig:pwc-illustration} might be useful). The oracle
estimator has the explicit formula
\begin{eqnarray}
\hat{x}_{A}^{r,k} & = & \frac{1}{n}\sum_{j=1}^{n}\frac{1}{b_{j}-a_{j}+1}\sum_{i=a_{j}}^{b_{j}}(v+z_{i}),\label{eq:pixel-oracle-est}
\end{eqnarray}
where $j=1,\dots,n$ corresponds to the index of the overlapping patch
containing the pixel, intersecting the constant segment on $\left[a_{j},b_{j}\right]$,
so that
\begin{align*}
a_{j} & =-\min\left(k,n-j\right),\\
b_{j} & =\min\left(\alpha-k-1,j-1\right).
\end{align*}

Now, the oracle error for the pixel is
\begin{align*}
\hat{x}_{A}^{r,k}-v & =\frac{1}{n}\sum_{j=1}^{n}\frac{1}{b_{j}-a_{j}+1}\sum_{i=a_{j}}^{b_{j}}z_{i}\\
 & =\sum_{i=-k}^{\alpha-k-1}c_{i,\alpha,n,k}z_{i},
\end{align*}
where the coefficients $c_{i,\alpha,n,k}$ are some \emph{positive}
rational numbers depending only on $i,\alpha,n$ and $k$. It is easy
to check by rearranging the above expression that
\begin{equation}
\sum_{i=-k}^{\alpha-k-1}c_{i,\alpha,n,k}=1,\label{eq:c-iank-sumto1}
\end{equation}
and furthermore, denoting $d_{i}:=c_{i,\alpha,n,k}$ for fixed $\alpha,n,k$,
we also have that
\begin{equation}
d_{-k}<d_{-k+1}<\dots d_{0}>d_{1}>\dots d_{\alpha-k-1}.\label{eq:c-iank-monotonic}
\end{equation}

\begin{example}
$n=4,\;\alpha=3$
\begin{itemize}
\item For $k=1$:{\footnotesize{}
\begin{eqnarray*}
\hat{x}_{A}^{r,k}-v & = & \frac{1}{4}\left(\frac{1}{2}+\frac{1}{3}+\frac{1}{3}+\frac{1}{2}\right)z_{0}+\frac{1}{4}\left(\frac{1}{2}+\frac{1}{3}+\frac{1}{3}\right)z_{-1}+\frac{1}{4}\left(\frac{1}{3}+\frac{1}{3}+\frac{1}{2}\right)z_{1}\\
 & = & \underbrace{\frac{7}{24}}_{d_{-1}}z_{-1}+\underbrace{\frac{5}{12}}_{d_{0}}z_{0}+\underbrace{\frac{7}{24}}_{d_{1}}z_{1}
\end{eqnarray*}
}{\footnotesize \par}
\item For $k=2$:{\footnotesize{}
\begin{eqnarray*}
\hat{x}_{A}^{r,k}-v & = & \frac{1}{4}\left(\frac{1}{3}+\frac{1}{3}+\frac{1}{2}+1\right)z_{0}+\frac{1}{4}\left(\frac{1}{3}+\frac{1}{3}+\frac{1}{2}\right)z_{-1}+\frac{1}{4}\left(\frac{1}{3}+\frac{1}{3}\right)z_{-2}\\
 & = & \frac{13}{24}z_{0}+\frac{7}{24}z_{-1}+\frac{1}{6}z_{-2}
\end{eqnarray*}
}{\footnotesize \par}
\end{itemize}
\end{example}
Now consider the optimization problem
\[
\min_{c\in\RR^{\alpha}}c^{T}c\quad\text{s.t}\;\mathbf{1}^{T}c=1.
\]
It can be easily verified that it has the optimal value $\frac{1}{\alpha}$,
attained at $c^{*}=\alpha\mathbf{1}$. From this, \eqref{eq:c-iank-sumto1}
and \eqref{eq:c-iank-monotonic} it follows that
\[
\sum_{i=-k}^{\alpha-k-1}c_{i,\alpha,n,k}^{2}>\frac{1}{\alpha}.
\]

Since the $z_{i}$ are i.i.d., we have
\[
\mathbb{E}\left(\hat{x}_{A}^{r,k}-v\right)^{2}=\sigma^{2}\sum_{i=-k}^{\alpha-k-1}c_{i,\alpha,n,k}^{2},
\]
while for the entire nonzero segment of length $\alpha=\ell_{r}$
\[
E_{r}:=\mathbb{E}\left(\sum_{k=0}^{\alpha-1}\left(\hat{x}_{A}^{r,k}-v\right)^{2}\right)=\sum_{k=0}^{\alpha-1}\mathbb{E}\left(\hat{x}_{A}^{r,k}-v\right)^{2}=\sigma^{2}\sum_{k=0}^{\alpha-1}\sum_{i=-k}^{\alpha-k-1}c_{i,\alpha,n,k}^{2}.
\]
Defining
\[
R\left(n,\alpha\right):=\sum_{k=0}^{\alpha-1}\sum_{i=-k}^{\alpha-k-1}c_{i,\alpha,n,k}^{2},
\]
we obtain that $R\left(n,\alpha\right)>1$ and furthermore
\[
\mathbb{E}\left\Vert \hat{x}_{A}-x\right\Vert ^{2}=\sum_{r=1}^{s}E_{r}=\sigma^{2}\sum_{r=1}^{s}R\left(n,\ell_{r}\right)>s\sigma^{2}.
\]
This proves item $\left(1\right)$ of \prettyref{thm:pwc-lpa-performance}.
For showing the explicit formulas for $R\left(n,\alpha\right)$ in
item $\left(2\right)$, we have used automatic symbolic simplification
software MAPLE \cite{maplesoft}. 

By construction \eqref{eq:pixel-oracle-est}, it is not difficult
to see that if $n\geqslant\alpha$ then
\begin{align*}
R(n,\alpha) & =\frac{1}{n^{2}}\sum_{k=0}^{\alpha-1}\Bigl(\sum_{j=0}^{k}\bigl(2H_{\alpha-1}-H_{k}+\frac{n-\alpha+1}{\alpha}-H_{\alpha-1-j}\bigr)^{2}\\
 & +\sum_{j=k+1}^{\alpha-1}\bigl(2H_{\alpha-1}-H_{\alpha-k-1}+\frac{n-\alpha+1}{\alpha}-H_{j}\bigr)^{2}\Bigr),
\end{align*}
where $H_{k}:=\sum_{i=1}^{k}\frac{1}{i}$ is the $k$-th harmonic
number. This simplifies to
\[
R(n,\alpha)=1+\frac{\alpha(2\alpha H_{\alpha}^{(2)}+2-3\alpha)-1}{n^{2}},
\]
where $H_{k}^{\left(2\right)}=\sum_{i=1}^{k}\frac{1}{i^{2}}$ is the
$k$-th harmonic number of the second kind.

On the other hand, for $n\leqslant\frac{\alpha}{2}$ we have
\[
R(n,\alpha)=\sum_{k=0}^{n-2}c_{n,k}^{(1)}+\sum_{k=n-1}^{\alpha-n}c_{n,k}^{(2)}+\sum_{k=\alpha-n+1}^{\alpha-1}c_{n,\alpha-1-k}^{(1)},
\]
where
\[
c_{n,k}^{(1)}=\frac{1}{n^{2}}\Biggl(\sum_{j=k}^{n-1}\bigl(H_{n-1}-H_{j}+\frac{k+1}{n}\bigr)^{2}+\sum_{i=n-k}^{n-1}\bigl(\frac{n-i}{n}\bigr)^{2}+\sum_{i=0}^{k-1}\bigl(H_{n-1}-H_{k}+\frac{k-i}{n}\bigr)^{2}\Biggr)
\]
 and
\[
c_{n,k}^{(2)}=\frac{1}{n^{2}}\Biggl(\sum_{j=k-n+1}^{k}\biggl(\frac{j-k+n}{n}\biggr)^{2}+\sum_{j=k+1}^{k+n-1}\biggl(\frac{k+n-j}{n}\biggr)^{2}\Biggr).
\]
Automatic symbolic simplification of the above gives
\[
R\left(n,\alpha\right)=\frac{11}{18}+\frac{2\alpha}{3n}-\frac{5}{18n^{2}}+\frac{\alpha-1}{3n^{3}}.
\]

\bibliographystyle{spmpsci}
\bibliography{local_global}

\end{document}